\documentclass[10pt]{iopart}
\usepackage{graphicx} 
\usepackage{xcolor} 
\usepackage{comment}
\usepackage{comment}
\usepackage{xcolor}

\begin{document}


\title{Green's Function Methods for Computing
  Supercurrents in Josephson Junctions}

\author{Eduardo R. Mucciolo\footnote[1]{corresponding author: eduardo.mucciolo@ucf.edu.}$^1$}

\author{Jouko Nieminen$^{2,3,4}$}

\author{Xiao Xiao$^{3,4}$}

\author{Wei-Chi Chiu$^{3,4}$}

\author{Michael N. Leuenberger$^{1,4,5,6}$}

\author{Arun Bansil$^{3,4}$}

\address{$^1$ Department of Physics, University of Central Florida, Orlando, FL 32789, USA}

\address{$^2$ Computational Physics Laboratory, Tampere University, Tampere 33014, Finland}

\address{$^3$ Department of Physics, Northeastern University, Boston, MA 02115, USA}

\address{$^4$ Quantum Materials and Sensing Institute, Northeastern University, Burlington, MA 01803, USA}

\address{$^5$ NanoScience Technology Center, University of Central Florida, Orlando, FL 32789, USA}

\address{$^6$ College of Optics and Photonics, University of Central Florida, Orlando, FL 32789, USA}

\begin{abstract}
Interest in Josephson junctions (JJs) has increased rapidly in recent
years not only because of their use in qubits and other quantum
devices but also due to the unique physics supported by the JJs. The
advent of various novel quantum materials for both the barrier region
and the superconducting leads has led to the possibility of adding new
functionalities to the JJs. Thus, there is a growing need for accurate
modeling of the JJs and related systems to enable their predictive
control and atomistic level understanding. This review presents an
in-depth discussion of a Green's-function-based formalism for
computing supercurrents in JJs. The formulation is tailored for
large-scale atomistic simulations and encompasses both dc and ac
supercurrents. We hope that this review will provide a timely and
comprehensive reference for researchers as well as beginning
practitioners interested in Green's-function-based methods to model
supercurrents in JJs.
\end{abstract}

\date{\today}


\maketitle
\ioptwocol


\section{Introduction}

In 1962, Josephson made the theoretical prediction that a
dissipationless supercurrent could tunnel through a thin barrier
separating two superconductors in the absence of a bias voltage
\cite{Josephson1962}. This phenomenon, known as the Josephson effect,
is the cornerstone of modern quantum devices ranging from SQUIDs
\cite{SQUID1964} and topological circuit elements \cite{Fu-Kane2008,
  Rokhinson2012}, to modern superconducting qubits
\cite{Shnirman1997,Makhlin2001,Koch2007}.  Despite decades of
theoretical effort, computing supercurrents in realistic Josephson
junctions (JJs) and related devices has remained a significant
challenge. In this context, a diverse array of theoretical and
computational tools involving analytical formulas valid in limiting
cases to sophisticated numerical methods for arbitrary geometries have
been developed. While these approaches have evolved from treating
single-channel tunnel junctions to handling multiple orbital systems,
realistic modeling of devices with realistic interfaces and practical
system sizes has essentially remained beyond practical reach in most
cases.

Modern JJs often involve complex materials, such as unconventional
superconductors \cite{Kashiwaya2000}, topological materials
\cite{Prada2020,Rokhinson2012} and semiconductors with strong
spin-orbit coupling \cite{Mourik2012}, along with multi-terminal
device geometries \cite{Riwar2016} and multilayer heterostructures
\cite{Krogstrup2015, Shabani2016}. Understanding these systems
requires robust computational methods suitable for {\it large-scale}
simulations that incorporate {\it atomic-level} details which are
essential for accurate modeling of transport properties. In the
following, we will first briefly review the progress that has been
made in addressing these challenges and then turn to discuss one of
the emerging Green's function methods that can not only handle
real-space atomistic details but also provide large-scale
computational capabilities.

The phenomenological descriptions of supercurrents began with
Ginzburg-Landau theory \cite{Ginzburg1950} in 1950, which provided a
macroscopic description of superconductivity and established a
framework for future understanding of Josephson effect. The
microscopic foundation of superconductivity was provided by
Bardeen-Cooper-Schrieffer (BCS) theory \cite{BCS} in 1957, which
explained superconductivity through Cooper pair formation. Following
Josephson's prediction in 1962 \cite{Josephson1962}, in 1963
Ambegaokar and Baratoff \cite{AmbegaokarBaratoff63} extended
Josephson's zero-temperature result to finite temperatures using
Gorkov's Green's functions \cite{Gorkov1958}. The supercurrent formula
derived by Ambegaokar and Baratoff appears as a special case of the
general superconductor-insulator-superconductor (SIS) junction theory
at arbitrary dc bias developed by Larkin and Ovchinnikov
\cite{LarkinOvchinnikov1967} and Werthamer \cite{Werthamer1966} in
1966. Subsequently, in 1969, Aslamazov and Larkin
\cite{AslamazovLarkin1969} demonstrated within the stationary
Ginzburg-Landau framework that a strong Josephson supercurrent can
flow through superconducting point contacts if its width remains on
the order of the coherence length and the temperature is close to
critical temperature.

Spatial inhomogeneities of the pairing potential can confine
quasiparticle excitations in much the same way as a normal-state
potential well \cite{SiprGyorffy1996}. Early theoretical examples
include subgap states in superconducting films on normal metals
\cite{deGennesSaintJames1963}, vortex-core bound states
\cite{CaroliDeGennesMatricon1964}, and discrete spectra in
superconductor-normal region-superconductor (SNS) systems
\cite{Andreev1966ElectronSpectrum,Kulik1969SurfaceSuperconductivity}. These
bound states are not merely a spectral detail: they provide an
essential contribution to Josephson current in mesoscopic weak links
\cite{BardeenJohnson1972,FurusakiTsukada1991}, strongly affect
tunneling densities of states and I-V characteristics of films with
superconducting surface sheaths
\cite{Kummel1977QuasiparticleScattering,EntinWohlmanBarSagi1978}, and
underlie prominent tunneling signatures near vortex cores (including
observables accessible via scanning tunneling microscopy)
\cite{ShoreHuangDorseySethna1989,GygiSchluter1991}. Later microscopic
work further clarified their role in transport through superconducting
mesoscopic weak links \cite{Martin-Rodero1994,Yeyati1995} and in
ballistic point contacts \cite{HurdWendin1994}. While Andreev
reflection at a normal region-superconductor boundary is a key
ingredient for understanding how such states can form
\cite{Andreev1964,Andreev1966ElectronSpectrum}, it does not, by
itself, account for every bound state in inhomogeneous
superconductors: additional scattering processes (including normal
reflection) can also be responsible, and the widely used semiclassical
(Andreev) approximations
\cite{Andreev1964,Bardeen1969} can miss important
classes of states (e.g., those associated with large momenta parallel
to the interface)
\cite{Kummel1974Spectrum,GygiSchluter1991,SiprGyorffy1996}. Although
early SIS tunneling theories predate the modern Andreev bound state
description, their supercurrent transfer mechanism can be
reinterpreted naturally within the bound-state picture
\cite{SiprGyorffy1996,Golubov2004}.

In parallel with microscopic advances, simplified circuit models were
developed to describe the dynamics of JJs in electrical circuits. The
resistively-and-capacitively-shunted junction (RCSJ) model introduced
independently by Stewart \cite{Stewart1968} and McCumber
\cite{McCumber1968} in 1968, offered a circuit-based phenomenological
framework that incorporates dissipation and capacitance. Although the
RCSJ model has been instrumental in describing the dynamics and
switching behavior of JJs at the circuit level, it is fundamentally a
classical phenomenological description which lacks microscopic details
\cite{Andreev1964}. For comprehensive reviews of early JJ modeling and
development, see the reviews by Likharev \cite{Likharev1979}, Golubov,
Kupriyanov and Il'ichev \cite{Golubov2004}, and textbooks by Barone
and Patern\`o, \cite{BaronePaterno1982}, Likharev \cite{Likharev1986},
Shmidt \cite{Shmidt1997}, and Tinkham \cite{Tinkham-book}.

By the late 1960s to early 1970s, new methods were developed to
address spatial inhomogeneities and non-equilibrium
effects. Quasi-classical Green's function methods, exemplified by the
Eilenberger equations (1968) \cite{Eilenberger1968} and Usadel
equations (1970) \cite{Usadel1970}, emerged to bridge microscopic
theory and practical simulation in ballistic and diffusive regimes
\cite{Ivanov1981, Zaitsev1984}. Within this quasiclassical framework,
a microscopic theory of short junctions with direct conduction was
developed in the diffusive \cite{Artemenko1979, AslamazovVolkov1986}
and ballistic \cite{Zaitsev1980a, Zaitsev1980b} regimes. However,
early applications of these approaches typically assume a simplified
geometry, such as one-dimensional (1D) uniform junctions, and neglect
atomistic details. Modern quasiclassical implementations can bo beyond
quasi-1D models \cite{Amundsen2016, Seja2022, Virtanen2025}, but
they still average over Fermi-wavelength-scale degrees of freedom and
therefore do not directly resolve atomistic interface structure,
orbital character, or material-specific multipband details, and they
remain limited in strongly correlated regimes. As device complexity
and material-specific effects became increasingly important, more
general quantum transport formalisms were developed. At the
noninteracting quasiparticle level, modern theoretical modeling of JJs
basically follows two frameworks: the scattering approach and the
Green's function method. These methods are mathematically equivalent
\cite{Khomyakov2005, Gaury2014, Waintal2024}, although with relative
practical advantages and disadvantages, depending on the
characteristics of the system under study.

Scattering approaches offer an intuitive and interface-focused
perspective of JJs. A prominent example is the Blonde-Tinkham-Klapwijk
(BTK) formalism \cite{Blonder1982} developed in 1982, which
successfully models the full crossover from tunneling to ballistic
transport by treating the interface as a scattering problem with
variable barrier strength, rooted in the fundamental process of
Andreev reflections. In the 1990s, Beenakker showed that the Josephson
current in a short weak link can be expressed entirely through its
normal-state transmission eigenvalues
\cite{Beenakker1991,Beenakker1992}. Furusaki and Tsukada
\cite{Furusaki_Tsukada1991} showed that the Josephson current can be
expressed directly in terms of Andreev reflection coefficients,
providing a transparent connection between the supercurrent and the
underlying Andreev bound states that form within the junction. Ando
\cite{Ando1991} formulated a lattice-mode-matching scattering scheme
that underpins mesoscopic conductance calculations.  Averin and Bardas
applied the scattering formalism to describe the ac Josephson effect
in a single quantum channel, linking the current oscillations to
multiple Andreev reflections \cite{Averin_Bardas_1995}. Beenakker's
scattering approach was later extended to mesoscopic JJs, including
disordered and chaotic geometries, which became a cornerstone for
random-matrix treatments of supercurrents \cite{Beenakker1997,
  Beenakker2015}. In the early 2000s, this framework was extended to
demonstrate that continuous-spectrum contributions are crucial even in
short junctions and that reflectionless tunneling with sharp
conductance features can occur in ballistic structures
\cite{Krichevsky2000, Schechter2001}.  Around the same time, Waintal
and Brouwer \cite{Waintal2002} developed a scattering-matrix framework
to study magnetic JJs, showing how spin-dependent scattering in JJs
links superconductivity with spintronics. More recent developments
include analyses of spin-orbit-coupled nanowires by Cheng and Lutchyn
\cite{ChengLutchyn2012} in 2012, building on scattering theory to
probe topological effects.

In 2014, Gaury {\it et al.} \cite{Gaury2014} introduced a scattering
wavefunction method for modeling large-scale transient transport,
enabling efficient simulations of time-resolved phenomena in JJs. In
2016, Weston and Waintal \cite{Weston2016} refined this method by
introducing a linear-scaling, source-sink algorithm that absorbs
electrons without requiring true semi-infinite leads, further
advancing the modeling of time-resolved superconducting transport.  A
steady-state implementation of the scattering wavefunction method is
provided by Kwant \cite{Kwant2014}, released in 2014, which combines
tight-binding models with scattering theory and the Bogoliubov-de
Gennes (BdG) \cite{deGennes1966} Hamiltonians to enable efficient
handling of complex device geometries. Kwant is widely used to compute
Andreev bound states and current-phase relations in JJs
\cite{Beenakker2023}, although it does not support time-dependent
phenomena such as the ac Josephson effect. In 2015, Weston {\it et
  al.} \cite{Weston2015} used this approach to model microwave control
of Andreev and Majorana bound states, and in 2015 Savinov
\cite{Savinov2015} generalized the scattering matrices to
multiterminal JJs. In 2017, Zhang {\it et al.} \cite{Zhang2017}
formulated transport in layered systems as a wave function propagation
problem for large-scale junction simulations, while Rossignol {\it et
  al.} \cite{Rossignol2019} incorporated quasiparticle dynamics
together with the surrounding circuit into a unified scattering
description of JJs. In 2021, the Tkwant software package
\cite{Kloss2021}, which is an extension of Kwant, was released to
enable time-dependent quantum transport simulations using the
scattering wavefunction formalism \cite{Gaury2014, Weston2016}. For a
comprehensive review of scattering theory in quantum transport, see
Beenakker \cite{Beenakker1997, Beenakker2015}, Lesovik and Sadovskyy
\cite{Lesovik2011}, and Waintal \cite{Waintal2024}.

Non-equilibrium Green's function (NEGF) methods introduced in the mid
1960s by Kadanoff and Baym \cite{Kadanoff-Baym} and Keldysh
\cite{Keldysh1965} offer a comprehensive microscopic framework for
handling equilibrium, finite bias, and time-dependent phenomena with
high spatial resolution, even if at the expense of increased
computational complexity \cite{Stefanucci2013, Camsari2023}. Building
on Keldysh's foundational work, Caroli {\it et al.} in 1971
\cite{Caroli1971} formulated electron transport through barriers using
NEGF techniques to obtain expressions for tunneling current that are
conceptually analogous to scattering approaches. This framework was
subsequently generalized by Meir and Wingreen in 1992
\cite{MeirWingreen1992} to incorporate many-body interactions in
nanostructures and extended by Jauho {\it et al.}  in 1994
\cite{Jauho1994} to capture time-dependent phenomena. Application of
the NEGF method to JJ systems was advanced through several key works
in the mid-1990s. In 1994, Furusaki \cite{Furusaki1994} applied NEGF
to study the dc Josephson effect in disordered junctions. In 1996,
Averin and Bardas utilized NEGF to investigate the adiabatic dynamics
of superconducting quantum point contacts, further establishing the
microscopic framework for time-dependent phenomena
\cite{Averin_Bardas_1996}.  Around the same time, a series of
foundational works by Martin-Rodero, Levy Yeyati, Cuevas, and
collaborators \cite{Martin-Rodero1994, Yeyati1995, Cuevas1996,
  Martin-Rodero1996, Yeyati1996} developed a
microscopic-Hamiltonian-based NEGF framework for superconducting weak
links and quantum point contacts. They solved the BdG equations
\cite{deGennes1966} self-consistently within the NEGF framework to
handle multiple coherent Andreev reflections, and laid the groundwork
for using NEGF as a practical and general formalism for modeling
JJs. Subsequently, in 2002, Sun {\it et al.}  \cite{SunGuoWang2002}
extended this framework to the ac Josephson effect in finite-sized
junctions, and Asano {\it et al.}  \cite{Asano2006} applied it to
diffusive junction configurations in 2006. Kazymyrenko and Waintal in
2008 \cite{Kazymyrenko2008} introduced the knitting algorithm that
accelerated NEGF calculations for multiterminal devices with arbitrary
geometries. Recent studies, such as those of San-Jose {\it et al.}
\cite{SanJose2013}, use the Floquet-Keldysh formalism to explore
topological JJs. In 2017, Teichert {\it et al.} \cite{Teichert2017}
improved recursively the performance of Green's functions for
quasi-one-dimensional conductors with realistic disorder, while Yap
{\it et al.}  \cite{Yap2017} formulated a recursive Floquet Green's
function scheme for periodically driven edge-state transport. In 2019,
Istas {\it et al.} \cite{Istas2019} proposed a pole-residue expansion
that effectively reaches the thermodynamic limit for nearly
translation-invariant structures.

Modern NEGF applications combined with density functional theory (DFT)
have enabled truly atomistic simulations of various device
configurations \cite{Taylor2001,Brandbyge2002,Papior2017,Nguyen2023},
although extending these methods to JJs remains an open challenge.
Nieminen {\it et al}. in 2023 \cite{Nieminen2023} introduced a new
recursive NEGF approach suitable for modeling JJs using realistic
tight-binding models, demonstrating its capability by revealing
spin-polarized ABS and triplet correlations in
Pb/MoS\textsubscript{2}/Pb junctions. Recent progress in implementing
BdG equations to spin-generalized DFT framework \cite{Reho2024,
  Russmann2022, Russmann2023} provides a promising prospect to combine
DFT-BdG calculations with large scale NEGF simulations in atomic
orbital basis.

The choice between the scattering-based and NEGF methods involves
tradeoffs between intuition, generality, and computational
cost. Scattering based approaches are often computationally more
efficient for transport in wide-channel geometries because they avoid
large matrix inversions, such as those needed to obtain a full
retarded Green's function. However, their reliance on matching
wavefunctions at interfaces implies that scattering methods are mainly
suitable for computing lead-to-lead equal-time (steady-state)
quantities \cite{Kloss2021}. Extracting local observables needed for
gaining insight into what happens inside the junction, such as the
local electron or current densities, therefore, requires additional
post-processing of the scattering wavefunctions
\cite{Kazymyrenko2008}. In contrast, the NEGF formalism offers a
comprehensive microscopic framework that naturally providing local
observables and handles time-dependent drives, fully non-equilibrium
conditions, and even many-body interaction effects.  Modern NEGF
implementations can also bypass the need to explicitly compute the
lesser Green's function by separating the leads from the barrier
region \cite{Nieminen2023}, requiring only retarded Green's functions
that can be obtained efficiently via recursive algorithms for the
junction region (see Ref.~\cite{Lewenkopf2013}) and via decimation for
the lead's surface (see Ref.~\cite{LopezSancho1985}). Recursive
methods are particularly suitable for long JJs, while the computation
of lead Green's functions can also be performed analytically in some
cases \cite{Umerski1997}. For instance,
Refs.~\cite{Kawai2017,Fukaya2022} used this method and a short
recursion to compute equilibrium Green's functions and supercurrent in
a JJ comprising s-wave and spin-triplet superconductors. A related
equilibrium Green's function implementation for tight-binding BdG
Hamiltonians has been developed to explain experimental current-phase
relation measurements in long ballistic graphene JJs
\cite{Rakyta2016}\cite{Nanda2017}.

A crucial distinction between the scattering-based and NEGF methods
arises in the presence of many-body interactions. The scattering
approach effectively describes noninteracting quasiparticle transport,
and it becomes inapplicable in the presence of dynamical electron
correlations. The Green's function formalism, however, remains viable,
as it can systematically incorporate correlations through additional
self-energy terms \cite{Miller2001,Freericks2002}. In the context of
superconductivity, the ability to incorporate many-body interactions
was first shown by Gorkov in his Green's function-based
self-consistent theory \cite{ABD}. Furthermore, self-consistent
Green's function approaches can capture the spatial variation of the
superconducting order parameter into the barrier region, instead of
assuming an abrupt step to zero at the interface. This capability is
essential for describing the suppression of the proximity gap in
correlated-metal-superconductor structures \cite{Nikolic2002} and the
intrinsic reduction of the critical current in short ballistic weak
links \cite{Nikolic2001}. Notably, recent developments have shown that
noninteracting time-dependent Keldysh Green's functions can serve as a
basis for achieving numerically exact results for transport through
strongly correlated junctions when combined with quantum Monte Carlo
algorithms and resummation techniques
\cite{Bertrand2019,Bertrand2019_2}.

An overview of the various methods for computing supercurrents in JJs,
with their relative pros and cons, is presented in
Table~\ref{table-1}.

The main motivation of this paper is to be a one-stop-shop reference
for those interested in learning and applying Green's function-based
methods to model and compute supercurrents in JJs.

The remainder of this paper is organized as follows. A brief
description of the Josephson effect and supercurrents in JJs is
provided in Sec.~\ref{sec:review-JJ}. In
Sec.~\ref{sec:Hamiltonian_form}, a microscopic Hamiltonian formulation
of a JJ is given, including the various steps employed in the
derivation of the expressions for the supercurrent.  NEGF method is
introduced in Sec.~\ref{sec:NEGF}, including 2- and 4-spinor
formulations. Section~\ref{sec:finite-T} contains a brief description
of finite-temperature (equilibrium) Green's functions. An efficient
method to compute supercurrents in the dc regime is developed in
Sec.~\ref{sec:dc-regime}. Efficacy of the method is illustrated by
application to the simple case of a quantum dot coupled to
one-dimensional leads, where it is shown to recover several well-known
results. Because the method relies heavily on a spatial representation
of the states in the underlying materials, in Sec.~\ref{sec:TB-model},
we present a description of the most important aspects of building a
realistic tight-binding model for JJs. Section~\ref{sec:ac-regime}
discusses a powerful formulation of the ac Josephson supercurrent in
voltage-biased junctions in terms of dressed tunneling
matrices. Finally, we conclude by summarizing our main results, along
with an outlook on the field in Sec.~\ref{sec:summary}.


\section{Brief Review of Josephson Junctions}
\label{sec:review-JJ}

JJs consist of two superconducting leads connected by a normal
(non-superconducting) medium, which is usually referred as the "weak
link". The junctions are typically classified as SNS, SIS, and ScS,
where S stands for superconductor, N for normal metal, I for
insulator, and c for constriction. Within the Green's function
methodology, there is no fundamental difference between the presence
of a metal or an insulator in the normal region between the
superconducting regions and, therefore, we merge SIS into SNS and
introduce SS to describe situations where the entire normal region is
represented by a single direct coupling between the superconductors.

The basic phenomenology of JJs is simple to describe
\cite{Tinkham-book}: In the absence of a bias voltage across the
junction, a dc supercurrent of magnitude
\begin{equation}
I = I_c\, \sin \varphi,
\label{eq:Josephson-dc}
\end{equation}
flows between the superconductors, where $I_c$ is the so-called
critical current and $\varphi$ denotes the difference in phase of the
superconductor order parameters. Upon applying a finite bias voltage
$V$, the phase difference gains a linear time dependence $d\varphi/dt
= 2eV/\hbar$, causing the appearance of an ac supercurrent with
angular frequency $\omega_J = 2eV/\hbar$ (which is called the
Josephson frequency).

Notably, Eq.~(\ref{eq:Josephson-dc}) is only approximately correct and
mainly valid when there is a low tunneling probability across the
barrier/lead interface. In general, more complex dependencies on
$\varphi$ are possible although they retain a $2\pi$
periodicity.\footnote{In topological materials, a $4\pi$
periodicity has been observed \cite{Wiedenmann2016}.}

JJs exhibit many different functional regimes, depending on the
junction length $L$ (distance between the superconducting electrodes
or the thickness of the normal region), the junction cross section
area ${\cal A}$ or transverse width $W$, the superconductor coherence
length $\xi$ and, for all-metallic junctions, the mean free path $l$
of electrons in the normal state \cite{Beenakker1992}. For short
junctions with a narrow constriction ($L,W \ll \xi \ll l$), the
critical current is quantized in units of $e|\Delta|/\hbar$, where $e$
is the electron charge and $|\Delta|$ is the magnitude of the
superconductor order parameter, independently of the nature of the
junction (metallic or insulating).

\begin{table*}[p]
\centering
\begin{tabular}{|p{2.5cm}|p{6cm}|p{6cm}|}
\hline
{\bf Methodology} & {\bf Pros} & {\bf Cons / Limitations} \\ \hline
Analytical approaches& 
\begin{itemize}
    \item Provides closed-form, explicit expressions (e.g., Ginzburg--Landau\cite{Ginzburg1950}, Ambegaokar--Baratoff \cite{AmbegaokarBaratoff63}.
    \item Computationally efficient.
    \item Provides fundamental physical insight.
\end{itemize} 
& 
\begin{itemize}
    \item Limited to idealized, simple geometries.
    \item Requires near-equilibrium conditions.
    \item Neglects atomistic detail and interface effects.
\end{itemize} \\ \hline
Quasiclassical Green's function & 
\begin{itemize}
    \item Captures spatial variations of the order parameter.
    \item Adaptable to both ballistic (Eilenberger \cite{Eilenberger1968}) and diffusive (Usadel \cite{Usadel1970}) regimes.
    \item Balance between microscopic detail and computational efficiency.
\end{itemize} 
& 
\begin{itemize}
    \item Averages out atomic-scale details due to momentum averaging.
    \item Neglects detailed interface structure
    \item Limited use in strongly correlated regimes
\end{itemize} \\ \hline

RCSJ & 
\begin{itemize}
    \item Simple circuit-based ODE model \cite{Stewart1968,McCumber1968}.
    \item Describes macroscopic dynamics.
    \item Key parameters ($I_c,R,C$) map directly to circuit design.
\end{itemize} 
& 
\begin{itemize}
    \item Purely phenomenological and classical.
    \item Lacks microscopic physics (e.g., ABS).
    \item Requires a phenomenological current-phase relation, often sinusoidal.
\end{itemize} 
\\ \hline

Scattering approaches& 
\begin{itemize}
    \item Intuitive interface-focused picture.
    \item Can yield analytical expressions in simple junction geometries.
    \item Ideal for terminal-to-terminal transport properties.
    \item Computationally efficient for wide channels.
\end{itemize} 
& 
\begin{itemize}
    \item Primarily for lead-to-lead quantities.
    \item Local observables require extra post-processing.
    \item Less suited for complex internal junction properties.
\end{itemize} \\ \hline
Non-equilibrium Green's function & 
\begin{itemize}
    \item Fully microscopic and capable of capturing detailed atomistic information.
    \item Applicable to non-equilibrium, finite bias, and time-dependent regimes.
    \item Naturally provides local observables (e.g., local current density).
    \item Can incorporate many-body interactions.
    \item Suitable for Long junctions.
\end{itemize} 
& 
\begin{itemize}
    \item Highest computational complexity.
    \item Implementation can be challenging for complex geometries or large systems.
    \item Can be less intuitive than scattering methods.
\end{itemize} \\ \hline

\end{tabular}
\caption{Overview of current framework for modeling Josephson junctions.}
\label{table-1}
\end{table*}
\clearpage

Here, we will work under the assumption that the so-called "rigid
boundary condition" holds \cite{Likharev1979}, namely, the
superconducting order parameter goes abruptly to zero in the normal
region of the junction. This approximation is justified when either
$W\ll \xi$ or the resistivity in the junction region is much higher
than the resistivity of the bulk superconductor.

In equilibrium, the dc supercurrent can be obtained by taking the
derivative of the free energy $F$ with respect to $\varphi$
\cite{Tinkham-book}:
\begin{equation}
I = \frac{2e}{\hbar} \frac{\partial F}{\partial\varphi}.
\label{eq:current-free-energy}
\end{equation}
This formula has been extensively used in the literature to generate
analytical expressions for the supercurrent in various regimes for
idealized situations. By writing the free energy in terms of the
energy eigenstates of the junction \cite{Bardeen1969}, 
Ref.~\cite{Beenakker1992} expresses the supercurrent as a
sum over the discrete and continuous parts of the energy spectrum,
with the discrete part consisting of ABSs. Their result can be cast as
\begin{eqnarray}
I & = & - \frac{2e}{\hbar} \left[ \sum_p
  \frac{d\varepsilon_p}{d\varphi} \tanh\left(
  \frac{\varepsilon_p}{2k_BT} \right) \right. \nonumber \\ & & \left.
  +\ 2k_BT \int_{\Delta}^\infty d\varepsilon\, \frac{d\rho}{d\varphi}
  \ln 2\cosh\left( \frac{\varepsilon}{2k_BT} \right) \right],
\label{eq:current-Beenakker}
\end{eqnarray}
where $\{\varepsilon_p\}$ are the discrete energy eigenvalues, $\rho$
is the density of states, $\Delta$ is the superconductor gap, and $T$
denotes temperature. This result clearly separates the contributions
from the discrete states lying in the superconductor gap from the
continuous states lying outside the gap. It also shows that the
supercurrent is proportional to the derivative of the eigenenergies
with respect to $\varphi$, so that one must know how the eigenenergies
depend on $\varphi$ for computing the supercurrent. Unfortunately,
reliance on the macroscopic free energy (or, equivalently, the
expectation value of the total Hamiltonian) makes this formula
impractical for computations where the atomic structure of the entire
junction needs to be taken into account or when the geometry is
irregular and multiple propagating channels are involved. Obtaining
the eigenenergies that go into Eq.~(\ref{eq:current-Beenakker})
requires exact diagonalization of excessively large matrices (see,
e.g., Appendix~B in Ref.~\cite{Beenakker2023}, where an estimate of
the required size is presented). In the following sections, we develop
an alternative formulation that is more suitable for numerical
computations in such cases.


\section{Hamiltonian Formulation}
\label{sec:Hamiltonian_form}


In this section, we derive a general tight-binding Hamiltonian for a
JJ and obtain an expression for the supercurrent. We begin by
simplifying the Hamiltonian via gauge transformations to gauge out the
bias voltage, then apply the BCS mean-field theory, and finally derive
a general expression for the charge current. We assume that no
magnetic field is present; a generalization to include Zeeman fields
is straightforward.

\begin{figure}[ht]
  \centering
  \includegraphics[width=0.48\textwidth]{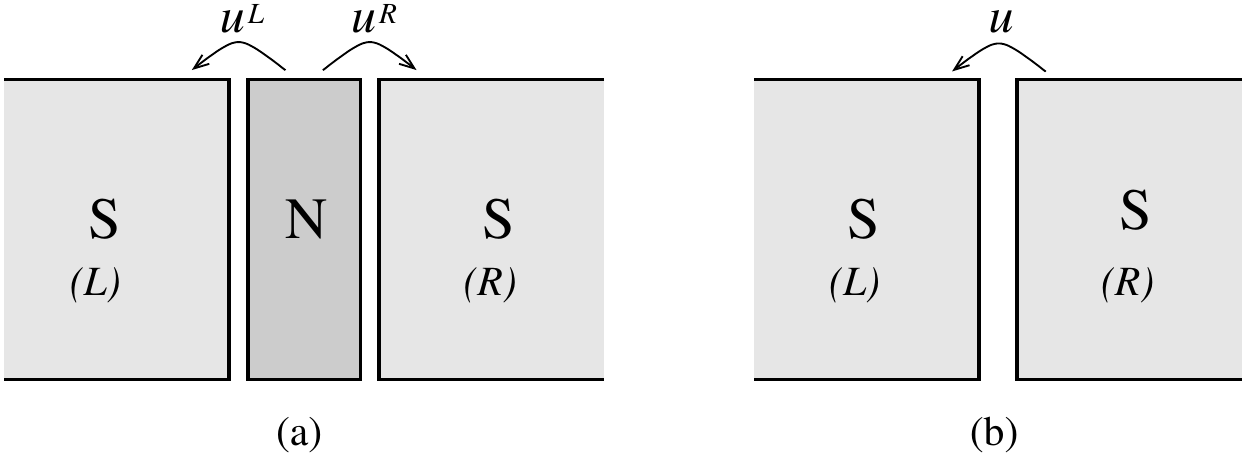}
  \caption{Schematic representations of (a) SNS and (b) SS
    junctions. $u$, $u_L$, and $u_R$ represent various couplings as
    shown in the figure. $L$ and $R$ refer to left and right,
    respectively.}
  \label{fig:system}
\end{figure}

The total tight-binding Hamiltonian for an SNS junction
comprises five terms, see Fig.~\ref{fig:system}a:
\begin{equation}
{\cal H} = {\cal H}_N + {\cal H}_L + {\cal H}_R + {\cal U}_L + {\cal
  U}_R.
  \label{eq:Hamilt_SNS}
\end{equation}
The Hamiltonian of the normal region (either a non-superconducting
metal or an insulator) can be expressed as
\begin{eqnarray}
{\cal H}_N & = & \sum_{a,a'\in N} \sum_{\sigma,\sigma'} h^N_{a \sigma,
  a' \sigma'} c_{a \sigma}^{\dagger} c_{a' \sigma'} - e V_N {\cal
  N}_N,
\end{eqnarray}
where $h^N_{a \sigma, a' \sigma'} = (h_{a' \sigma', a
  \sigma}^N)^{\ast}$ includes both hopping and on-site energies, along
with coupling terms to a magnetic field. Indices $a$ and $a'$ denote
lattice sites and $\sigma$ and $\sigma'$ denote spin components
$\uparrow,\downarrow$. $V_N$ represents an applied voltage and ${\cal
  N}_N$ denotes the total electron-number operator for the normal
region,
\begin{equation}
{\cal N}_N = \sum_{a \in N} \sum_{\sigma} c_{a \sigma}^{\dagger} c_{a
  \sigma}.
\end{equation}
${\cal H}_L$ and ${\cal H}_R$ are the voltage-dependent Hamiltonians
of the left and right superconducting leads, respectively (additional
leads can be added straightforwardly):
\begin{equation}
  {\cal H}_\alpha = {\cal H}_\alpha^{(0)} - e V_\alpha\,  {\cal N}_\alpha \\
\end{equation}
where $\alpha=L,R$. The voltage-independent partial Hamiltonians,
\begin{eqnarray}
\label{eq:H0alpha}
{\cal H}_\alpha^{(0)} & = & \sum_{a, a' \in \alpha} \sum_{\sigma,
  \sigma'} h^\alpha_{a \sigma, a' \sigma'} c_{a \sigma}^{\dagger}
c_{a' \sigma'} \nonumber \\ & & - \Lambda_\alpha \sum_{a \in \alpha}
c_{a \uparrow}^{\dagger} c_{a \downarrow}^{\dagger} c_{a \downarrow}
c_{a \uparrow},
\end{eqnarray}
also include both hopping and on-site amplitudes satisfying
$h^{\alpha}_{a \sigma, a' \sigma'} = (h_{a' \sigma', a
  \sigma}^{\alpha})^{\ast}$. $V_\alpha$ is the applied voltage on the
superconducting lead $\alpha$ and ${\cal N}_\alpha$ denotes the
associated electron number operators,
\begin{equation}
{\cal N}_\alpha = \sum_{a \in \alpha} \sum_{\sigma} c_{a
  \sigma}^{\dagger} c_{a \sigma}.
\end{equation}
%

The coupling constant $\Lambda_\alpha$ in Eq.~(\ref{eq:H0alpha})
accounts for the pairing interaction in the $\alpha$ superconducting
lead. Finally, ${\cal U}_\alpha$ is the Hamiltonian that describes
electron hopping between the $\alpha$ superconducting lead and the
normal region,
\begin{eqnarray}
  {\cal U}_\alpha & = & \sum_{a \in \alpha, a' \in N}
  \sum_{\sigma,\sigma'} \left( u^\alpha_{a\sigma, a'\sigma'} c_{a
    \sigma}^{\dagger} c_{a' \sigma'} \right. \nonumber \\ & &
  \left. +\ u^{\alpha \ast}_{a\sigma, a'\sigma'} c_{a'
    \sigma'}^{\dagger} c_{a \sigma} \right).
\end{eqnarray}

It is useful to also consider a related system in which the
superconducting leads are coupled directly, with coupling amplitude
that accounts for the insulating barrier, namely, an SS junction, so
that the total Hamiltonian now contains only three terms
(Fig.~\ref{fig:system}b):
\begin{equation}
{\cal H} = {\cal H}_L + {\cal H}_R + {\cal U}_{LR},
\end{equation}
where
\begin{eqnarray}
{\cal U}_{LR} & = & \sum_{a \in L, a' \in R} \sum_{\sigma,\sigma'}
\left( u_{a\sigma,a'\sigma'} c_{a \sigma}^{\dagger} c_{a' \sigma'}
\right. \nonumber \\ & & \left. +\ u^\ast_{a\sigma, a'\sigma'} c_{a'
  \sigma'}^{\dagger} c_{a\sigma} \right).
\end{eqnarray}

Note that ${\cal H}_L$ and ${\cal H}_R$ above are many-body
Hamiltonians. It is important to keep two-body interactions explicitly
to properly gauge out the applied voltages from the superconducting
leads. A mean-field approximation is made only after the voltages have
been gauged out, as we show next.

The matrices representing Hamiltonians for the lead and normal regions
($h_{a\sigma,a'\sigma'}$) will typically contain intersite hopping
terms $t_{a\sigma,a'\sigma'}$, which may become spin dependent in the
presence of spin-orbit coupling, on-site spin-independent potential
terms $v_{a}\, \delta_{a,a'} \delta_{\sigma,\sigma'}$, and the Zeeman
term $E_Z\, \zeta_\sigma\, \delta_{\sigma,\sigma'} \delta_{a,a'}$,
where $\zeta_\uparrow = -\zeta_\downarrow = 1$.

\subsection{Superconducting leads: Gauging out applied voltages}
\label{sec:gaugeout}

Consider the Heisenberg equation of motion for an electron
annihilation operator on the $\alpha$ superconducting lead ($a\in
\alpha)$,
\begin{eqnarray*}
  \frac{d}{d t} c_{a \sigma} (t) & = & \frac{i}{\hbar} \left[ {\cal
      H}, c_{a \sigma} (t) \right] \nonumber\\ & = & \frac{i}{\hbar}
  \left[ {\cal H}_\alpha^{(0)} + {\cal U}_\alpha, c_{a \sigma} (t)\right]  -
    \frac{i e V_\alpha}{\hbar} c_{a \sigma} (t),
\end{eqnarray*}
where we employed the anticommutation relations
\begin{eqnarray*}
\{ c_{a \sigma}^{\dagger}, c_{a' \sigma'} \} & = & \delta_{a, a'}
\delta_{\sigma, \sigma'} \\ \{ c_{a \sigma}, c_{a' \sigma'} \} & = & \{
c_{a \sigma}^{\dagger}, c_{a' \sigma'}^{\dagger} \} = 0.
\end{eqnarray*}
Transforming the annihilation operators using
\begin{equation}
c_{a \sigma} = e^{- i e V_\alpha t / \hbar}\, \tilde{c}_{a
  \sigma},
\label{eq:gaugeout}
\end{equation}
yields the transformed operators $\tilde{c}_{a\sigma}$ that satisfy an
equation of motion where the voltage $V_\alpha$ is absent:
\begin{equation}
\frac{d}{d t} \tilde{c}_{a \sigma}(t) = \frac{i}{\hbar} \left[ {\cal
    H}_\alpha^{(0)} + {\cal U}_\alpha, \tilde{c}_{a \sigma} (t)
  \right] .
\label{eq:ctilde_evol}
\end{equation}
Notice that the transformation in Eq.~(\ref{eq:gaugeout}) does not
affect the standard fermionic anticommutation relations for the
$\tilde{c}_{a \sigma}$ and $\tilde{c}_{a \sigma}^{\dagger}$ operators.

Following the preceding steps for the electron-annihilation operator
in the normal region allows us to simply replace the operators $c_{a
  \sigma}$ and $c_{a \sigma}^{\dagger}$ with $\tilde{c}_{a \sigma}$
and $\tilde{c}_{a \sigma}^{\dagger}$ in ${\cal H}_N^{(0)}$, ${\cal
  H}_L^{(0)}$, and ${\cal H}_R^{(0)}$ without introducing a
voltage-dependent phase factor since these Hamiltonians are
particle-number conserving. The only Hamiltonian terms in which
voltage dependencies appear are now the couplings, namely,
\begin{eqnarray}
{\cal H}(V_L, V_N, V_R ; t) & = & {\cal H}_N^{(0)} + {\cal H}_L^{(0)}
+ {\cal H}_R^{(0)} \nonumber \\ & & + {\cal U}_L (V_L - V_N ; t)
\nonumber \\ & & + {\cal U}_R (V_R - V_N ; t),
\end{eqnarray}
where
\begin{equation}
 {\cal H}_N^{(0)} = \sum_{a,a'\in N} \sum_{\sigma,\sigma'}
 h^N_{a\sigma,a'\sigma'}\, \tilde{c}^\dagger_{a\sigma}
 \tilde{c}_{a'\sigma'},
\end{equation}
\begin{eqnarray}
  {\cal U}_\alpha (V_\alpha - V_N ; t) & = & \sum_{a \in \alpha, a'
    \in N} \sum_{\sigma,\sigma'} \left[ u^\alpha_{a\sigma, a'\sigma'}
    (t)\, \tilde{c}_{a \sigma}^{\dagger} \tilde{c}_{a' \sigma'}
    \right. \nonumber \\ & & \left. +\ u^{\alpha \ast}_{a\sigma,
      a'\sigma'} (t)\, \tilde{c}_{a' \sigma'}^{\dagger} \tilde{c}_{a
      \sigma} \right],
\end{eqnarray}
and
\begin{equation}
u^\alpha_{a\sigma, a'\sigma'} (t) = e^{i e (V_\alpha - V_N) t /
  \hbar}\, u^\alpha_{a\sigma,a'\sigma'},
\end{equation}
where $\alpha=L,R$. Thus, the voltage dependencies become embedded in
the phases of the coupling amplitudes. The total Hamiltonian is
explicitly time dependent now due to the voltage-dependent coupling
terms.

Along the preceding lines, in the case of the SS junction, where the
normal region is replaced by direct coupling between the
superconductors, it can be shown straightforwardly that the result is
a single-coupling Hamiltonian term which depends only on the
difference between the left and right voltages,
\begin{eqnarray}
{\cal U}_{LR} (V_L - V_R ; t) & = & \sum_{a \in L, a' \in R}
\sum_{\sigma,\sigma'} \left[ u_{a\sigma, a'\sigma'} (t)\, \tilde{c}_{a
    \sigma}^{\dagger} \tilde{c}_{a' \sigma'} \nonumber \right. \\ & &
  \left. +\ u^{\ast}_{a\sigma, a'\sigma'} (t)\, \tilde{c}_{a'
    \sigma'}^{\dagger} \tilde{c}_{a \sigma} \right],
\end{eqnarray}
where
\begin{equation}
u_{a\sigma, a'\sigma'} (t) = e^{i e (V_L - V_R) t / \hbar}\,
u_{a\sigma, a'\sigma'} .
\end{equation}
Hereafter, to simplify the notation, we will denote the zero-voltage
superconducting lead Hamiltonian terms as ${\cal H}_N$, ${\cal H}_L$,
and ${\cal H}_R$, making the voltage and time dependencies implicit;
we will also drop the tilde from the electron creation and
annihilation operators.

Incorporation of the bias voltage into the creation and annihilation
operators was first invoked in connection with superconductivity by
Cohen, Falicov, and Phillips in 1962 \cite{Cohen1962} and later
featured in Rickayzen's 1965 book on superconductivity
\cite{Rickayzen1965}. In 1995, it was reintroduced by Levy Yeyati,
Martin-Rodero, and Garcia-Vidal in the context of mesoscopic transport
\cite{Yeyati1995}.

\subsection{Mean-field approximation (BCS theory)}

After gauging the voltages out of the Hamiltonians ${\cal H}_N$,
${\cal H}_L$, and ${\cal H}_R$, we can use the BCS theory \cite{BCS}
and write the Hamiltonian of the superconducting leads in the
mean-field approximation as
\begin{eqnarray}
{\cal H}_\alpha & = & \sum_{a, a' \in \alpha} \sum_{\sigma, \sigma'}
h^\alpha_{a \sigma, a' \sigma'} c_{a \sigma}^{\dagger} c_{a' \sigma'}
\nonumber \\ & & + \sum_{a \in \alpha} \left( \Delta_a^\alpha\, c_{a
  \uparrow}^{\dagger} c_{a \downarrow}^{\dagger} + \Delta_a^{\alpha
  \ast}\, c_{a \downarrow} c_{a \uparrow} \right),
\end{eqnarray}
where the superconductor order parameter is defined as
%
\begin{equation}
\Delta_a^{\alpha} = - \Lambda \langle c_{a \downarrow} c_{a
  \uparrow} \rangle,
\label{eq:mean-field-approx}
\end{equation}
with $a \in \alpha$ and $\alpha = L,R$. For practical purposes, it is
useful to separate the phase from the magnitude in the superconductor
order parameters:
\begin{equation}
 \Delta_a^{\alpha} = e^{i \phi_a^{\alpha}} | \Delta^{\alpha}_a|.
\end{equation}
Hereafter, we will assume that the order parameter is homogeneous and
equal in magnitude on both superconducting leads: $|\Delta_a^{\alpha}|
= \Delta$.

In Eq.~(\ref{eq:mean-field-approx}), it is implicitly assumed that the
order parameter $\Delta_\alpha$ drops sharply to zero once the
interface between the $\alpha$ lead and the normal region is
crossed. This may not be a realistic assumption for certain systems.
The self-consistency condition leading to the mean-field approximation
may then need to also include the normal region
\cite{Black-Schaffer2008}, and the order parameter may leak into the
normal region.

Notably, the phase of the order parameters must vary spatially to
allow for a nonzero current in the superconductors. This point is
further addressed in the following section. For a uniform current, it
is sufficient to assume a linear dependence: $\phi_a^\alpha =
\phi_\alpha + q\, \hat{u} \cdot (a-a_0)$, where $q$ represents the net
wave number of Cooper pairs that sets the current amplitude, $a_0$ is
a reference coordinate, and $\hat{u}$ is a unit vector along the
current (longitudinal) direction (assuming $a$ to be a position
vector). It is convenient to locate $a_0$ at the superconductor side
of the junction, such that $\phi_\alpha$ is the associated phase.

\subsection{Transferring superconductor phases to
couplings}\label{sec:transferphase}

It is convenient to transfer the phase of the superconductor order
parameter to the coupling amplitudes. For the electron operators on
the $\alpha$ superconducting lead, consider the transformation
\begin{eqnarray*}
\bar{c}_{a\sigma} & = & e^{i \phi_{a}^{\alpha} / 2} c_{a}
\nonumber \\ \bar{c}_{a\sigma}^{\dagger} & = & e^{- i \phi_{a}^{\alpha} / 2}
c_{a\sigma}^{\dagger}
\end{eqnarray*}
for $a \in \alpha= L,R$ only. Notice that this transformation does not
affect the standard fermionic anticommutation relations. In terms of
the new creation and annihilation operators, we thus have
\begin{eqnarray}
{\cal H}_{\alpha} & = & \sum_{a, a' \in \alpha} \sum_{\sigma, \sigma'}
\bar{h}^{\alpha}_{a \sigma, a' \sigma'}\, \bar{c}_{a \sigma}^{\dagger}
\bar{c}_{a' \sigma'} \nonumber \\& & +\, \Delta \sum_{a \in \alpha}
\left( \bar{c}_{a \uparrow}^{\dagger} \bar{c}_{a \downarrow}^{\dagger}
+ \bar{c}_{a \downarrow} \bar{c}_{a \uparrow} \right)
\end{eqnarray}
and
\begin{eqnarray}
{\cal U}_{\alpha} (t) & = & \sum_{a \in \alpha, a' \in N}
\sum_{\sigma,\sigma'} \left[ \bar{u}^\alpha_{a\sigma, a'\sigma'} (t)\,
  \bar{c}_{a \sigma}^{\dagger} c_{a' \sigma'} \right. \nonumber \\ & &
  \left. +\, \bar{u}^{\alpha \ast}_{a\sigma, a'\sigma'} (t)\, c_{a'
    \sigma'}^{\dagger} \bar{c}_{a \sigma} \right],
    \label{eq:U_alpha_t}
\end{eqnarray}
where we have introduced the modified hopping amplitude
\begin{equation}
\bar{h}^\alpha_{a \sigma, a' \sigma'} = h^{\alpha}_{a \sigma, a'
  \sigma'}\, e^{i(q/2) \hat{u} \cdot(a-a')},
  \label{eq:hopping_mod}
\end{equation}
as well as the modified coupling amplitudes
\begin{equation}
  \bar{u}_{a\sigma, a'\sigma'}^{\alpha} (t) = e^{- i \varphi_{\alpha}
    (t)\, / 2}\, u_{a\sigma, a'\sigma'}^{\alpha}
  \label{eq:phase-transfer}
\end{equation}
with the time-dependent phase defined as
\begin{equation}
  \varphi_{\alpha} (t) = \phi_{\alpha} - 2 e\, (V_{\alpha} - V_N)\, t
  / \hbar .
  \label{eq:varphi}
\end{equation}
Here, we assume that the coupling amplitudes
$u_{a\sigma,a'\sigma'}^\alpha$ are restricted to superconductor sites
at the interface with the normal region. Moreover, since the hopping
amplitudes $h_{a\sigma,a'\sigma'}^\alpha$ usually only depend on the
difference $a-a'$, the addition of the phase factor in
Eq.~(\ref{eq:hopping_mod}) does not bring significant changes.

Now, all the information about voltages and superconductor phases is
embedded in the coupling amplitudes, which have become time dependent,
i.e., harmonic functions with frequency $\omega_{\alpha} = 2 e\,
(V_{\alpha} - V_N) / \hbar$, where $\alpha = L$ or $R$. After this
transformation, the mean-field self-consistency condition becomes
phase independent,
\begin{equation}
  - \Lambda \langle \bar{c}_{a \uparrow}^{\dagger} \bar{c}_{a
    \downarrow}^{\dagger} \rangle = - \Lambda_{} \langle \bar{c}_{a
    \downarrow} \bar{c}_{a \uparrow} \rangle = \Delta
  . \label{eq:new-order-par}
\end{equation}

In an SS junction, the transformation still applies but we
have instead
\begin{eqnarray}
  {\cal U}_{LR} (t) & = & \sum_{a \in L, a' \in R}
  \sum_{\sigma,\sigma'} \left[ \bar{u}_{a\sigma, a'\sigma'} (t)\,
    \bar{c}_{a \sigma}^{\dagger} \bar{c}_{a' \sigma'}
    \right. \nonumber \\ && \left. +\, \bar{u}^{\ast}_{a\sigma,
      a'\sigma'} (t)\, \bar{c}_{a' \sigma'}^{\dagger} \bar{c}_{a
      \sigma} \right],
      \label{eq:U_LR_t}
\end{eqnarray}
where
\begin{equation}
  \bar{u}_{a\sigma, a'\sigma'} (t) = e^{- i \varphi (t) / 2}\,
  u_{a\sigma, a'\sigma'}
  \label{eq:phase-transfer-S-S}
\end{equation}
and
\begin{equation}
  \varphi (t) = \phi_L - \phi_R - 2 e\, (V_L - V_R)\, t / \hbar .
  \label{eq:varphi-S-S}
\end{equation}

Once the transformations of Eqs.~(\ref{eq:phase-transfer}) and
(\ref{eq:phase-transfer-S-S}) are implemented in the corresponding
Hamiltonians, we can remove the bar from the lead electron creation
and annihilation operators to simplify the notation.

\subsection{Fundamental current expression}

Charge and spin transport through the junction corresponds to
electrons hopping in and out of the superconducting leads. The charge
current emanating from the $\alpha$ lead can be computed from the
expectation value of the rate of change of the associated electron
number,
\begin{equation}
I_{\alpha} = - e \left\langle \frac{d {\cal N}_{\alpha}}{d t}
\right\rangle = - \frac{i e}{\hbar} \langle [{\cal H}, {\cal
    N}_{\alpha}] \rangle,
    \label{eq:I-H-N-comm}
\end{equation}
where all operators are assumed to be in the Heisenberg picture
(omitting time variable). Expectation value of the commutator can be
readily computed:\footnote{The expectation value can include thermal
averaging.}
\begin{equation}
\langle [{\cal H}, {\cal N}_{\alpha}] \rangle = \langle [{\cal
    H}_{\alpha}, {\cal N}_{\alpha}] \rangle + \langle [{\cal
    U}_{\alpha}, {\cal N}_{\alpha}] \rangle,
    \label{eq:H-N-decomp}
\end{equation}
where
\begin{eqnarray}
\langle [{\cal H}_{\alpha}, {\cal N}_{\alpha}] \rangle & = & \sum_{a,
  a' \in \alpha} \sum_{\sigma, \sigma'} \nonumber \\ & &
\bar{h}^{\alpha}_{a \sigma, a' \sigma'} \left( \left\langle c_{a
  \sigma}^{\dagger} c_{a' \sigma'} \right\rangle - \left\langle c_{a
  \sigma}^{\dagger} c_{a' \sigma'} \right\rangle \right) \nonumber
\\ & & +\, 2\Delta \sum_{a \in \alpha} \left( - \left\langle c_{a
  \uparrow}^{\dagger} c_{a \downarrow}^{\dagger} \right\rangle +
\left\langle c_{a \downarrow} c_{a \uparrow} \right\rangle \right)
\nonumber\\ & = & 0 \label{eq:H-N-commut}
\end{eqnarray}
and
\begin{eqnarray}
\langle [{\cal U}_{\alpha}, {\cal N}_{\alpha}] \rangle & = & \sum_{a
  \in \alpha, a' \in N} \sum_{\sigma,\sigma'} \nonumber \\ & & \left[
  - \bar{u}^{\alpha}_{a\sigma, a'\sigma'} (t) \langle c_{a
    \sigma}^{\dagger} c_{a' \sigma'} \rangle \right. \nonumber \\ & &
  \left. +\, \bar{u}^{\alpha \ast}_{a\sigma, a'\sigma'} (t) \langle
  c_{a' \sigma'}^{\dagger} c_{a \sigma} \rangle \right] .
  \label{eq:U-N-comm}
\end{eqnarray}
In Eq.~(\ref{eq:H-N-commut}), we employed the self-consistency
condition defined by Eq.~(\ref{eq:new-order-par}) to cancel 
contributions from the pairing terms. We thus find
\begin{eqnarray}
I_{\alpha} (t) & = & \frac{i e}{\hbar} \sum_{a \in \alpha, a' \in N}
\sum_{\sigma,\sigma'} \left[ \bar{u}^{\alpha}_{a\sigma, a'\sigma'} (t)
  \langle c_{a \sigma}^{\dagger} c_{a' \sigma'} \rangle
  \right. \nonumber \\ & & \left. -\, \bar{u}^{\alpha \ast}_{a\sigma,
    a'\sigma'} (t) \langle c_{a' \sigma'}^{\dagger} c_{a \sigma}
  \rangle \right] .
\label{eq:current-SNS}
\end{eqnarray}
For the SS junction, it is straightforward to obtain along the
preceding lines:
\begin{eqnarray}
I_{LR} (t) & = & \frac{i e}{\hbar} \sum_{a \in L, a' \in R}
\sum_{\sigma,\sigma'} \left[ \bar{u}_{a,\sigma a'\sigma'} (t) \langle
  c_{a \sigma}^{\dagger} c_{a' \sigma'} \rangle \right. \nonumber
  \\ && \left. -\, \bar{u}^{\ast}_{a\sigma, a'\sigma'} (t) \langle
  c_{a' \sigma'}^{\dagger} c_{a \sigma} \rangle \right]
\label{eq:current-SS}
\end{eqnarray}
for the charge current flowing from the left to the right lead.

Equations~(\ref{eq:current-SNS}) and (\ref{eq:current-SS}) are related
to Eq.~(\ref{eq:current-free-energy}). This connection can be seen by
considering the case of an SNS. Assuming thermal equilibrium, we can
write
\begin{eqnarray}
\frac{\partial F}{\partial \varphi_\alpha} & = & \frac{1}{\beta}
\frac{\partial \ln(Z)}{\partial \varphi_\alpha} = \frac{1}{\beta Z}
\frac{\partial Z}{\partial \varphi_\alpha} \nonumber \\ & = &
\frac{1}{\beta Z} {\rm tr} \left[ \frac{\partial}{\partial
    \varphi_\alpha} e^{-\beta {\cal H}} \right] = -\frac{1}{Z} {\rm
  tr} \left[ e^{-\beta {\cal H}} \frac{\partial {\cal H}} {\partial
    \varphi_\alpha} \right] \nonumber \\ & = & - \left\langle
\frac{\partial {\cal H}}{\partial \varphi_\alpha} \right\rangle =
-\left\langle \frac{\partial {\cal U}_\alpha}{\partial \varphi_\alpha}
\right\rangle,
\label{eq:F-to-U}
\end{eqnarray}
where $\langle \cdots \rangle = {\rm tr}[ e^{-\beta {\cal H}}
  \cdots]/Z$ and $Z={\rm tr}[e^{-\beta{\cal H}}]$. In the last
equality of Eq.~(\ref{eq:F-to-U}), we used the fact that ${\cal H}$
depends on the phase $\varphi_\alpha$ only through ${\cal
  U}_\alpha$. Note that
\begin{equation}
\left\langle \frac{\partial {\cal U}_\alpha}{\partial \varphi_\alpha}
\right\rangle = \frac{i}{2} \langle [{\cal U}_\alpha, {\cal N}_\alpha]
\rangle = \frac{i}{2} \langle [{\cal H}, {\cal N}_\alpha] \rangle,
\end{equation}
as Eqs.~(\ref{eq:H-N-decomp}), (\ref{eq:H-N-commut}), and
(\ref{eq:U-N-comm}) imply. Finally, using Eq.~(\ref{eq:I-H-N-comm}),
we arrive at Eq.~(\ref{eq:current-free-energy}). A similar sequence of
steps can be applied to the SS case. Of course,
Eqs.~(\ref{eq:current-SNS}) and (\ref{eq:current-SS}) are more general
than Eq.~(\ref{eq:current-free-energy}) since they are also valid
in non-equilibrium situations.



\section{Non-Equilibrium Green's Function Formulation}
\label{sec:NEGF}

Non-equilibrium Green's functions (NEGFs) offer a convenient and
practical way to express the current across the junction
\cite{Haug-Jauho2008}. Consider the one-particle lesser Green's
function
\begin{equation}
G_{a \sigma, a' \sigma'}^{<} (t, t') \equiv i \langle c_{a'
  \sigma'}^{\dagger} (t')\, c_{a \sigma} (t) \rangle,
\label{eq:Glesser}
\end{equation}
where we made explicit the time dependencies of the operators. The
retarded and advanced one-particle Green's functions are defined as
\begin{equation}
G_{a \sigma, a' \sigma'}^{\rm r} (t, t') \equiv -i \theta(t-t') \langle
\{ c_{a \sigma}^\dagger (t), c_{a' \sigma'} (t') \} \rangle
\end{equation}
and
\begin{equation}
G_{a \sigma, a' \sigma'}^{\rm a} (t, t') \equiv i \theta(t'-t) \langle
\{ c_{a \sigma}^\dagger (t), c_{a' \sigma'} (t') \} \rangle,
\end{equation}
resulting in $[G^{\rm r}(t,t')]^\dagger = G^{\rm a}(t',t)$. It is also
useful to define the time-ordered Green’s function,
\begin{equation}
G^{\rm t}_{a\sigma,a'\sigma'} \left( t,t' \right) \equiv
-i\left\langle T_t\, c_{a\sigma}(t)\, c^\dagger_{a'\sigma'}(t')
\right\rangle,
\label{eq:GF_time_order}
\end{equation}
where $T_t$ is the time-ordering operator.

\subsection{Time-dependent charge current}

Combining Eq.~(\ref{eq:Glesser}) with Eqs.~(\ref{eq:current-SNS}) and
(\ref{eq:current-SS}), we can write currents in terms of lesser
Green's functions,
\begin{equation}
  I_\alpha(t) = \frac{2e}{\hbar} {\rm Re} \sum_{a\in\alpha,a'\in N}
  \sum_{\sigma,\sigma'} \bar{u}^\alpha_{a\sigma,a'\sigma'} (t)\,
  G^{<}_{a'\sigma',a\sigma}(t,t)
  \label{eq:I-GF-SNS}
\end{equation}
for $\alpha=L,R$ and
\begin{equation}
  I_{LR}(t) = \frac{2e}{\hbar} {\rm Re} \sum_{a\in L,a'\in R}
  \sum_{\sigma,\sigma'} \bar{u}_{a\sigma,a'\sigma'}(t)\,
  G^{<}_{a'\sigma',a\sigma}(t,t).
  \label{eq:I-GF-SS}
\end{equation}
%

\subsection{Frequency domain}

For most calculations, including those for equilibrium or stationary
conditions, it is useful to work with Green's functions in the
frequency or energy domain. This is usually done in two steps. First,
a Fourier transform is taken with respect to the time difference:
\begin{equation}
\tilde{G}(\bar{t}, \varepsilon) \equiv \int_{-\infty}^{\infty} ds\,
e^{i \varepsilon s}\, G(\bar{t}+s/2, \bar{t}-s/2),
\label{eq:mixed-domain}
\end{equation}
where $\hbar$ can be introduced to switch $\varepsilon$ from a
frequency to an energy scale. In equilibrium, since the Green's
function in the time domain depends only on the time difference, there
is no dependence on the average time $\bar{t}$. For other situations,
an additional Fourier transform on $\bar{t}$ may be needed. When the
Green's function oscillates periodically in time, it is better to use
a discrete Fourier transform to remove $\bar{t}$ and introduce Floquet
frequencies, see Sec.~\ref{sec:ac-regime}.

\subsection{Equilibrium: Time-independent charge current}

In the stationary dc regime (zero bias, $V_L=V_R$), when
$G(t,t')=G(t-t')$, the current is time independent and one can switch
to an energy (or frequency) representation for all three Green's
functions, and the fluctuation-dissipation theorem applies. One can
then easily obtain the following relationship \cite{Haug-Jauho2008}:
\begin{equation}
\tilde{G}^{<}(\varepsilon) = f(\varepsilon)\, [\tilde{G}^{\rm
    a}(\varepsilon) - \tilde{G}^{\rm r}(\varepsilon) ],
    \label{eq:fluct-dissip-theor}
\end{equation}
where $f(\varepsilon)$ is the Fermi-Dirac distribution. The current
expressions can then be written in terms of energy integrals with time
dropping out,
\begin{eqnarray}
I_\alpha & = & \frac{2e}{\hbar} {\rm Re} \sum_{a\in\alpha,a'\in N}
\sum_{\sigma,\sigma'} \bar{u}^\alpha_{a\sigma,a'\sigma'} \nonumber
\\ & & \times \int \frac{d\varepsilon}{2\pi}\, f(\varepsilon)
   [\tilde{G}^{\rm a}_{a'\sigma',a\sigma}(\varepsilon) -
     \tilde{G}^{\rm r}_{a'\sigma',a\sigma}(\varepsilon)]
   \label{eq:I-SNS-dc}
\end{eqnarray}
and
\begin{eqnarray}
I_{LR} & = & \frac{2e}{\hbar} {\rm Re} \sum_{a\in L,a'\in R}
\sum_{\sigma,\sigma'} \bar{u}_{a\sigma,a'\sigma'} \nonumber \\ & &
\times \int \frac{d\varepsilon}{2\pi}\, f(\varepsilon) [\tilde{G}^{\rm
    a}_{a'\sigma',a\sigma}(\varepsilon) - \tilde{G}^{\rm
    r}_{a'\sigma',a\sigma}(\varepsilon)].
\label{eq:I-SS-dc}
\end{eqnarray}
Hereafter, for simplicity of notation, we will drop the tilde from the
Green's function in the energy domain by adopting only $\varepsilon$
and $\omega$ to denote energy and frequency variables, respectively.

\subsection{Bogoliubov-de~Gennes-Nambu formulation}

When treating superconductors in the self-consistent mean-field
approximation, it is very useful to adopt the Bogoliubov-de Gennes
formalism \cite{deGennes1966} and Nambu's spinors
\cite{Schrieffer1964} for a more compact formulation. The operator
$c_{a \uparrow}$ then annihilates an {\it electron} with spin up while
the operator $d_{a \downarrow} = c_{a \downarrow}^{\dagger}$
annihilates a {\it hole} with spin down. Similarly, $c_{a
  \uparrow}^{\dagger}$ creates an {\it electron} with spin up while
$d_{a \downarrow}^{\dagger} = c_{a \downarrow}$ creates a {\it hole}
with spin down. In the following, we will make use of these recast
fermionic operators in two distinct ways, depending on whether or not
spin-orbit interactions are present.

\subsubsection{2-spinor formulation.}

We introduce 2-spinor fermionic operators,
\begin{equation}
\psi_a = \left(\begin{array}{c} c_{a \uparrow} \\ d_{a
    \downarrow}^{}
   \end{array}\right) \quad {\rm and} \quad \psi_a^{\dagger} =
   \left(\begin{array}{cc} c_{a \uparrow}^{\dagger} & d_{a
       \downarrow}^{\dagger}
   \end{array}\right).
\end{equation}
It is straightforward to show that these operators satisfy standard
anticommutation relations:
\begin{equation}
  \{ (\psi_a)_j, (\psi^{\dagger}_{a'})_{j'} \} = \delta_{j,j'}\,
  \delta_{a,a'},
\end{equation}
\begin{equation}
  \{ (\psi_a)_j, (\psi_{a'})_{j'} \} = 0,
\end{equation}
and
\begin{equation}
  \{ (\psi^\dagger_a)_j, (\psi^\dagger_{a'})_{j'} \} = 0,
\end{equation}
where $j,j'=1,2$. 

The 2-spinors are useful when spin-orbit coupling is not
present. Consider the Hamiltonian for the $\alpha$ lead stripped from
spin-flipping terms. It is then straightforward to show that
\begin{eqnarray}
{\cal H}_\alpha & = & \sum_{a, a' \in \alpha} \sum_{j,j'} (\psi_a^\dagger)_j\,
(H_{a,a'}^\alpha)_{j,j'}\, (\psi_a')_{j'} \nonumber \\ & = & {\rm Tr} \left[
  \psi^\dagger\, H^\alpha\, \psi \right],
\label{eq:Hcal_alpha}
\end{eqnarray}
where we introduced the $2\times 2$ matrix
\begin{equation}
  H_{a, a'}^{\alpha} = \left(\begin{array}{cc} h_{a \uparrow, a'
      \uparrow}^{\alpha} & \Delta\, \delta_{a, a'}\\ \Delta\,
    \delta_{a, a'} & - h^{\alpha \ast}_{a \downarrow, a' \downarrow}
  \end{array}\right) 
  \label{eq:Hph}
\end{equation}
for $a,a'\in \alpha$. The trace is over the site and spinor
electron-hole indices. Similarly, for the other terms of the total
Hamiltonian, we have
\begin{eqnarray}
{\cal H}_N & = & \sum_{a, a' \in N} \sum_{j,j'} (\psi_a^{\dagger})_j\,
(H_{a,a'}^N)_{j,j'}\, (\psi_{a'})_{j'} \nonumber \\ & = & {\rm Tr}
\left[ \psi^\dagger\, H^N\, \psi \right]
    \label{eq:Hcal_N}
\end{eqnarray}
and
\begin{eqnarray}
{\cal U}_\alpha(t) & = & \sum_{a \in \alpha} \sum_{a' \in N}
\sum_{j,j'} \left[ (\psi_a^{\dagger})_j\, \left( U_{a,a'}^\alpha(t)
  \right)_{j,j'}\, (\psi_{a'})_{j'} \right. \nonumber \\ & &
  \left. +\ (\psi_{a'}^{\dagger})_{j'}\, \left( U_{a,a'}^{\alpha}(t)
  \right)_{j,j'}^\ast\, (\psi_a)_j \right] \nonumber \\ & = & {\rm Tr}
\left[ \psi^\dagger \, U^\alpha(t)\, \psi + \psi^\dagger\,
  U^{\alpha\dagger}(t)\, \psi \right],
    \label{eq:Ucal_alpha}
\end{eqnarray}
where
\begin{equation}
  H_{a,a'}^N = \left( \begin{array}{cc} h_{a\uparrow,a'\uparrow}^N & 0
    \\ 0 & - h^{N\ast}_{a\downarrow,a'\downarrow} \end{array} \right)
\end{equation}
for $a,a'\in N$ and
\begin{equation}
  U_{a,a'}^\alpha(t) = \left( \begin{array}{cc}
    \bar{u}^\alpha_{a\uparrow,a'\uparrow}(t) & 0 \\ 0 & -
    \bar{u}^{\alpha\ast}_{a\downarrow,a'\downarrow}(t) \end{array}
  \right)
\end{equation}
for $a\in\alpha$ and $a'\in N$. Note that the matrices $H^\alpha$,
$H^N$, and $U^\alpha$ restrict which site components of $\psi$ and
$\psi^\dagger$ contribute to the trace. Employing
Eq.~(\ref{eq:ctilde_evol}) for $a\in\alpha$ yields the following
Heisenberg equation for the 2-spinor:
\begin{eqnarray}
\frac{d}{dt} \left( \psi^\dagger_a(t) \right)_j & = & \frac{i}{\hbar}
\left[ \sum_{a'\in \alpha,j'} \, \left( \psi^\dagger_{a'}(t)
  \right)_{j'}\, ( H^\alpha_{a',a})_{j',j} \right. \nonumber \\ & &
  \left. +\ \sum_{a'\in N, j'} \left( \psi^\dagger_{a'}(t)
  \right)_{j'}\, \left( U^{\alpha}_{a,a'}(t) \right)_{j,j'}^\ast
  \right].
\label{eq:psi_heisenberg}
\end{eqnarray}

For the case of an SS junction, we have instead
\begin{eqnarray}
{\cal U}_{LR}(t) & = & \sum_{a\in R} \sum_{a'\in L} \left[
  \psi_a^\dagger\, U_{a,a'}(t)\, \psi_{a'} + \psi_{a'}^\dagger\,
  U_{a,a'}^\dagger(t)\, \psi_a \right] \nonumber \\ & = & {\rm Tr} \left[
  \psi^\dagger\, U(t)\, \psi + \psi^\dagger\, U^\dagger(t)
  \, \psi \right]
\label{eq:Ucal_LR}
\end{eqnarray}
where the direct coupling matrix is defined as
\begin{equation}
U_{a,a'}(t) = \left( \begin{array}{cc}
  \bar{u}_{a\uparrow,a'\uparrow}(t) & 0 \\ 0 & -
  \bar{u}^\ast_{a\downarrow,a'\downarrow}(t) \end{array} \right).
\end{equation}

We can use 2-spinors to build a lesser Nambu-Gorkov Green's function
in the form of a $2\times 2$ matrix acting on the electron-hole spinor
space. The matrix elements are
\begin{equation}
\left[ G^<_{a,a'}(t,t') \right]_{j,j'} = i\left\langle \left[
  \psi^\dagger_{a'}(t') \right]_{j'}\, \left[ \psi_a(t) \right]_j
\right\rangle,
\end{equation}
where $j,j'=1,2$. Using this matrix, we can rewrite
Eq.~(\ref{eq:I-GF-SNS}) as
\begin{equation}
I_\alpha(t) = \frac{2e}{\hbar} {\rm Re}\, {\rm Tr} \left[
  U^\alpha (t)\, \tau_3\, G^<(t,t) \right]
    \label{eq:current-GF}
\end{equation}
and Eq.~(\ref{eq:I-GF-SS}) as
\begin{equation}
I_{LR}(t) = \frac{2e}{\hbar} {\rm Re}\, {\rm Tr} \left[ U(t)\,
  \tau_3\, G^<(t,t) \right],
    \label{eq:current-GF-LR}
\end{equation}
where $\tau_3$ is a Pauli matrix acting in the two-dimensional
particle-hole space.

As for the other Green's functions, we have
\begin{equation}
\left[ G^{\rm r}_{a,a'}(t,t') \right]_{j,j'} = -i\theta(t-t')
\left\langle \left\{ \left[ \psi_a^\dagger(t) \right]_j, \left[
  \psi_{a'}(t')\right]_{j'} \right\} \right\rangle,
\end{equation}
\begin{equation}
\left[ G^{\rm a}_{a,a'}(t,t') \right]_{j,j'} = i\theta(t'-t)
\left\langle \left\{ \left[ \psi_a^\dagger(t) \right]_j, \left[
  \psi_{a'}(t') \right]_{j'} \right\} \right\rangle,
\end{equation}
and
\begin{equation}
\left[ G^{\rm t}_{a,a'}(t,t') \right]_{j,j'} = -i \left\langle T_t\,
\left[\psi_a(t)\right]_j \left[ \psi_{a'}^\dagger(t') \right]_{j'}
\right\rangle .
\end{equation}
%

\subsubsection{4-spinor formulation.}

A 4-spinor formulation is needed in the presence of spin-orbit
coupling when Hamiltonian amplitudes have non-zero off-diagonal spin
components (e.g., $h_{a\uparrow,a'\downarrow},
h_{a\downarrow,a'\uparrow} \neq 0$ and $u_{a\uparrow,a'\downarrow},
u_{a\downarrow,a'\uparrow} \neq 0$). For this purpose, we introduce
the spinor
\begin{equation}
\Psi_a = \left(\begin{array}{c} c_{a \uparrow}\\ c_{a
    \downarrow}\\ d_{a \downarrow}\\ -d_{a \uparrow}
   \end{array}\right)
\end{equation}
and its adjoint
\begin{equation}
\Psi_a^{\dagger} = \left(\begin{array}{cccc} c_{a \uparrow}^{\dagger}
  & c_{a \downarrow}^{\dagger} & d_{a \downarrow}^\dagger & -d_{a
    \uparrow}^{\dagger} \end{array}\right).
\end{equation}
These 4-spinors satisfy the following anticommutation relations:
\begin{equation}
\{ (\Psi_a)_j, (\Psi^{\dagger}_{a'})_{j'} \} = \delta_{j,j'}\, \delta_{a,
  a'},
\end{equation}
\begin{equation}
\{ (\Psi_a)_j, (\Psi_{a'})_{j'} \} = (\sigma_2\, \tau_2)_{j,j'}\,
\delta_{a,a'},
\end{equation}
and 
\begin{equation}
\{ (\Psi^{\dagger}_a)_j, (\Psi^\dagger_{a'})_{j'} \} = (\sigma_2\,
\tau_2)_{j,j'}\, \delta_{a,a'},
\end{equation}
where $j,j'=1,2,3,4$. In the equations above, we have made use of
Pauli matrices that act on the spin space, $\{\sigma_0, \sigma_1,
\sigma_2, \sigma_3\}$. The last two anticommutation relations differ
from the standard form for fermions. The reason is the symmetry in the
composition of the 4-spinor (first two and last two components are not
independent). In fact, $C\, \Psi_a^\ast = \Psi_a$, where $C =
\sigma_2\, \tau_2$. This anomalous anticommutator, which is known as
the Majorana form in contrast to the standard Dirac form
\cite{Chamon2010}, however, does not impact on the Green's function
formulation of the current substantially, as we will show below.

Equations~(\ref{eq:Hcal_alpha}), (\ref{eq:Hcal_N},
(\ref{eq:Ucal_alpha}), and (\ref{eq:Ucal_LR}) are still valid once we
substitute the 2-spinor operators $\psi$ by the 4-spinor $\Psi$, if we
substitute the $2\times 2$ matrices by $4\times 4$ matrices as well:
\begin{equation}
H_{a,a'}^\alpha = \frac{1}{2} \left( \begin{array}{cc}
  \hat{h}^\alpha_{a,a'} & \delta_{a,a'}\, \Delta_a\, \sigma_0
  \\ \delta_{a,a'}\, \Delta_a\, \sigma_0 & - \sigma_2\, \left(
  h_{a,a'}^{\alpha} \right)^\ast\, \sigma_2 \end{array} \right),
\end{equation}
where
\begin{equation}
 h^\alpha_{a,a'} = \left( \begin{array}{cc}
   h^\alpha_{a\uparrow,a'\uparrow} & h^\alpha_{a\uparrow,a'\downarrow}
   \\ h^\alpha_{a\downarrow,a'\uparrow} &
   h^\alpha_{a\downarrow,a'\downarrow} \end{array} \right),
\end{equation}
\begin{equation}
  H_{a,a'}^N = \frac{1}{2} \left( \begin{array}{cc} h^N_{a,a'} &
    0 \\ 0 & - \sigma_2\, \left( h^N_{a,a'} \right)^\ast\,
    \sigma_2 \end{array} \right),
\end{equation}
where
\begin{equation}
  h^N_{a,a'} = \left( \begin{array}{cc}
    h^N_{a\uparrow,a'\uparrow} & h^N_{a\uparrow,a'\downarrow}
    \\ h^N_{a\downarrow,a'\uparrow} &
    h^N_{a\downarrow,a'\downarrow} \end{array} \right),
\end{equation}
\begin{eqnarray}
  U_{a,a'}^\alpha(t) & = & \left( \begin{array}{cc}
    \bar{u}^\alpha_{a,a'}(t) & 0 \\ 0 & -\sigma_2\, \left(
    \bar{u}^\alpha_{a,a'}(t) \right)^\ast\, \sigma_2 \end{array}
  \right)
\end{eqnarray}
for the SNS junction, where
\begin{equation}
\bar{u}^\alpha_{a,a'}(t) = \left( \begin{array}{cc}
  \bar{u}^\alpha_{a\uparrow,a'\uparrow}(t) &
  \bar{u}^\alpha_{a\uparrow,a'\downarrow}(t)
  \\ \bar{u}^\alpha_{a\downarrow,a'\uparrow}(t) &
  \bar{u}^\alpha_{a\downarrow,a'\downarrow}(t) \end{array} \right),
\end{equation}
and
\begin{eqnarray}
  U_{a,a'}(t) & = & \left( \begin{array}{cc} \bar{u}_{a,a'}(t) & 0
    \\ 0 & - \sigma_2\, \left( \bar{u}_{a,a'}(t) \right)^\ast
    \sigma_2 \end{array} \right)
\end{eqnarray}
for the SS junction, where
\begin{equation}
\bar{u}_{a,a'}(t) = \left( \begin{array}{cc}
  \bar{u}_{a\uparrow,a'\uparrow}(t) &
  \bar{u}_{a\uparrow,a'\downarrow}(t)
  \\ \bar{u}_{a\downarrow,a'\uparrow}(t) &
  \bar{u}_{a\downarrow,a'\downarrow}(t) \end{array} \right).
\end{equation}
The analog of Eq.~(\ref{eq:psi_heisenberg}) reads\footnote{Terms
due to the anomalous anticommutators recombine with regular terms.}
\begin{eqnarray}
\frac{d}{dt} \left( \Psi^\dagger_a(t) \right)_j & = & \frac{i}{\hbar}
\left[ \sum_{a'\in \alpha,j'} \left( \Psi^\dagger_{a'}(t)
  \right)_{j'}\, (H_{a',a}^\alpha)_{j',j} \right. \nonumber \\ & &
  \left. +\ \sum_{a'\in N,j'} \left( \Psi^\dagger_{a'}(t)
  \right)_{{j'}} \, \left( U^\alpha_{a,a'}(t) \right)_{j,j'}^\ast
  \right]
\label{eq:Psi_dagger_Heisenberg}
\end{eqnarray}
for $a\in\alpha$.

We can define the 4-spinor one-particle lesser Green's function matrix
in analogy to the 2-spinor case:
\begin{equation}
\left[ G^<_{a,a'}(t,t') \right]_{j,j'} \equiv i \left\langle
\left[ \Psi^\dagger_{a'}(t') \right]_{j'}\, \left[ \Psi_a(t) \right]_j
\right\rangle,
\end{equation}
where $j,j'=1,2,3,4$. Using this definition, we can rewrite the current
expressions as
\begin{equation}
I_\alpha(t) = \frac{e}{\hbar} {\rm Tr} \left[ U^\alpha(t)\,
  {\cal T}_3\, G^<(t,t) \right]
  \label{eq:I_alpha-general}
\end{equation}
for the SNS junction ($\alpha=L,R$), and
\begin{equation}
I(t) = \frac{e}{\hbar} {\rm Tr} \left[ U(t)\, {\cal T}_3\,
  G^<(t,t) \right]
\end{equation}
for the SS junction, where ${\cal T}_3 = \sigma_0 \tau_3$. The trace
runs over both the site and spinor indices. As in the case of
2-spinors, the matrices $U^\alpha$ and $U$ enforce the appropriate
sums over the site indices.

\subsection{Other quantities of interest}

Although not the focus of this review, note that other quantities of
interest beyond the supercurrent can also be obtained from the Green's
function formulation, especially when we switch from the time to the
frequency domain under equilibrium conditions.

The equilibrium local density of states can be computed using, 
\begin{equation}
\rho_a(\varepsilon) = -\frac{1}{\pi} {\rm Im}\, \sum_j \left[ G^{\rm
    r}_{a,a} (\varepsilon) \right]_{j,j}\, ,
\end{equation}
where $j$ runs over all 2-spinor indices. For the 4-spinor case, $j$
also runs over all indices but there is a prefactor of 1/2 to avoid
overcounting.

Anomalous (pairing) part of the Green's function in equilibrium
conditions is another quantity of interest. For the 2-spinors, we can
express the singlet pairing as
\begin{eqnarray}
\langle c_{a\downarrow}\, c_{a'\uparrow} \rangle & = & \langle
(\psi_{a}^\dagger)_2\, (\psi_{a'})_1 \rangle \nonumber \\ & = & -i [
  G^<_{a',a}(t,t) ]_{1,2} \nonumber \\ & = & \int
\frac{d\varepsilon}{2\pi} f(\varepsilon) \, F_{a,a'}(\varepsilon),
\end{eqnarray}
where
\begin{equation}
F_{a,a'}(\varepsilon) = -i \left[ G^{\rm r}_{a',a}(\varepsilon) -
  G^{\rm a}_{a',a}(\varepsilon) \right]_{1,2}.
\end{equation}
For the 4-spinor, we can add pairing channels beyond the singlet case:
\begin{eqnarray}
\langle c_{a\sigma}\, c_{a'\sigma'} \rangle & = & \int
\frac{d\varepsilon}{2\pi} f(\varepsilon) \,
F_{a\sigma,a'\sigma'}(\varepsilon),
\label{eq:sc_correlation}
\end{eqnarray}
where
\begin{equation}
F_{a\sigma,a'\sigma'}(\varepsilon) = -i \eta \left[ G^{\rm
    r}_{a',a}(\varepsilon) - G^{\rm a}_{a',a}(\varepsilon)
  \right]_{j,j'}
\label{eq:anomalous-GF}
\end{equation}
and
\begin{eqnarray}
& & j=1,\quad j=4,\quad \eta=1 \quad {\rm for}\quad \sigma=\sigma'=\uparrow \\
& & j=2,\quad j=3, \quad \eta=-1 \quad {\rm for}\quad \sigma=\sigma'=\downarrow \\
& & j=1,\quad j=3,\quad \eta=-1 \quad {\rm for}\quad \sigma=\uparrow, \ \sigma'=\downarrow \\
& & j=2,\quad j=4 \quad \eta=1 \quad {\rm for}\quad \sigma=\downarrow, \ \sigma'=\uparrow.
\end{eqnarray}
When the anomalous part of the Green's function is found to be nonzero
for sites in the barrier (normal) region, it indicates that there is
nonzero pairing in that region induced by the proximity to the
superconducting leads [see. Eq.~(\ref{eq:mean-field-approx})].


\section{Finite-Temperature Equilibrium Green's Functions}
\label{sec:finite-T}

In the absence of bias and at finite temperatures, it is convenient to
work with imaginary-time Green's functions and their Matsubara
imaginary-frequency counterparts. We define the imaginary time-ordered
Green's function as \cite{ABD}
\begin{equation}
G_{a\sigma,a'\sigma'}(\tau-\tau') \equiv -\left\langle T_\tau\,
c_{a\sigma}(\tau)\, c_{a'\sigma'}^\dagger(\tau') \right\rangle_\beta,
\end{equation}
where
\begin{equation}
c_{a\sigma}(\tau) = e^{\tau({\cal H} - \mu{\cal N})}\, c_{a\sigma}\,
e^{-\tau({\cal H} - \mu{\cal N})},
\end{equation}
\begin{equation}
c_{a\sigma}^\dagger(\tau) = e^{\tau({\cal H} - \mu{\cal N})}\,
c_{a\sigma}^\dagger\, e^{-\tau({\cal H} - \mu{\cal N})},
\end{equation}
and
\begin{equation}
\langle \cdots \rangle_\beta = {\rm Tr}\left[ e^{-\beta({\cal H} -
    \mu{\cal N} - \Omega)} \cdots \right].
\end{equation}
Note that $c_{a\sigma}(\tau)$ and $c_{a\sigma}^\dagger(\tau)$ are
not Hermitian conjugates. Here, $\mu$ denotes the chemical potential,
$\beta$ is the inverse temperature, and $\Omega$ is the grand
canonical free energy,
\begin{equation}
e^{-\beta\Omega} = {\rm Tr} \left[e^{-\beta({\cal H}-\mu{\cal N})} \right].
\end{equation}
In the imaginary-frequency domain, we have
\begin{equation}
G_{a\sigma,a'\sigma'}(i\omega_n) = \int_0^\beta d\tau\,
e^{i\omega_n\tau} G_{a\sigma,a'\sigma'}(\tau),
\label{eq:GF_Matusbara_freq}
\end{equation}
where $\omega_n = (2n+1)\pi/\beta$, with $n=0,\pm 1, \ldots$
(fermionic case). One can generate the zero-temperature retarded and
advanced equilibrium Green's functions in the energy representation by
performing an analytical continuation in the Matsubara Green's
function:
\begin{equation}
G_{a\sigma,a'\sigma'}(i\omega_n \rightarrow \omega \pm i0^+) =
G_{a\sigma,a'\sigma'}^{\rm r,a}(\omega).
\label{eq:analytical-cont}
\end{equation}
%


\section{dc Regime (Zero Bias)}
\label{sec:dc-regime}

In most numerical computations of Josephson currents with Green's
functions, Eqs.~(\ref{eq:I-GF-SNS}) and (\ref{eq:I-GF-SS}) and their
2- and 4-spinor counterparts are utilized for obtaining the
supercurrent. In the dc stationary regime at zero bias, the lesser
Green's function is replaced by an integration over energy weighed by
the Fermi-Dirac distribution, see Eqs.~(\ref{eq:I-SNS-dc}) and
(\ref{eq:I-SS-dc}). Obtaining the fully dressed Green's functions
appearing in these equations can be nontrivial, particularly when
considering SNS junctions with numerous underlying single-particle
atomistic basis states. As shown in Ref.~\cite{Nieminen2023}, an
equivalent but alternative expression can be derived, which involves
separate contributions from the equilibrium Green's functions of the
leads and the normal region in analogy to the Caroli formula employed
in coherent mesoscopic electronics and originally derived for
metal-insulator-metal junctions \cite{Caroli1971}. In the following,
we will provide a concise derivation of this expression and then apply
it to two extreme situations: a single-orbital normal region (i.e., a
quantum dot with no spin-orbit coupling) and an extended
dichalcogenide insulator which involves strong spin-orbit coupling.

\subsection{A compact Josephson dc current expression}
\label{sec:JJ-compact}

For deriving a compact expression for the dc current of SNS junctions
following Ref.~\cite{Nieminen2023}, we start with the time-ordered
Green's function
\begin{equation}
  \left[ G_{a,a'}^{{\rm t}} (t - t') \right]_{j,j'} \equiv - i
  \left\langle T_t \left\{ [\Psi_a (t)]_j [\Psi_{a'}^{\dagger}
    (t')]_{j'} \right\} \right\rangle,
\end{equation}
Here, to make the notation more compact, we have lumped indices
as $a, j \rightarrow \alpha$.

Using this definition and Eq.~(\ref{eq:Psi_dagger_Heisenberg}), an
equation of motion for the time-ordered Green's function that connects
the normal region to the left lead can be derived ($a \in N$ and $a'
\in L$):
\begin{eqnarray}
  & - & i \hbar \frac{\partial}{\partial t'} G_{\alpha,
    \alpha'}^{{\rm t}} (t - t') = - \frac{1}{\hbar} \left\langle T_t
  \left\{ \Psi_{\alpha} (t) \frac{d}{d t'} \Psi_{\alpha'}^{\dagger}
  (t') \right\} \right\rangle \nonumber\\ & = & - i \sum_{a''\in L,
    j''} \langle T_t \{ \Psi_{\alpha} (t)\, \Psi_{\alpha''}^{\dagger}
  (t') \} \rangle\, H_{\alpha'',\alpha'}^L \nonumber\\ & & - i
  \sum_{a''\in N, j''} \langle T_t \{ \Psi_{\alpha} (t)\,
  \Psi_{\alpha''}^{\dagger} (t') \} \rangle\,
  U^{L\dagger}_{\alpha'',\alpha'} \nonumber \\ & = & \sum_{a''\in L,
    j''} G^{\rm t}_{\alpha,\alpha''}(t-t')\, H_{\alpha'',\alpha}^L
  \nonumber \\ & & +\ \sum_{a''\in N, j''} G^{\rm
    t}_{\alpha,\alpha''}(t-t')\, U^{L\dagger}_{\alpha'',\alpha'}.
  \label{eq:eq-of-m} 
\end{eqnarray}
Both $H^L$ and $U^{L\dagger}$ here are time independent. This
expression can be rewritten in a more suitable form as
\begin{equation}
  \left[ G_{\alpha, \alpha'}^{{\rm t}} \left( g^{{\rm t}} \right)^{-
      1}\right] (t - t') = \sum_{a'' \in N, j''} G_{\alpha,
    \alpha''}^{{\rm t}} (t - t')\, U_{\alpha'', \alpha'}^{L\dagger},
    \label{eq:Gt-eq}
\end{equation}
where $g^{{\rm t}}$ is the time-ordered Green's function of the left
superconductor in isolation, and satisfies the equation
\begin{equation}
- i \hbar \frac{\partial}{\partial t'} g_{\alpha,
    \alpha'}^{{\rm t}} (t - t') = \sum_{a''\in L, j''} g_{\alpha,
    \alpha''}^{{\rm t}} (t - t')\, H^L_{\alpha'',\alpha'}.
\end{equation}
Operating on Eq.~(\ref{eq:Gt-eq}) with $g^{{\rm t}}$ from the right,
we obtain
\begin{eqnarray}
  G_{\alpha, \alpha'}^{{\rm t}} (t - t') & = & \sum_{a'' \in L, j''}
  \sum_{a''' \in N, j'''} \int d t_1 G_{\alpha, \alpha'''}^{{\rm t}}
  (t - t_1)\nonumber \\ & & \times U_{\alpha''', \alpha'}^{L\dagger}
  \, g_{\alpha'', \alpha'}^{{\rm t}} (t_1 - t'),
\end{eqnarray}
or, more compactly,
\begin{equation}
  G^{{\rm t}} (t - t') = \int d t_1 G^{{\rm t}} (t - t_1)\, U^{L\dagger}\,
  g^{{\rm t}} (t_1 - t'),
\end{equation}
where all summations have turned into matrix
multiplications. Applying Langreth rules \cite{Haug-Jauho2008} yields
\begin{eqnarray}
  G^{<} (t - t') & = & \int d t_1 \left[ G^{{\rm r}} (t - t_1)\,
    U^{L\dagger}\, g^{<} (t_1 - t') \right. \nonumber \\ &&
    \left. +\ G^{<} (t - t_1)\, U^{L\dagger}\, g^{{\rm a}} (t_1 - t')
    \right] .
\end{eqnarray}
Switching to the energy (frequency) representation, we find
\begin{equation}
  G^{<} (\varepsilon) =  G^{{\rm r}} (\varepsilon)\,
    U^{L\dagger}\, g^{<} (\varepsilon) + G^{<} (\varepsilon)\, U^{L\dagger}\,
    g^{{\rm a}} (\varepsilon) .
\end{equation}
Finally, substituting this result into Eq.~(\ref{eq:I_alpha-general}),
we obtain the following expression for the dc current:\footnote{For
the case of 2-spinors, the prefactor is $2e/\hbar$ and only the real
part of the integrand contributes.}
\begin{eqnarray}
  I_L & = & \frac{e}{\hbar} \int \frac{d \varepsilon}{2
    \pi} {\rm Tr} \left\{ U^L\, {\cal T}_3 \left[ G^{{\rm r}}
    (\varepsilon)\, U^{L\dagger}\, g^{<} (\varepsilon)
    \right. \right. \nonumber \\ & & \left. \left. +\ G^{<}
    (\varepsilon)\, U^{L\dagger}\, g^{{\rm a}} (\varepsilon) \right]
  \right\}.
  \label{eq:IL-intermediate}
\end{eqnarray}
Here, the trace acts on the site-spinor space. Notice that the Green's
functions with capital letters correspond to the normal region, while
those in lowercase are for the left superconductor in isolation.

It is convenient to rewrite the decoupled lead Green's function
in an energy eigenbasis,
\begin{equation}
  [g (\varepsilon)]_{\alpha'', \alpha} = \sum_{\kappa} (O^{- 1})_{\alpha'',
  \kappa}\,  \left[ g_{{\rm e}} (\varepsilon) \right]_{\kappa}\, O_{\kappa,
  \alpha},
\end{equation}
where $O_{\kappa,\alpha}$ are matrix elements of the basis
transformation. Then,
\begin{equation}
I_L = \frac{e}{\hbar} \int \frac{d \varepsilon}{2
    \pi} \left[ F_1(\varepsilon) + F_2(\varepsilon) \right]
    \label{eq:I_L-breakdown}
\end{equation}
where
\begin{equation}
  F_1(\varepsilon) =  {\rm Tr} \left\{ {\cal T}_3 \left[ G^{{\rm r}}
    (\varepsilon)\, U^{L\dagger}_h\, g_{{\rm e}}^{<} (\varepsilon)\,
    U^L_h \right] \right\}
  \label{eq:IL-complete1}
\end{equation}
and
\begin{equation}
F_2(\varepsilon) = {\rm Tr} \left\{ {\cal T}_3 \left[ G^{<}
  (\varepsilon)\, U^{L\dagger}_h\, g^{{\rm a}}_{{\rm e}}
  (\varepsilon)\, U^L_h \right]
      \label{eq:IL-complete2} \right\},
\end{equation}
where we have introduced hybrid-coupling matrices
\begin{eqnarray*}
\left[ U^L_h \right]_{\kappa, \alpha'} & = & \sum_{a \in L,j}
O_{\kappa, \alpha} U^L_{\alpha,\alpha'} \\ \left[ U^{L\dagger}_h
  \right]_{\alpha''' , \kappa} & = & \sum_{a'' \in L,j''}
U^{L\dagger}_{\alpha''', \alpha''} (O^{-1})_{\alpha'', \kappa}
\end{eqnarray*}
and added the subscript ''e'' to differentiate the lead's Green's
function in the eigenenergy basis representation, where it is
diagonal, from the one in the site-spinor basis. When expressed in the
energy eigenbasis, the lead's Green's function depends only on the
lead's energy eigenvalues $\varepsilon_\kappa$, namely,
\begin{equation}
\left[ g^{{\rm r,a}}_{\rm e} (\varepsilon) \right]_{\kappa} =
     [\varepsilon - \varepsilon_{\kappa} \pm i 0^+ ]^{-1}
\end{equation}
and, using Eq.~(\ref{eq:fluct-dissip-theor}),
\begin{equation}
  \left[ g^{<}_{{\rm e}} (\varepsilon) \right]_{\kappa} = 2 \pi i
  f_L (\varepsilon)\, \delta (\varepsilon -
  \varepsilon_{\kappa}).
\end{equation}
With these expressions in hand, consider
Eq.~(\ref{eq:IL-complete1}):
\begin{equation}
F_1(\varepsilon) = i\, f_L (\varepsilon)\, {\rm Tr} \left[ {\cal T}_3\,
  G^{{\rm r}} (\varepsilon)\, \Gamma_L (\varepsilon) \right],
\end{equation}
where we introduced the level-width matrix
\begin{eqnarray}
  [\Gamma_L (\varepsilon)]_{\alpha''', \alpha'} & = & 2 \pi
  \sum_{\kappa} \delta (\varepsilon - \varepsilon_{\kappa})\, \left[
    U^{L\dagger}_h \right]_{\alpha''',\kappa}\, \left[ U_h^L
    \right]_{\kappa,\alpha'} \nonumber \\ & = & i \left[
    U^{L\dagger}\, [ g_L^{\rm r}(\varepsilon)-g_L^{\rm a}(\varepsilon)
    ]\, U_L \right]_{\alpha''',\alpha'} .
\end{eqnarray}
Moreover,
\begin{eqnarray}
& & {\rm Tr} \left[
     {\cal T}_3\, G^{{\rm r}} (\varepsilon)\, \Gamma_L
     (\varepsilon) \right] \nonumber \\& & = \frac{1}{2}
   {\rm Tr} \left\{ \Gamma_L (\varepsilon)\, \left[ {\cal T}_3\,
     G^{{\rm r}} (\varepsilon) - G^{{\rm a}} (\varepsilon)\,
     {\cal T}_3 \right] \right\}.
  \label{eq:1stterm} 
\end{eqnarray}
We can similarly obtain a compact form for the second term in the
integrand of Eq.~(\ref{eq:I_L-breakdown}):
\begin{eqnarray}
F_2 & = & \frac{1}{4} \Big\{ i\, {\rm Tr} \left[ \left( {\cal T}_3\,
  G^{<} + G^{<}\, {\cal T}_3 \right) \Gamma_L \right] \nonumber \\ & &
+\ {\rm Tr} \left[ \left( {\cal T}_3\, G^{<} - G^{<}\, {\cal T}_3
  \right)\, U_h^{L\dagger} \left( g^{{\rm a}}_{{\rm e}} +
  g^{{\rm r}}_{{\rm e}} \right)\, U_h^L \right] \Big\}. \nonumber \\
\label{eq:I_coh_part}
\end{eqnarray}
Separating the first and second terms in the curly brackets on the
r.h.s. of Eq. (\ref{eq:I_coh_part}) and absorbing the first term into
$F_1$, we can write
\begin{equation}
I_L = I_L^{(1)} + I_L^{(2)},
\end{equation}
where
\begin{eqnarray}
  I_L^{(1)} & = & \frac{i e}{2\hbar} \int \frac{d \varepsilon}{2 \pi} {\rm
    Tr} \Big( \Gamma_L (\varepsilon) \Big\{ \frac{1}{2} \left[
    {\cal T}_3\, G^{<} (\varepsilon) + G^{<} (\varepsilon)\,
    {\cal T}_3 \right] \nonumber \\ & & +\ f_L (\varepsilon) \left[
    {\cal T}_3\, G^{{\rm r}} (\varepsilon) - G^{{\rm a}}
    (\varepsilon)\, {\cal T}_3 \right] \Big\} \Big) .
  \label{eq:IL}
\end{eqnarray}
and
\begin{eqnarray}
  I_L^{(2)} & = & \frac{e}{4 \hbar} \int \frac{d \varepsilon}{2 \pi} {\rm
    Tr} \left\{ [{\cal T}_3, G^{<} (\varepsilon)] U_h^{L\dagger}
  \left[ g^{{\rm a}}_{{\rm e}} (\varepsilon) + g^{{\rm r}}_{{\rm e}}
    (\varepsilon) \right] U_h^L \right\}. \nonumber \\ & &
  \label{eq:IL-2}
\end{eqnarray}

At zero bias, assuming that both leads are at the same temperature, we
can write $f_R(\varepsilon) = f_L(\varepsilon) =
f(\varepsilon)$. Moreover, in equilibrium, $G^<(\varepsilon) =
f(\varepsilon) [G^{\rm a}(\varepsilon)-G^{\rm r}(\varepsilon)]$,
yielding
\begin{eqnarray}
  I_L^{(1)} & = & \frac{i e}{4\hbar} \int \frac{d \varepsilon}{2 \pi}
  f(\varepsilon)\, {\rm Tr} \Big\{ \Gamma_L (\varepsilon) \left[ {\cal
      T}_3, G^{{\rm r}} (\varepsilon) + G^{{\rm a}}(\varepsilon)
    \right] \Big\} . \nonumber \\ & & 
  \label{eq:IL-1-cont}
\end{eqnarray}
Because $U^\alpha(\varphi_\alpha) = e^{-i \varphi_\alpha \tau_3 / 2}\,
U^\alpha(0)$, it is straightforward to show that the level-width
matrices $\Gamma_L$ and $\Gamma_R$ do not depend on the phases
$\varphi_L$ and $\varphi_R$. Hence, for a symmetric junction, when
$u^L = u^R$ (identical superconductor-normal region couplings),
Eq.~(\ref{eq:IL-1-cont}) implies that $I^{(1)}_L =
I^{(1)}_R$. However, in the dc stationary regime, we expect $I^{(1)}_L
= -I^{(1)}_R$. Therefore, $I^{(1)}_L = I^{(1)}_R = 0$ and we can drop
the $I^{(1)}_L$ contribution to $I_L$.\footnote{It can be shown that
even for a nonsymmetric junction, provided that there is time-reversal
symmetry such that the couplings $u^{L,R}$ can be made real,
$I^{(1)}_L$ vanishes.}  This reasoning, however, does not apply to
$I^{(2)}_L$ since the integrand in Eq.~(\ref{eq:IL-2}) is phase
dependent and we cannot conclude that $I_L^{(2)}$ and $I_R^{(2)}$ are
equal even for a symmetric junction.

Switching back to the site-only representation for the coupling
matrices and the lead's surface Green's functions, we can rewrite
Eq~(\ref{eq:IL-2}) as
\begin{eqnarray}
  I_L^{(2)} & = & \frac{e}{4 \hbar} \int \frac{d \varepsilon}{2 \pi}
  {\rm Tr} \left\{ [{\cal T}_3, G^{<} (\varepsilon)] U^{L\dagger}
  \left[ g_L^{{\rm a}} (\varepsilon) + g_L^{{\rm r}} (\varepsilon)
    \right] U^L \right\} \nonumber \\ & = & \frac{e}{4\hbar} \int
  \frac{d\varepsilon}{2\pi} f(\varepsilon)\, {\rm Tr} \Big\{ \left[
    {\cal T}_3, G^{\rm r}(\varepsilon) - G^{\rm
      a}(\varepsilon) \right] \nonumber \\ & & \times U^{L\dagger}
  \left[ g_L^{\rm r}(\varepsilon) + g_L^{\rm a}(\varepsilon) \right]\,
  U^L \Big\}.
    \label{eq:supercurrent-4}
\end{eqnarray}
The current from the right superconductor has an analogous
expression. Recall that $G^{\rm r (a)}(\varepsilon)$ is the
fully-dressed retarded (advanced) Green's function of the normal
region only and $g_L^{\rm r(a)}(\varepsilon)$ is the retarded
(advanced) surface Green's function of the left lead when decoupled
from the normal region. The separation of the integrand into two
distinct factors makes Eq.~(\ref{eq:supercurrent-4}) very convenient
for computations. $g_L^{\rm r,a}$ can be readily computed numerically
using decimation methods \cite{LopezSancho1985}, while $G^{\rm r,a}$
can be computed numerically either by exact diagonalization or
recursive iterations.

The expression corresponding to Eq.~(\ref{eq:supercurrent-4}) for the
2-spinor case is
\begin{eqnarray}
\label{eq:supercurrent-2}
  I_L & = & \frac{e}{2\hbar} \int \frac{d\varepsilon}{2\pi}
  f(\varepsilon)\, {\rm Tr} \Big\{ \left[ \tau_3, G^{\rm
      r}(\varepsilon) - G^{\rm a}(\varepsilon) \right] \nonumber \\ &
  & \times U^{L\dagger} \left[ g_L^{\rm r}(\varepsilon) + g_L^{\rm
      a}(\varepsilon) \right] U^L \Big\},
\end{eqnarray}

The factor in the brackets on the second line of
Eqs.~(\ref{eq:supercurrent-4}) and (\ref{eq:supercurrent-2}) vanishes
when $|\varepsilon| \geq |\Delta|.$ Therefore, this expression only
captures the contribution to the supercurrent from resonant ABSs
confined to the barrier region and within the superconducting
gap. However, contributions from extended states, which are not
captured by Eqs.~(\ref{eq:supercurrent-4}) and
(\ref{eq:supercurrent-2}), may be relevant in certain situations, as
we explain below \cite{Levchenko2006}.

A hallmark of Josephson currents is their sensitivity to the phase
difference $\varphi$, which is caused by electrons traversing the
barrier region and being reflected at least once at the barrier-lead
interfaces. This requires the electron's propagation time $t_{\rm
  prop} = \hbar/\varepsilon$ to be larger than the traversal time
across the barrier, $t_{\rm trav}$. For ballistic barriers, $t_{\rm
  trav} = v_F/L$, where $v_F$ is the electron's Fermi velocity, and
one arrives at the condition $\varepsilon < (\xi/L) \Delta$, where
$\xi=\hbar v_F/L$ is the ballistic superconductor coherence
length.\footnote{We have made the simplifying assumption that the
Fermi velocity is the same in the barrier and leads.} For ballistic
barriers, $t_{\rm trav} = L/v_F$, where $v_F$ is the electron's Fermi
velocity, and one arrives at the condition $\varepsilon < \hbar
v_F/L=(\xi_{\rm BCS}/L)\pi \Delta$, where $\xi_{\rm BCS}=\hbar
v_F/\pi\Delta$ is the ballistic superconductor coherence length. Thus,
in a fully ballistic SNS junction and for $\xi \leq L$, the states
contributing to the current's phase sensitivity predominantly reside
in the gap and, therefore, they must be ABSs. However, for short
barriers, this is not guaranteed.

For diffusive barriers, $t_{\rm trav} = L^2/D = \hbar/E_{\rm Th}$
instead, where $D$ is the diffusion constant and $E_{\rm Th}$ is the
so-called Thouless energy. When the superconducting leads are also
diffusive, since the diffusive coherence length $\xi=\sqrt{\hbar
  D/\Delta}$, one arrives at the condition $\varepsilon < E_{\rm Th} =
(\xi/L)^2 \Delta$. Therefore, for long barriers ($L>\xi$), the
predominant contribution to the Josephson current also comes from
ABSs.\footnote{We assume that barrier and lead have the same diffusion
constant.} As pointed out in Ref.~\cite{Levchenko2006}, for SNS
systems where $L<\xi$, the analysis is more complex and results in
both confined (i.e., ABS) and extended states contributing to the
Josephson current.

\subsection{Application: Quantum dot}

As an illustrative example, we discuss the application of the
formulation developed in Sec.~\ref{sec:JJ-compact} to the well-known
system of one-dimensional superconductors coupled through a quantum
dot with a single resonant level, where the current can be computed
analytically in the absence of a Coulomb blockade.\footnote{This
problem can also be readily solved by other methods, including using
Tkwant, see package tutorial.}

The Hamiltonian for a quantum dot (QD) comprising a single
(spin-degenerate) orbital is
\begin{equation}
{\cal H}_{D} = \sum_{\sigma} \varepsilon_d\, c_{d\sigma}^\dagger c_{d\sigma}.
\label{eq:H_dot}
\end{equation}
Here, $c_{d\sigma}$ is the annihilation operator for an electron with
spin $\sigma$ in the QD. We neglect capacitive effects in the junction
and Coulomb interactions in the QD. Assuming homogeneous,
translation-invariant semi-infinite one-dimensional leads, it is
convenient to switch from a spatial to a momentum representation, so
we have
\begin{eqnarray}
{\cal H}_{\alpha} & = & \sum_{{\bf k},\sigma} \varepsilon_{\bf
  k}^{\alpha}\, c_{{\bf k}\sigma, \alpha}^\dagger\, c_{{\bf
    k}\sigma,\alpha} \nonumber \\ & & +\ \sum_{\bf k} \left(
\Delta_{\bf k}^\alpha\, c_{{\bf k}\uparrow,\alpha}^\dagger\, c_{-{\bf
    k}\downarrow,\alpha}^\dagger + {\rm H.c.} \right)
\label{eq:QF-full-Hamilt}
\end{eqnarray}
and
\begin{equation}
{\cal U}_\alpha = \sum_{{\bf k},\sigma} \left( u_\alpha\,
e^{-i\varphi_\alpha/2}\, c_{d\sigma}^\dagger c_{{\bf k}\sigma,\alpha}
+ {\rm H.c.} \right),
\label{eq:U-alpha}
\end{equation}
where $\alpha=L,R$. Here, $c_{{\bf k}\alpha}$ is the annihilation
operator for an electron with spin $\sigma$ and momentum ${\bf
  k}$. The coupling amplitudes are assumed to be momentum and spin
independent and $\varepsilon_{\bf k}^\alpha$ and $\Delta_{\bf
  k}^\alpha$ are assumed to be even functions of ${\bf k}$.

In the absence of spin-orbit coupling, the total Hamiltonian can be
written in the 2-spinor formulation (up to an irrelevant additive
constant) as
\begin{eqnarray}
{\cal H} & = & \sum_{\alpha = L,R} \sum_{\bf k} \psi_{{\bf
    k}\alpha}^\dagger H_{\bf k}^{\alpha}\, \psi_{{\bf k}\alpha} +
\psi_d^\dagger\, H_{d}\, \psi_d \nonumber\\ & & +\ \sum_{\alpha = L,R}
\sum_{\bf k} \left( \psi_d^\dagger\, U_{\alpha}\, \psi_{{\bf k}\alpha}
+ \psi_{{\bf k}\alpha}^\dagger\, U_{\alpha}^\dagger\, \psi_d \right),
\label{eq:Hamilt-nonint}
\end{eqnarray}
where
\begin{equation}
\psi_{{\bf k},\alpha} = \left( \begin{array}{c}
c_{{\bf k} \uparrow ,\alpha} \\ c_{ -{\bf k} \downarrow ,\alpha}^\dagger 
\end{array} \right)
\end{equation}
and
\begin{equation}
\psi_d = \left( \begin{array}{c} c_{d\uparrow} \\ c_{d\downarrow}^\dagger
\end{array} \right).
\end{equation}
The $2\times 2$ matrices corresponding to the leads, QD, and lead-QD
couplings are given by
\begin{equation}
H_{{\bf k},\alpha} = \left( \begin{array}{cc}
\varepsilon_{\bf k}^{\alpha} & \Delta_{\bf k}^\alpha \\
\Delta_{\bf k}^{\alpha} & -\varepsilon_{\bf k}^{\alpha}
\end{array} \right) = \varepsilon_{\bf k}^\alpha\, \tau_3 + \Delta_{\bf k}^\alpha\, \tau_1,
\end{equation}
\begin{equation}
H_{d} = \left( \begin{array}{cc}
\varepsilon_d & 0 \\ 0 & -\varepsilon_d \end{array} \right) = \varepsilon_d\, \tau_3,
\end{equation}
and
\begin{eqnarray}
U_\alpha & = & \left( \begin{array}{cc} u_\alpha\,
  e^{-i\varphi_\alpha/2} & 0 \\ 0 & - u_\alpha\,
  e^{i\varphi_\alpha/2} \end{array} \right) \nonumber \\ & = &
u_\alpha \left[ \cos(\varphi_\alpha/2)\, \tau_3 - i
  \sin(\varphi_\alpha/2)\, \tau_0 \right],
\end{eqnarray}
respectively.

\subsubsection{Quantum-dot Green's functions and self-energy.}

The 2-spinor thermal Green's functions are defined as:
\begin{equation}
\left[ G_{d}(\tau,\tau') \right]_{j,j'} = -\left\langle T_\tau\,
\left[ \psi_d^\dagger(\tau') \right]_{j'}\, \left[ \psi_d(\tau)
  \right]_j \right\rangle_\beta,
\end{equation}
\begin{equation}
\left[ G_{{\bf k}\alpha,d}(\tau,\tau') \right]_{j,j'} = -\left\langle
T_\tau\, \left[ \psi_d^\dagger(\tau') \right]_{j'}\, \left[ \psi_{{\bf
      k}\alpha} (\tau) \right]_j \right\rangle_\beta,
\end{equation}
and
\begin{equation}
\left[ G_{{\bf k}\alpha,{\bf k}'\alpha}(\tau,\tau') \right]_{j,j'} =
-\left\langle T_\tau\, \left[ \psi_{{\bf k}'\alpha}^\dagger(\tau')
  \right]_{j'}\, \left[ \psi_{{\bf k}\alpha}(\tau) \right]_j
\right\rangle_\beta,
\end{equation}
where $j,j'=1,2$.

Recall that in the dc (zero-bias) regime we can assume equilibrium
and, therefore, the Green's functions depend only on time
differences. Thus, starting from the Heisenberg equations of motion
for the spinor creation and annihilation operators, it is possible to
derive
\begin{eqnarray}
\frac{d}{d\tau} G_{d}(\tau) & = & - \delta (\tau )\, \tau_0 - H_{d}\,
G_{d}(\tau) \nonumber \\ & & - \sum_{\alpha=L,R} \sum_{\bf k}
U_{\alpha}\, G_{{\bf k}\alpha,d}(\tau),
\end{eqnarray}
We also need the Heisenberg equation for the lead-QD Green's function,
\begin{equation}
\frac{d}{d\tau} G_{{\bf k}\alpha,d}(\tau) = - H_{\bf k}^\alpha\,
G_{{\bf k}\alpha,d}(\tau) - U_{\alpha}^\dagger \, G_{d}(\tau).
\end{equation}
After Fourier transforming the Green's functions according to
Eq.~(\ref{eq:GF_Matusbara_freq}), and combining the two equations
above, one obtains
\begin{equation}
G_{{\bf k}\alpha,d}(i\omega_n) = G^{(0)}_{{\bf k}\alpha,{\bf
    k}\alpha}(i\omega_n)\, U_\alpha^\dagger\, G_{d}(i\omega_n)
\label{eq:G_lead_d}
\end{equation}
and
\begin{equation}
G_{D}(i\omega_n) = \left[ i\omega_n\, \tau_0 - H_{d} -
  \Sigma_{d}(i\omega_n) \right]^{-1},
\label{eq:G_dd}
\end{equation}
where the self-energy is
\begin{equation}
\Sigma_{d}(i\omega_n) = \sum_{\alpha=L,R} U_{\alpha} \left[ \sum_{\bf
    k} g_{{\bf k}\alpha}(i\omega_n) \right] U_{\alpha}^\dagger.
\label{eq:self-energy-imag}
\end{equation}
and the lead's decoupled Green's function is
\begin{equation}
g_{{\bf k}\alpha}(i\omega_n) = \left( i\omega_n\, \tau_0 - H_{\bf
  k}^\alpha \right)^{-1}.
\end{equation}
The term in the square brackets in Eq.~(\ref{eq:self-energy-imag}) is
the lead Green's function projected at the interface with the QD and
can be readily computed:
\begin{eqnarray}
g_\alpha(i\omega_n) & = & \sum_{\bf k} g_{{\bf k}\alpha}(i\omega_n)
\nonumber\\ & = & \sum_{\bf k} \frac{i\omega_n\, \tau_0 +
  \varepsilon_{\bf k}^\alpha\, \tau_3 + \Delta_{\bf k}^\alpha\,
  \tau_1} {\left( i\omega_n \right)^2 - \left( \varepsilon_{\bf
    k}^{\alpha} \right)^2 - (\Delta^\alpha_{\bf k})^2} \nonumber \\ \,
& \to & \int_{-\infty }^\infty d\varepsilon
\frac{\rho_\alpha(\varepsilon) [i\omega_n \tau_0 + \varepsilon \tau_3
    + \Delta_\alpha \tau_1]} {\left( i\omega_n \right)^2 -
  \varepsilon^2 - \Delta_\alpha^2},
\label{eq:g_alphas}
\end{eqnarray}
where $\rho_\alpha(\varepsilon)$ is the single particle,
spin-independent, density of states of the $\alpha$ lead and, in the
last step, we assumed that $\Delta_\alpha=\Delta^\alpha_{\bf k}$
(i.e., momentum independence). After some algebraic manipulations, we
arrive at
\begin{eqnarray}
\Sigma_d(i\omega_n) & = & \sum_{\alpha=L,R} u_\alpha^2
\int_{-\infty}^\infty d\varepsilon
\frac{\rho_\alpha(\varepsilon)}{\left( i\omega_n \right)^2 -
  \varepsilon^2 - \Delta_\alpha^2} \nonumber \\ & & \times \left\{
i\omega_n\, \tau_0 + \varepsilon\, \tau_3 \right. \nonumber \\ & &
\left. -\ \Delta_\alpha [\cos(\varphi_\alpha) \tau_1
  +\sin(\varphi_\alpha) \tau_2 ] \right\}.
\label{eq:Sigma_D}
\end{eqnarray}
These expressions can be simplified using the so-called wide-band
approximation ($\rho_\alpha(\varepsilon) \approx \rho_\alpha(0)$) to
obtain
\begin{equation}
g_\alpha(i\omega_n) \approx - \frac{\pi\, \rho_\alpha(0)}
{\sqrt{\Delta_\alpha^2 - \left( i\omega_n \right)^2}} \left(
i\omega_n\, \tau_0 + \Delta_\alpha\, \tau_1 \right)
\label{eq:GF-surface}
\end{equation}
and
\begin{equation}
\Sigma_d(i\omega_n) = -\left[ i\omega_n\, a(i\omega_n)\, \tau_0 -
  c(i\omega_n)\, \tau_1 - s(i\omega_n)\, \tau_2 \right],
\end{equation}
where
\begin{eqnarray}
a(i\omega_n) & = & \frac{1}{2} \sum_{\alpha=L,R} \eta_\alpha(i\omega_n) \\
c(i\omega_n) & = & \frac{1}{2} \sum_{\alpha=L,R} \eta_\alpha(i\omega_n)\, \Delta_\alpha\, \cos \varphi_\alpha , \\
s(i\omega_n) & = & \frac{1}{2} \sum_{\alpha=L,R} \eta_\alpha(i\omega_n)\, \Delta_\alpha\, \sin \varphi_\alpha ,
\end{eqnarray}
and
\begin{equation}
\eta_\alpha(i\omega_n) = \frac{\Gamma_\alpha}{\sqrt{\Delta_\alpha^2 -
    (i\omega_n)^2}}.
\end{equation}
and $\Gamma_\alpha = 2\pi \rho_\alpha(0)\, u_\alpha^2$. The QD Green's
function can then be written as
\begin{eqnarray}
G_d(i\omega_n) & = & \frac{1}{{\rm Det}(i\omega_n)} \left\{ i\omega_n
[1+a(i\omega_n)] \tau_0 + \varepsilon_d\, \tau_3 \right. \nonumber
\\ & & \left. -\ c(i\omega_n)\, \tau_1 - s(i\omega_n)\, \tau_2
\right\},
\label{eq:GD_simple}
\end{eqnarray}
where
\begin{eqnarray}
{\rm Det}(i\omega_n) & = & \{i\omega_n[ 1 + a(i\omega_n)]\}^2 -
\varepsilon_d^2 - [c(i\omega_n)]^2 \nonumber \\ & &
-\ [s(i\omega_n)]^2.
\end{eqnarray}
%

\begin{figure}[ht]
  \centering
  \includegraphics[width=0.48\textwidth]{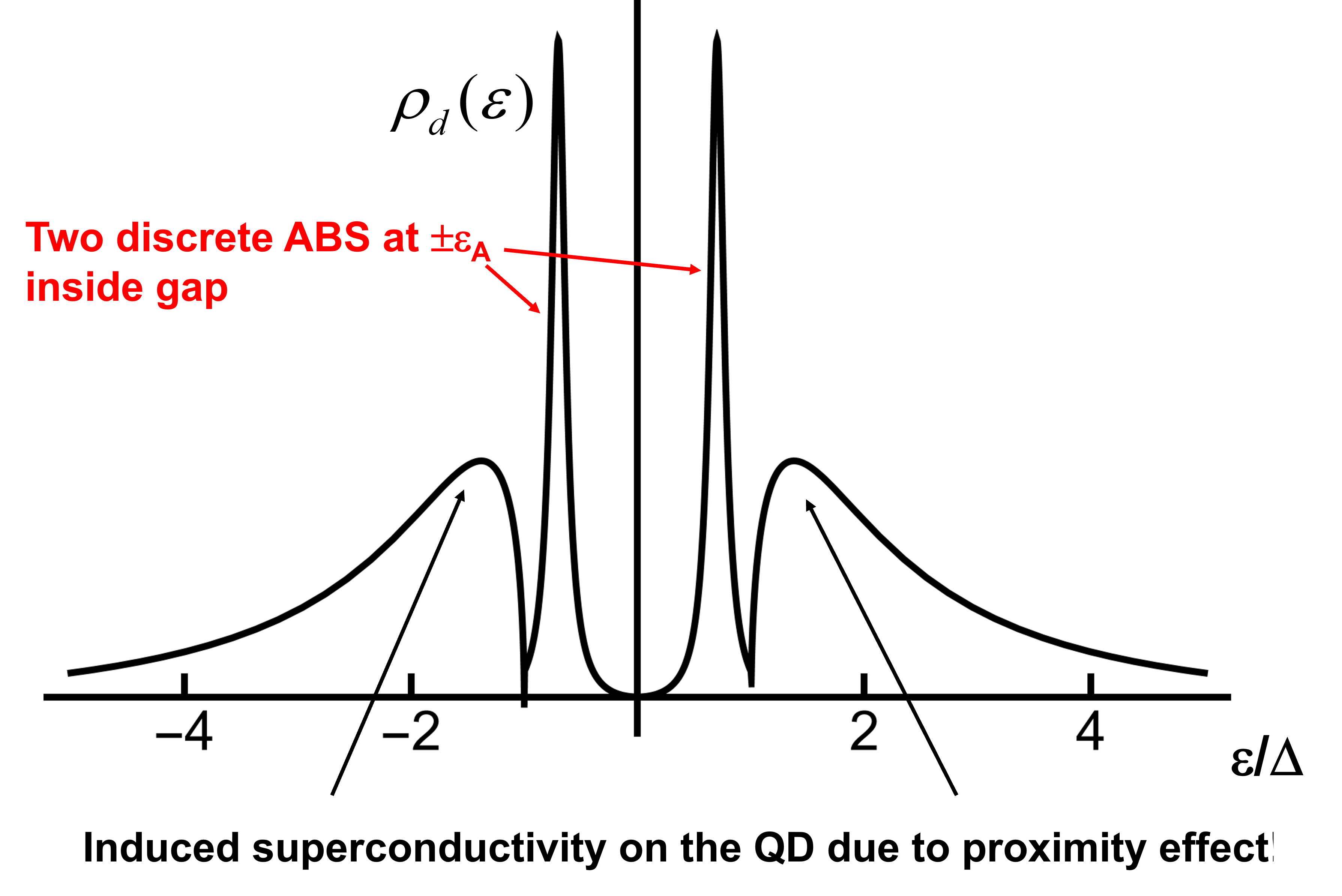}
  \caption{Density of states of the QD in the non-interacting limit when $U=0$.}
  \label{fig:DOS_U=0}
\end{figure}

\subsubsection{Density of states and Andreev bound states}

The density of states on the QD can be computed through the expression
\begin{equation}
\rho_d(\varepsilon) = -\frac{1}{\pi} {\rm Im}\, {\rm Tr}\left[ G^{\rm
    r}_{d}(\varepsilon) \right],
\end{equation}
Assuming identical superconducting leads ($\Gamma_\alpha=\Gamma$ and
$\Delta_\alpha=\Delta$) and performing the analytic continuation of
the Matsubara Green's function, see Eq.~(\ref{eq:analytical-cont}), we
obtain
\begin{equation}
{\rm Tr} \left[G_{d}^{\rm r}(\varepsilon) \right] = \frac{2}{{\rm
    Det}(\varepsilon_+)} \left\{ \varepsilon_+ [ 1+
  \eta(\varepsilon_+)] + {\varepsilon_d} \right\},
\end{equation}
where
\begin{equation}
{\rm Det}(\varepsilon_\pm) = \varepsilon_\pm^2 [1 +
  \eta(\varepsilon_\pm)]^2 - \varepsilon_d^2 - \left[
  \eta(\varepsilon_\pm) \Delta \cos \left( {\frac{\varphi}{2}} \right)
  \right]^2,
\label{eq:determinant}
\end{equation}
$\varphi  = \varphi_R - \varphi_L$, $\varepsilon_\pm = \varepsilon \pm i0^+$, and
\begin{equation}
\eta(\varepsilon_\pm) = \left\{ \begin{array}{ll} \Gamma/\sqrt{\Delta
      ^2 - \varepsilon^2} \mp i0^+, & |\varepsilon| < \Delta, \\ \mp
    i\, \Gamma\, {\rm sgn}(\varepsilon) / \sqrt{\varepsilon^2 - \Delta
      ^2}, & |\varepsilon| > \Delta.
\end{array} \right.
\end{equation}
The density of states can be rewritten as \cite{Meng_thesis}
\begin{equation}
\rho_d(\varepsilon) = W_+ \delta(\varepsilon - \varepsilon_A) + W_-
\delta (\varepsilon + \varepsilon_A)
\end{equation}
for $|\varepsilon| < \Delta$ and
\begin{eqnarray}
\rho_d(\varepsilon) & = &
\frac{2|\varepsilon|\Gamma/\pi}{\sqrt{\varepsilon^2 - \Delta^2}}
\nonumber \\ & & \times \left[ \varepsilon^2 + \varepsilon_d^2 +
  \frac{\Gamma^2}{\varepsilon^2 - \Delta^2} \left( \varepsilon^2 -
  \Delta^2 \cos^2 \left( \frac{\varphi}{2} \right) \right) \right]
\nonumber \\ & & \times \left\{\left[ \varepsilon^2 - \varepsilon_d^2
  + \frac{\Gamma^2}{\varepsilon^2 - \Delta^2} \left( \Delta^2
  \cos^2\left( \frac{\varphi}{2} \right) - \varepsilon^2 \right)
  \right]^2 \right. \nonumber \\ & & \left. +\ \frac{4\Gamma^2
  \varepsilon^4} {\varepsilon^2 - \Delta^2} \right\}^{-1}
\label{eq:DOS_U=0}
\end{eqnarray}
for $|\varepsilon| > \Delta$. Here, $\pm\varepsilon_A$ represent the
$\varphi$-dependent ABS energies, which can be obtained by setting
${\rm Det}(\varepsilon_A)=0$ for $|\varepsilon_A|<\Delta$; $W_\pm$ are
the associated spectral weights. For $|\varepsilon| > \Delta$, there
is a continuum of states outside the superconducting gap, see
Fig.~\ref{fig:DOS_U=0}.

\subsubsection{Josephson current.}

For computing the dc Josephson current through the QD, it is also
convenient to work with real-time Green's functions. Starting from
Eq.~(\ref{eq:supercurrent-2}) and noticing that $I_L = -I_R = I$, we
have
\begin{equation}
I = \frac{e}{2\hbar} \int \frac{d\varepsilon}{2\pi}\, f(\varepsilon)\, T(\varepsilon),
\end{equation}
where
\begin{eqnarray}
T(\varepsilon) & = & {\rm Tr} \left\{ \left[ \tau_3, G^{\rm
    a}_{d}(\varepsilon) - G^{\rm r}_{d}(\varepsilon) \right]
U_L^\dagger [g_L^{\rm a}(\varepsilon) + g_L^{\rm r}(\varepsilon) ]
U_L\right\}. \nonumber \\
\end{eqnarray}
We will first compute the contribution from the lead's Green's
functions, then from the dot's Green's functions, and finally insert
these contributions in the preceding expression.

For the lead's surface Green's functions, in the wide-band
approximation [Eq.~(\ref{eq:GF-surface})],
\begin{eqnarray}
g_L^{\rm a}(\varepsilon) + g_L^{\rm r}(\varepsilon) & \approx &
-2\pi\rho(0) \left( \varepsilon \tau_0 + \Delta \tau_1 \right)
\nonumber \\ & & \times \left\{ \begin{array}{l} 1/\sqrt{\Delta^2 -
    \varepsilon^2}, \ \ |\varepsilon| < \Delta \\ -i\,{\rm
    sgn}(\varepsilon)/\sqrt{\varepsilon^2 - \Delta^2},
  \ |\varepsilon|>\Delta \end{array} \right. ,
\label{eq:current-trace}
\end{eqnarray}
resulting in
\begin{eqnarray}
& & U_L^\dagger [g_L^{\rm a}(\varepsilon) + g_L^{\rm r}(\varepsilon) ]
  U_L \nonumber \\ & \approx & - \eta(\varepsilon) \left[
    \varepsilon\, \tau_0 - \Delta_L ( \tau_1 \cos\varphi_L + \tau_2
    \sin\varphi_L ) \right] \nonumber \\ && \times
  \left\{ \begin{array}{l} 1, \quad |\varepsilon| < \Delta_L \\ 0,
    \quad |\varepsilon|>\Delta_L \end{array} \right. .
\label{eq:leadfactor}
\end{eqnarray}
Note that the contribution to the current from states outside the
superconductor gap is not included in the coherent part of the
Josephson current here. Using Eq.~(\ref{eq:GD_simple}) and analytical
continuation, we obtain the QD Green's functions:
\begin{eqnarray}
G_{d}^{\rm r,a} (\varepsilon) & = & \frac{1}{{\rm
    Det}(\varepsilon_\pm)} \left\{ \varepsilon_\pm
[1+\eta(\varepsilon_\pm)] \tau_0 + \varepsilon_d\, \tau_3
\right. \nonumber \\ & & \left. - \Delta\, \eta(\varepsilon_\pm) (\cos
\varphi_L + \cos \varphi_R)\, \tau_1 \right. \nonumber \\ & & \left. -
\Delta\, \eta(\varepsilon_\pm) (\sin \varphi_L + \sin \varphi_R )\,
\tau_2 \right\},
\end{eqnarray}
It is then straightforward to show that
\begin{eqnarray}
& & \left[ \tau_3, G^{\rm a}_{d}(\varepsilon) - G^{\rm
      r}_{d}(\varepsilon) \right] \nonumber =
  2i\Delta\eta(\varepsilon) \nonumber \\ & & \times \left[ (\sin
    \varphi_L + \sin \varphi_R)\, \tau_1 - (\cos\varphi_L +
    \cos\varphi_R)\, \tau_2 \right] \nonumber \\ & & \times \left[
    \frac{1}{{\rm Det}(\varepsilon_-)} - \frac{1}{{\rm
        Det}(\varepsilon_+)} \right].
\label{eq:commutator}
\end{eqnarray}
Combining Eqs.~(\ref{eq:leadfactor}) and (\ref{eq:commutator}) and
taking the trace yields
\begin{equation}
T(\varepsilon) = 2i \Delta^2 [\eta(\varepsilon)]^2 \sin(\varphi)
\left[ \frac{1}{{\rm Det}(\varepsilon_-)} - \frac{1}{{\rm
      Det}(\varepsilon_+)} \right].
\end{equation}

To proceed further, we need to consider the zeros of ${\rm
  Det}(\varepsilon_\pm)$. We rewrite Eq.~(\ref{eq:determinant}) as
\begin{equation}
{\rm Det}(\varepsilon_\pm) = [1+\eta(\varepsilon)]^2 [
  \varepsilon_\pm^2 - \varepsilon_A^2(\varphi)],
\end{equation}
where the ABS energies satisfy
\begin{equation}
\varepsilon_A^2(\varphi) =
\frac{\varepsilon_d^2}{[1+\eta(\varepsilon_A(\varphi))]^2} + \left[
  \frac{\eta(\varepsilon_A(\varphi)) \Delta
    \cos(\varphi/2)}{1+\eta(\varepsilon_A(\varphi))} \right]^2.
\end{equation}
Therefore,
\begin{eqnarray}
T(\varepsilon) & = & 8i \Delta^2 \sin(\varphi) \left[
  \frac{\eta(\varepsilon)} {1 + \eta(\varepsilon)} \right]^2 \nonumber
\\ & & \times \left[ \frac{1}{\varepsilon_+^2 -
    \varepsilon_A^2(\varphi)} - \frac{1}{\varepsilon_-^2 -
    \varepsilon_A^2(\varphi)} \right] \nonumber \\ & = & -2\pi
\Delta^2 \frac{\sin(\varphi)} {\varepsilon_A(\varphi)} \left[
  \frac{\Gamma} {\sqrt{\Delta^2-\varepsilon_A^2(\varphi)}+\Gamma}
  \right]^2 \nonumber \\ & & \times \left[ \delta(\varepsilon -
  \varepsilon_A(\varphi)) - \delta(\varepsilon +
  \varepsilon_A(\varphi))\right],
\label{eq:trace-final}
\end{eqnarray}
where we have used the relationship
\begin{equation}
\frac{1}{\varepsilon_+^2 - \varepsilon_A^2} - \frac{1}{\varepsilon_-^2
  - \varepsilon_A^2} = \frac{i\pi}{2\varepsilon_A} \left[
  \delta(\varepsilon - \varepsilon_A) - \delta(\varepsilon +
  \varepsilon_A) \right].
\end{equation}
Plugging Eq.~(\ref{eq:trace-final}) into (\ref{eq:current-trace}), we
finally obtain a concise expression for the Josephson current,
\begin{eqnarray}
I & = & \frac{\pi e \Delta^2}{h}
\frac{\sin(\varphi)}{\varepsilon_A(\varphi)} \tanh \left(
\frac{\varepsilon_A(\varphi)}{2k_BT} \right) \left[ \frac{\Gamma}
  {\sqrt{\Delta^2-\varepsilon_A(\varphi)^2}+\Gamma} \right]^2. \nonumber \\ & & 
\end{eqnarray}

It is interesting to consider a few special cases.

\begin{itemize}
\item For $\Gamma \ll \Delta$ and $\varepsilon_d=0$,
\begin{equation}
\varepsilon_A(\varphi) = \pm\Gamma\, \cos(\varphi/2) + O(\Gamma^3)
\end{equation}
and
\begin{equation}
I \approx \frac{2\pi e\Gamma}{h} \sin(\varphi/2)\, \tanh\left(
\frac{\Gamma \cos(\varphi/2)} {2k_BT} \right).
\end{equation}
\item For $\Gamma \gg \Delta$ and $\varepsilon_d=0$,
\begin{equation}
\varepsilon_A(\varphi) = \pm \Delta\, \cos(\varphi/2) + O(\Gamma^{-1})
\end{equation}
and
\begin{equation}
I = \frac{2\pi e\Delta}{h} \sin(\varphi/2)\, \tanh\left( \frac{\Delta
  \cos(\varphi/2)} {2k_BT} \right).
\end{equation}
This matches the expression derived by Beenakker and van Houten for a
short ballistic junction ($L \ll \xi, l$) with a single propagating
channel \cite{BEENAKKER19911}. It also matches the result obtained by
Kulik and Omel'yanchuk for short and narrow ballistic junctions when
$T=0$ \cite{Kulik1978}. The appropriateness of the ballistic regime
makes sense since transport across the dot is resonant
($\varepsilon_d=0)$ and the coupling is strong ($\Gamma \gg \Delta)$,
implying that the QD does not mix the lead modes.
\item For $\Gamma \rightarrow 0$ and $0 < |\varepsilon_d| < \Delta$,
\begin{equation}
\varepsilon_A(\varphi) = \pm \varepsilon_d \left( 1 -
\frac{\Gamma}{\sqrt{\Delta^2 - \varepsilon_d^2}}\right) +
O(\Gamma^{2})
\end{equation}
and
\begin{eqnarray}
I & = & \frac{\pi e\Delta^2\Gamma^2}{h} \frac{\sin
  \varphi}{\varepsilon_d(\Delta^2-\varepsilon_d^2)}\, \tanh\left(
\frac{\varepsilon_d} {2k_BT} \right) + O(\Gamma^4). \nonumber \\ & & 
\end{eqnarray}
This result differs from the classical expression derived by
Ambegaokar and Baratoff for a one-dimensional weak link
\cite{AmbegaokarBaratoff63}. The reason is that the energy-dependent
normal-state resistance across the dot and the ABS are neglected in
Ref.~\cite{AmbegaokarBaratoff63}.
\end{itemize}

\section{Tight-Binding Modeling of Josephson Junctions}
\label{sec:TB-model}

This section goes beyond the idealized systems considered in
Sec.~\ref{sec:JJ-compact} and discusses realistic atomistic-level
modeling of systems. Here, explicit analytical expressions are not
possible, and efficient numerical implementations are required. We
will start by considering the construction of an appropriate
Hamiltonian. This will be followed by a discussion of the capabilities
of the methodology of Sec.~\ref{sec:dc-regime} with reference to
results of Ref.~\cite{Nieminen2023}. Finally, we will give an in-depth
account of how material-specific spin-orbit coupling (SOC) matrix
elements can be constructed, along with the construction of elements
of superconducting-order-parameter matrix beyond the on-site singlet
terms.

The advantage of using Eq.~(\ref{eq:supercurrent-4}) in
materials-specific atomistic-scale modeling is that it allows one to
calculate the Green's functions of parts of the system
independently. The superconducting leads are incorporated as
self-energy terms in the barrier Green's function, so that a
Hamiltonian written on a large atomic-orbital-basis set can be divided
into smaller blocks. Furthermore, because of the locality of this
Hamiltonian, one can use recursive methods to solve Dyson's equation
for constructing Green's functions for the semi-infinite lead
structures, as well as for the barrier region.

The tight-binding Hamiltonians for the barrier region and the
superconducting electrodes must, of course, yield electronic
structures that respect the corresponding first-principles results or
are based on experimental data (e.g., photoemission spectra). A useful
starting point for building such a Hamiltonian is the Slater-Koster
(SK) approach \cite{Slater_Koster, Harrison}, which gives correct
angular dependencies of the hopping integrals between the atomic
orbitals and their relative magnitudes for different types of bonding
($\sigma$, $\pi$, and $\delta$). Hence, the task is to fit the on-site
matrix elements and the amplitudes of the relevant overlap terms to
realistic electronic structures. We emphasize that the barrier region
plays a critical role because this is where the electronic spectrum is
modified by the proximity effect and the ABSs are formed. Therefore,
details of the tunneling barrier and the symmetry of the order
parameter and the singlet or triplet nature of the Cooper pairs must
be incorporated properly. However, we do not expect the results to be
sensitive to the details of the electronic structure of the
superconducting electrodes.

The matrix elements coupling the leads and the tunneling barrier
require special attention because of their key role in the proximity
effect. In generic one-dimensional junctions, a single parameter is
usually sufficient to characterize the interaction between the leads
and the barrier. But, in more realistic junctions, the two surfaces of
the barrier will, in general, have different orientations and lattice
constants, so that multiple parameters and large simulation cells will
be required to capture the contributions of hoppings between the
various interfacial orbitals; here, the most important are the surface
orbitals of the barrier, which are intimately involved in the
proximity effect.

\begin{figure*}
  \centering
  \includegraphics[width=\textwidth]{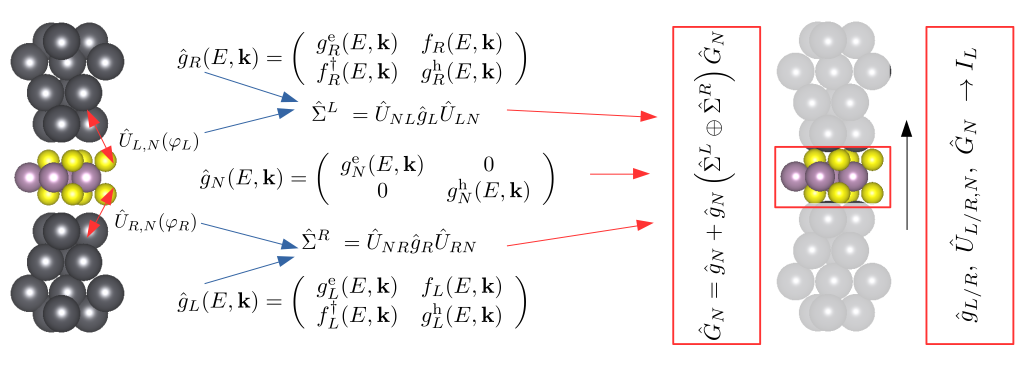}
  \caption{Steps involved in supercurrent calculations, shown
    schematically, are: (i) Green's functions for the non-interacting
    leads and tunneling barrier are obtained; (ii) Self-energy
    matrices for the junction are calculated using the lead Green's
    functions and the interaction matrices; and, (iii) $\hat{G}_{N}$
    is used to obtain ABSs and other features such as the real-space
    mappings of the anomalous matrix elements of the Nambu-Gorkov
    Green's function.}
  \label{fig:parts_of_junction}
\end{figure*}

In the calculations of Ref.~\cite{Nieminen2023}, the methodology of
Sec.~\ref{sec:JJ-compact} was applied to a vertical JJ where the
scattering region consisted of one or more MoS$_{2}$ layers sandwiched
between two generic s-wave symmetric semi-infinite superconducting
leads that mimicked the fcc(111)-structure of Pb
(Fig.~\ref{fig:parts_of_junction}). The system Hamiltonian was written
in the form
\begin{equation}   
\label{eq:TB_hamiltonian}
        {\cal H}={\cal H}_{N}+{\cal H}_{L} + {\cal H}_{R}+{\cal
          U}_{N,L} +{\cal U}_{N,R}.
\end{equation}
Three parts of this Hamiltonian can be written in terms of on-site
energies and SK hopping integrals and augmented with spin-orbit
coupling and matrix elements for superconducting pairing:
\begin{equation*}
    {\cal H}_{N} = \sum_{a,b \in \rm{N},
      \sigma=\uparrow,\downarrow}\left( \varepsilon_{a}\,
    c^{\dagger}_{a\sigma} c_{a \sigma} + V_{a b}\, c^{\dagger}_{a
      \sigma} c_{b \sigma} \right) + {\cal H}_{\rm SOC} \nonumber
\end{equation*}
\begin{equation*}
    {\cal H}_{L/R}= \sum_{a,b \in \rm{L/R},
      \sigma=\uparrow,\downarrow} \left( \varepsilon_{a}\,
    c^{\dagger}_{a\sigma} c_{a \sigma} + V_{a b}\, c^{\dagger}_{a
      \sigma} c_{b \sigma} \right) + {\cal H}_{\rm SC} \nonumber
\end{equation*}
\begin{equation*}
    {\cal U}_{N,L/R}= \sum_{a \in \rm{N} ,b \in \rm{L/R},
      \sigma=\uparrow,\downarrow} V_{a b}\, c^{\dagger}_{a \sigma}
    c_{b \sigma}. \nonumber
\end{equation*}
Here, $N$ refers to the barrier, $L/R$ refers to the left/right
electrodes, $a$ and $b$ are orbital indices, and $\varepsilon_a$ and
$V_{ab}$ are tight-binding hopping parameters. ${\cal H}_{N}$ and
${\cal H}_{L/R}$ are used to calculate the decoupled block Green's
function $\hat{g}_{N}$ and $\hat{g}_{L/R}$, respectively
(Fig.~\ref{fig:parts_of_junction}). ${\cal H}_{\rm SOC}$ codifies SOC
contributions, and ${\cal H}_{\rm SC}$ encodes the superconducting
leads. The tight-binding parameters were obtained using SK hopping
integrals \cite{Slater_Koster} with fitted hopping amplitudes and
on-site energies; see also Ref.~\cite{Harrison} on how multiorbital
tight-binding Hamiltonians can be constructed. In the barrier part,
the basis consisted of a set of $\{s,p_{x},p_{y},p_{z}\}$ orbitals of
sulphur and $\{s,d_{z^{2}},d_{xz},d_{yz},d_{xy},d_{x^{2}-y^{2}}\}$
orbitals of molybdenium. SOC matrix elements for the $d$ orbitals of
Mo atoms can be obtained following Ref.~\cite{Konschuh} (see also
below). An interesting consequence of the SOC for an odd number of TMD
layers is spin-valley coupling, which is reflected in the
$k$-dependence of the spin-resolved ABSs.

The Hamiltonian for the superconducting leads is constructed to
reproduce the most important features of the electronic bands of Pb in
the vicinity of Fermi energy based on $\{s,p_{x},p_{y},p_{z}\}$
orbitals. There is a substantial lattice mismatch between the fcc(111)
surfaces of Pb and MoS$_{2}$, but this is eased by rotating the
orientation of the Pb surfaces by $30^{\circ}$ and slightly tuning the
lattice constant of Pb. Matrix elements of the order parameter are
compatible with a singlet $s$-wave symmetry and modeled with anomalous
matrix elements between the $p$ orbitals of the Pb atoms. The
parameterization used in Ref.~\cite{Nieminen2023} was based on
Refs.~\cite{Trainer2020} and
~\cite{Trainer2022}. Reference~\cite{Trainer2022} unfolded the bands
into the primitive cell of an overlayer (equivalent to the barrier
material in a JJ) to deconstruct the contribution of the substrate
(electrode) to the electronic structure of the overlayer, which proved
useful in fitting parameters of the interaction hopping integrals.

As depicted in Fig.~\ref{fig:parts_of_junction}, the Green's functions
for the superconducting leads and the barrier are calculated
separately. Surface Green's functions of the semi-infinite leads are
computed with a fast-converging method based on coupling replicas of a
minimal slab of the leads and solving Dyson's equation recursively for
the interfacial orbitals between the slab and its replica only. Since
the system is doubled for each recursion without increasing the size
of the matrix involved, the size of the structure modeled in this way
increases by $2^{n+1}$ after $n$ recursion steps.

\begin{figure*}
  \centering
  \includegraphics[width=\textwidth]{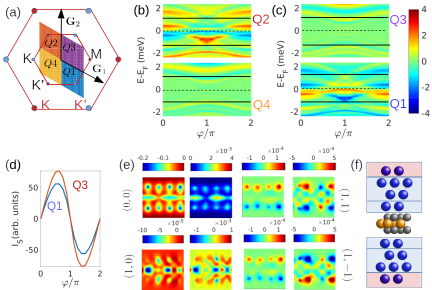}
  \caption{Collection of quantities that can be calculated using the
    Green's function method. (a) The horizontal Brillouin zone
    (BZ$_{p}$) of MoS$_{2}$ (outer hexagon), and the Brillouin
    zone (BZ$_{s}$) (inner hexagon) for the $3\times 3$
    computational supercell used in Ref.~\cite{Nieminen2023}. Two
    non-equivalent valley points K and K' are indicated, as well as
    the four quadrants used to reveal spin-valley coupling in ABSs. (b)
    and (c) Spin-polarized dispersion of ABSs projected on different
    quadrants of the Brillouin zone indicated in (a): quadrants Q2 and
    Q4 (b), and quadrants Q3 and Q1 (c). Note the spin-valley coupling
    in the quadrants containing the $K$ and $K'$ points. (d)
    Supercurrent across a monolayer projected in two non-equivalent
    quadrants (Q1 and Q3). (e) Projection of singlet ($s=0,~m_{s}=0$)
    and triplet ($s=1,~m_{s}=-1,~0,~1$) components of the real and
    imaginary parts of the anomalous Green's function at the barrier
    location. (f) Side view of the atomic configuration of the
    junction. See Ref.~\cite{Nieminen2023} for details.}
  \label{fig:TMD_Josephson}
\end{figure*}

Figure~\ref{fig:TMD_Josephson} presents a representative collection of
properties that can be computed as reported in
Ref.~\cite{Nieminen2023}. Notably, the Green's function formalism
allows one to straightforwardly parse various contributions to the
supercurrent. For instance, one can confine the calculations to
different parts of the Brillouin zone to obtain $k$-dependent results,
which give insight into how breaking the horizontal symmetry of the
barrier can affect the spin polarization of the supercurrent. The
method also allows the computation of spin-resolved dispersion of
ABSs. Furthermore, one can produce a real-space projection of the
proximity-induced superconductivity in the barrier region and
decompose it into singlet and triplet components of the anomalous part
of the Green's function. Another important result is the current-phase
relation for junctions of different thicknesses, which closely
resembled the available experimental data \cite{Island_2016}, see
Fig.~3c of Ref.~\cite{Nieminen2023}.

\subsection{Further insights into tight-binding parameterization}

In Ref.~\cite{Nieminen2023}, the Hamiltonian terms ${\cal H}_{\rm
  SOC}$ and ${\cal H}_{\rm SC}$ were constructed in a relatively
simple way. For the SOC, the matrix elements derived for $d$ orbitals
in Ref.~\cite{Konschuh} were employed. In the following, we discuss in
depth how the SOC term can be systematically built within a
tight-binding framework including the beyond-on-site contributions.

We start with the expression for on-site elements used in
Ref.~\cite{Konschuh}, which was also used for the
next-to-nearest-neighbor matrix elements in Ref.~\cite{Liu2011}:
\begin{equation}
  \label{eq:SOCtheory}
  U_{\rm SOC} = -\frac{g \mu_{B}}{2mc^{2}} \mathbf{\sigma}
  \cdot\mathbf{p} \times {\nabla V}.
\end{equation}
For the on-site case, Eq.~(\ref{eq:SOCtheory}) can be rewritten as
\begin{equation}
  \label{eq:SOCnext}
  U_{\rm SOC} = \frac{g \mu_{B}}{2rmc^{2}}\frac{\partial V}{\partial
    r}\mathbf{L}\cdot {\bf \sigma}.
\end{equation}
In Ref.~\cite{Konschuh}, this expression led to
\begin{equation}
  \label{eq:SOCKonschuh}
  {\cal H}_{\rm SOC} =\lambda \sum_{a,b}
  \sum_{\sigma,\sigma'=\uparrow,\downarrow} \langle a, \sigma \vert
  \mathbf{L}\cdot {\bf \sigma} \vert b, \sigma'\rangle
  c^{\dagger}_{a \sigma} c_{b \sigma'},
\end{equation}
where $a$ and $b$ are atomic-orbital indices. In going from
Eq.~(\ref{eq:SOCnext}) to (\ref{eq:SOCKonschuh}), the prefactor in
Eq.~(\ref{eq:SOCnext}) is assumed to be the same for all orbitals with
quantum number $l$ and approximately equal to the parameter $\lambda$
in Eq.~(\ref{eq:SOCKonschuh}), which is obtained by fitting the band
structure. Matrix elements of $\mathbf{L}\cdot {\bf \sigma}$ are
straightforward to calculate since
\begin{equation}
  \label{eq:angular_and_spin}
 \mathbf{L}\cdot {\bf \sigma} = \frac{\hbar}{2} \left(
\begin{array}{cc} L_{z} & L_{-} \\
L_{+} & -L_{z}
\end{array}
\right).
\end{equation}
Note that the matrix elements in Eq.~(\ref{eq:SOCKonschuh})
automatically include the regular SOC as well as the Rashba terms, as
tabulated in Refs.~\cite{Konschuh, Nieminen2023}. For transition-metal
atoms with strong $d$-orbital contributions,
Eq.~(\ref{eq:angular_and_spin}) gives on-site terms such as
$H_{SOC}^{z^{2},yz} = \lambda i \sqrt{3} \sigma_{x}$ and
$H_{SOC}^{xy,yz} = \lambda i \sigma_{y}.$

In materials with weak on-site contributions to SOC, the
next-to-nearest-neighbor matrix elements can play an important
role. An example is a silicene ribbon \cite{Saari1, Saari2}. To obtain
this contribution to SOC, we go back to Eq.~(\ref{eq:SOCtheory}) and
approximate the direction of the momentum operator by the unit vector
connecting the two next-to-nearest neighbors when there is a non-zero
potential gradient pointing away from the intermediate atom in the
plane spanned by the three atoms. Hamiltonian matrix elements such as
those obtained in Refs.~\cite{Saari1, Saari2,Liu2011} are necessary
when considering the Josephson effect through barriers with a weak
on-site contribution to SOC.

The superconducting leads in JJs are traditionally made of
conventional superconductors, and thus simple parameterized s-wave
symmetric Hamiltonian matrix elements are sufficient for modeling
transport and other properties. For example, in
Ref.~\cite{Nieminen2023}, the matrix elements of the SC order
parameter were parameterized by intra-orbital terms with singlet
pairing. However, it is not uncommon for electrodes to be made of
materials with unconventional superconductors such as NbSe$_{2}$. An
example is reported in Ref.~\cite{Kim2017}, which considers vertically
stacked NbSe$_2$-graphene-NbSe$_2$ van der Waals junctions. A simple
s-wave parameterization is not appropriate for superconducting
variants of graphene such as twisted layers, metal-decorated
monolayers or intercalated multilayered graphene structures, or
high-$T_c$ (HTS) cuprates with d-wave superconductivity.

In the tight-binding formulation, the superconducting part of the
Hamiltonian can be generally written as
\begin{eqnarray}
{\cal H}_{\rm SC} & = & \sum_{a,b}
\sum_{\sigma,\sigma'=\uparrow,\downarrow} \sum_\mu \left[ \Delta_{a
    \sigma b \sigma'}(\mu)\, c^{\dagger}_{a \sigma}\, c^{\dagger}_{b
    \sigma'} \right. \nonumber \\ & & \left.  +\ \Delta_{b \sigma' a
    \sigma}^{\dagger}(\mu)\, c_{b \sigma'}\, c_{a \sigma} \right]
 \label{eq:anomalous_hamiltonian}
\end{eqnarray}
in the basis $\{| a \uparrow \rangle, | b \uparrow \rangle, | a
\downarrow \rangle, | b\downarrow \rangle\}$, with $\mu$ referring to
the symmetry of the order parameter. The latter can be determined
using symmetry arguments or, in some cases, self-consistently, as
discussed below.

While the superconductor order parameter was modeled in
Ref.~\cite{Nieminen2023} with generic s-wave on-site matrix elements,
a more sophisticated approach is needed for materials which present
several $d$ orbitals and strong SOC~\cite{Chiu2025}. Also, the
superconductivity in TMDs such as NbSe$_2$ is not conventional, which
along with a relatively strong SOC, make modeling more demanding (see,
e.g., Refs.~\cite{Margalit, Mockli}). Realistic modeling of the
electronic structures of TMDs requires a relatively large orbital
basis set~\cite{Kadek2023}, so that a sensible strategy is to start
with a limited set of orbitals for the matrix elements of the
superconductor order parameter.

An elegant example of a multiorbital tight-binding Hamiltonian for
superconducting TMDs is presented in Ref.~\cite{Mockli}, where the
outcome of a self-consistent calculation for a three-orbital
Hamiltonian for NbSe$_{2}$ is found to yield mixed-parity
superconductivity with singlet and triplet matrix elements. An
advantage of this approach is that the matrix elements related to
possible unconventional superconductivity can also emerge
naturally. The normal state electronic structure is modeled with a
minimal set of $d$ orbitals of the transition-metal atom, consisting
of the basis of $\{d_{z^{2}},d_{xy},d_{x^{2}-y^{2}}\}$ orbitals. A
possible basis for the Cooper pairs is built using a group-theoretical
method including SOC. Interestingly, singlet-parity superconductivity
is related only to the $d_{z^{2}}$-orbitals, whereas a mixed-parity
results from the combinations of the $d_{xy}$ and $d_{x^{2}-y^{2}}$
orbitals. The same kind of matrix elements for the order parameter
were obtained from symmetry arguments in Ref.~\cite{Margalit}. In its
general form, the self-consistent order parameter yields matrix
elements such as
\begin{equation}
    \label{eq:order_parameter}
    \Delta_{a \sigma b \sigma'} = (i \sigma_{y})_{\sigma \sigma'}\,
    \sum_{c,d} U^{cd}_{ a b} \langle\, c_{c \sigma} c_{d \sigma'}
    \rangle,
\end{equation}
which is a generalization of Eq.~(\ref{eq:mean-field-approx}). The
electron-electron interaction is, in general, a fourth-rank tensor
$U^{c d}_{a b}$ in the atomic-orbital basis since it is a two-site
function in real space. To simplify the calculations,
Ref.~\cite{Mockli} uses only diagonal on-site elements of the pairing
interactions $U^{d_{z^{2}}}, U^{d_{xy}},~U^{d_{x^{2}-y^{2}}}$ for the
transition metal, but Ref. ~\cite{Margalit} uses an interaction
tensor, which also leads to inter-orbital terms. The pairing
correlation term is, in principle, straightforward to calculate from
the anomalous part of the Nambu-Gorkov Green's function, see
Eq.~(\ref{eq:anomalous-GF}). The order parameter of
Eq.~(\ref{eq:order_parameter}) can then be obtained self-consistently
by solving the Nambu-Gorkov Green's function for the Hamiltonian with
the pairing correlations computed from Eq.~(\ref{eq:sc_correlation})
to get the next-order approximation for the order parameter. Notably,
due to SOC, mixed-parity pairing can be generated in this way despite
a conventional pairing interaction between the electrons.

Reference~\cite{Kim2017} shows the relevance of TMDs as a material for
superconducting leads in JJs. In addition to this specific example,
there is a wide variety of materials that could be used as
superconducting leads, but the relevant features are very system
specific. Graphene, for example, may not only act as a barrier
material in the normal state, but it can also assume a superconducting
state for an electrode material; superconductivity in graphene can be
induced via methods such as the proximity effect, metal decoration of
monolayers, intercalating multilayer films \cite{Kotov2012}, and
creating twisted multilayers \cite{Cao2018}. To illustrate the
tight-binding modeling methods here, we note a few different
cases. For JJs with s-wave superconducting leads, Black-Schaffer and
Doniach \cite{Black-Schaffer2008} report a tight-binding model for an
experimentally feasible graphene-based SNS junction where the
superconducting leads consist of heavily doped graphene layers
attached to superconducting metal electrodes. They derive
self-consistent equations for the on-site matrix elements of the
superconducting order parameter. This approach has been extended by
Linder {\it et al.} \cite{Linder_2009} to unconventional
superconductivity with s and d wave symmetries with nearest-neighbor
matrix elements of the order parameter. Another example of
tight-binding modeling of graphene-based systems is a JJ involving
twisted graphene bilayers studied by Munoz {\it et al.}
\cite{Munoz_2012}. As in Ref.~\cite{Black-Schaffer2008}, the matrix
elements of the anomalous part of the Hamiltonian in
Ref.~\cite{Munoz_2012} are obtained from self-consistent calculations
of pairing correlation. A special methodological aspect is the
computation of Green's functions using Chebyshev-Bogoliubov-de-Gennes
method. Yet another class of superconducting graphene consists of
intercalated graphene layers and metal-decorated graphene
monolayers. Uchoa and Castro Neto studied this system using a
different methodology \cite{Uchoa2007}. Their derivation of the matrix
elements for the order parameter is based on a self-consistent
calculation of the pairing correlation. However, in the interaction
mechanism they also considered electron-phonon and electron-plasmon
couplings.

The cuprates provide a different materials family which exhibits
high-temperature superconductivity (HTS) with a d-wave superconducting
order parameter \cite{Ding1996,Tsuei2000}. Bi$_{2}$Sr$_{2}$Ca
Cu$_2$O$_{8+x}$ (BSCCO) which is an extensively studied HTS material
is also relevant for JJ systems. Recent experiments involving JJs with
twisted BSCCO flakes \cite{Zhu_2021} suggest that there may also be
s-wave contributions to the order parameter. Accurate
material-specific modeling of cuprate-based JJs would require a
relatively large orbital-basis set. A useful starting point could be
the three-band model, see, e.g, Ref. \cite{Dagotto1994}, which takes
into account the $d_{x^{2}-y^{2}}$ orbitals of Cu and the $p_{x}$ and
$p_{y}$ orbitals of the O atoms connecting the adjacent Cu
atoms. Since the tunneling barrier in the junction would not
necessarily be in direct contact with the CuO$_{2}$ layer,
material-specific modeling would require a tight-binding basis that
includes the relevant atomic orbital of all other atoms. This was done
in interpreting scanning tunneling spectroscopy and photoemission
experiments from BSCCO in Refs.~\cite{Nieminen2009, Nieminen2012} by
using a full set of $d$ orbitals of Cu atoms, and $s$ and $p$ orbitals
of all other atoms, where the superconductor-order-parameter matrix
elements were implemented in a parameterized form as d-wave-symmetric
matrix elements between the adjacent Cu atoms. A similar multi-orbital
model would be appropriate for modeling JJs involving the cuprates.


\subsection{Beyond Slater-Koster parameterization}
\label{sec:beyondSK}

It is clear that there are several challenges in constructing a
realistic material-specific Hamiltonian for NEGF simulations using
parameter fitting in atomic-orbital basis. These challenges include
(i) the crucial role of the interface matrix elements in the proximity
effect, (ii) materials with complicated electronic structure, where
the role of individual atomic orbitals to the relevant eigenstates is
not obvious, and (iii) the parameterization of pairing matrix elements
in the case of unconventional superconductivity.  Although
Slater-Koster overlap integrals can correctly incorporate the symmetry
properties inherited from the atomic orbitals, it can be tedious to
derive a parameterization that can faithfully reproduce the electronic
structures of various parts of the junction and the interface. A
significant improvement would be to connect tight-binding NEGF
simulations more closely to DFT calculations, e.g., by utilizing DFT
calculations based on linear combinations of atomic orbitals (LCAO),
or by using projection methods to convert plane-wave or grid-based DFT
calculations to obtain effective tight-binding Hamiltonians.

A route beyond semi-empirical Slater-Koster fitting is to derive a
fully \textit{ab initio} tight-binding Hamiltonian directly from the
DFT results using Wannier functions \cite{Wannier1937}, which provide
a bridge between the delocalized Bloch representation of the DFT based
wavefunctions and the localized orbital picture underlying
tight-binding models \cite{ Marzari2012}. The key idea here is to use
the Wannier transformation to map Bloch functions onto a set of
localized, orthonormal functions in real space, which can then serve
as the basis for an accurate tight-binding Hamiltonian. This
projection method offers a practical route for constructing
well-localized Wannier functions, where the target Bloch states are
projected onto a set of trial orbitals (typically atomic-like $s$,
$p$, or $d$ orbitals) to determine both the relevant subspace and the
appropriate gauge \cite{Souza2001}. Further optimization can be
achieved through the maximally-localized Wannier function approach
\cite{Marzari1997}, which yields Wannier functions with minimal
spatial extent. A standard workflow uses plane-wave DFT codes such as
VASP \cite{Kresse1996,Kresse1999} or Quantum
Espresso~\cite{Giannozzi2017} followed by projection onto Wannier
orbitals using Wannier90 \cite{wannier90_2008, wannier90_2020}.

Once the Wannier tight-binding parameters $t_{ij}(R)=\langle
w_{i0}|H|w_{jR}\rangle$ are obtained on a coarse reciprocal-space
mesh, the Hamiltonian can be efficiently interpolated to arbitrary
$k$-points \cite{Yates2007}. This enables the calculation of transport
properties that require dense reciprocal-space sampling, such as the
intrinsic anomalous Hall conductivity \cite{Wang2006, Yang2021,
  Zhu2020, Yang2020} and intrinsic spin-Hall conductivity
\cite{Qiao2018,Ryoo2019}. In addition, Wannier tight-binding models
serve as compact first-principles inputs for Green's function based
methods, including the iterative decimation schemes for surface and
edge spectral functions \cite{LopezSancho1985, Wang2019, Dhakal2021,
  Chiu2023}, and coherent quantum transport calculations
\cite{Calzolari2004, Lee2005, Thygesen2005}. Fully self-consistent
first-principles calculations of transport in devices connect
naturally to DFT-based NEGF implementations and the related transport
solvers \cite{Strange2008,Thygesen2007,Thygesen2008, Afzalian2021}. In
a recent work by Bekaert et al.~\cite{Bekaert2025}, real-space
superconducting properties of NbSe$_{2}$ have been simulated using
Migdal-Eliashberg equations from the electron-phonon Wannier code
based on ab initio Quantum Espresso~\cite{Giannozzi2017, Ponce2016}
calculations.

Complementary to Wannier downfolding, tight-binding Hamiltonians can
also be exported directly from first-principles calculations performed
in a localized orbital basis. This approach is particularly robust for
modeling complex, large-scale heterostructures where capturing the
precise geometric and electronic structure of interfaces is
critical. A prominent example is the projector expansion method using
pseudo-atomic localized basis functions (PAOs), as implemented in the
OpenMX package by Ozaki and Kino \cite{Ozaki2003, Ozaki2005}. As an
example of an ab initio simulation of a complex large-scale system in
an atomic orbital basis, the recent work by Mahfouzi {\it et al.}
\cite{Mahfouzi2020} considers the spin-orbit torque in ferromagnetic
heterostructures. The PAO approach has also been implemented for NEGF
in electronic transport calculations ~\cite{Ozaki2010}.

The accuracy of any ab initio tight-binding model, whether constructed
via Wannierization or another projection scheme, inherits the accuracy
of the underlying exchange-correlation functional used to generate the
DFT electronic structure. Here, recent progress in constructing the
strongly-constrained-and-appropriately-normed (SCAN)
meta-generalized-gradient-approximation (meta-GGA) functional
\cite{SCAN2015}, has demonstrated significant improvement over the
standard GGA \cite{Perdew1996} across diverse material systems
\cite{Chiu2025,Patra2017, Furness2018, Lane2018, Zhang2020, Chiu2020}.
For JJ modeling, first-principles tight-binding methods are attractive
for constructing the normal-region Hamiltonian with accurate,
material-specific electronic structure. However, for the lead region,
this approach typically yields only the normal-state Hamiltonian; the
superconducting order parameter must still be added phenomenologically
or determined via a separate self-consistent gap equation, as standard
DFT does not natively capture the Cooper pairing instability. This
motivates the development of fully self-consistent DFT-based
approaches that treat superconductivity on an equal footing.

A further advance would involve the development of a combined DFT-BdG
approach for NEGF simulations. Theoretical formulations that combine
DFT and BdG are already several decades old, starting with the work of
Oliveira {\it et al.}  \cite{Oliveira1988}, where Kohn-Sham-like
equations describe the coupled electron and hole degrees of freedom
and incorporate exchange and correlation terms into the BdG
equations. An important intermediate implementation of
Kohn-Sham-Bogoliubov-de Gennes (KS-BdG) equations is the work by
Suvasini {\it et al.}  \cite{Suvasini1993}, who adapt a local
approximation of the electron-electron interaction to simplify the
self-consistent calculation of the superconducting order
parameter. Recent generalizations of the formulation by Suvasini {\it
  et al.} utilize Nambu spinors to incorporate spin-dependent
features\cite{Reho2024, Russmann2022, Russmann2023}. An interesting
recent implementation of DFT-BdG is by Reho {\it et al.}
\cite{Reho2024}, who incorporate DFT-BdG into the SIESTA package
\cite{Soler2002, Garcia2020}, which uses LCAO basis. This suggests
that the Hamiltonian matrix elements utilized in SIESTA-BdG could be
directly utilized in tight-binding NEGF simulations. In the following,
therefore, we discuss the DFT-BdG approach with focus on SIESTA-BdG.

In order to incorporate SOC and magnetic interactions, we start by
writing the eigenfunctions in terms of the Nambu spinors
$\Psi_{i}(\mathbf{r}) = (u^{\uparrow}_{i}(\mathbf{r}),
u^{\downarrow}_{i}(\mathbf{r}), v^{\uparrow}_{i}(\mathbf{r}),
v^{\downarrow}_{i}(\mathbf{r})),$ where $i$ is the band index. For
diagonalizing the spin-generalized KS-BdG equations, the
eigenfunctions are written in a LCAO basis:
$u^{\alpha}_{i}(\mathbf{r}) = \sum_{\mu} \varphi_{\mu}(\mathbf{r})
u^{\alpha}_{i\mu},$ and $v^{\alpha}_{i}(\mathbf{r}) = \sum_{\mu}
\varphi^{*}_{\mu}(\mathbf{r}) v^{\alpha}_{i\mu}.$ Here, $\alpha$ is
the spin index and $\mu = \{I,n,l,m_{l}\}$ is a composite index that
contains atomic labels and the quantum numbers of eigenfunctions for a
spherical potential. The functionals in KS-BdG equations depend on the
normal electron density, which is defined as
\begin{equation}
    \rho(\mathbf{r}) = \sum_{\alpha,i} \left\{ \vert
    u^{\alpha}_{i}(\mathbf{r}) \vert^{2}f(E_{i})+\vert
    v^{\alpha}_{i}(\mathbf{r}) \vert^{2}\left[ 1-f(E_{i}) \right]
    \right\},
\end{equation}
and on the anomalous density, 
\begin{eqnarray}
    \chi^{\alpha \beta}(\mathbf{r},\mathbf{r'}) & = & \sum_{i} \left\{
    u^{\beta}_{i}(\mathbf{r'})v^{\alpha*}_{i}(\mathbf{r})f(E_{n})
    \right. \nonumber \\ & &
    \left. +\ u^{\alpha}_{i}(\mathbf{r})v^{\beta*}_{i}(\mathbf{r'})
    \left[ 1-f(E_{n}) \right] \right\},
\end{eqnarray}
which is a second-rank $2\times 2$ tensor generalization introduced in
Ref.~\cite{Suvasini1993}. A further generalization is the tensor form
of the effective single-particle $V^{\alpha}_{\beta, {\rm eff}}$ and
the pairing $\Delta^{\alpha, {\rm eff}}_{\beta}$ potentials. In the
single-particle potential, SOC is included in a $2\times 2$ spinor
form. in the pairing potential The possibility of triplet pairing, in
addition to the singlet pairing, is included in the pairing potential,
which results in four components of the pairing matrix.

For spin-generalized KS-BdG, the Nambu-Gorkov Hamiltonian is:
\begin{eqnarray}
(T-\mu+V^{\alpha}_{\beta, {\rm eff}})u^{\beta}_{i}(\mathbf{r}) + \int
  d^{3}r'\, \Delta^{\alpha, {\rm eff}}_
  {\beta}(\mathbf{r},\mathbf{r'})v^{\beta}_{i}(\mathbf{r')}
  \nonumber\\ = E_{i}\, u^{\alpha}_{i}(\mathbf{r})
  \\ (T-\mu+V^{\alpha*}_{\beta, {\rm eff}})v^{\beta}_{i}(\mathbf{r}) -
  \int d^{3}r'\, \Delta_{\beta}^{\alpha, {\rm
      eff}*}(\mathbf{r},\mathbf{r'})\, u^{\beta}_{i}(\mathbf{r')}
  \nonumber\\ = -E_{i}\, v^{\alpha}_{i}(\mathbf{r})
\end{eqnarray}
where
\begin{eqnarray}
    V_{\rm eff}(\mathbf{r}) & = & V_{\rm ext}(\mathbf{r}) + V_{\rm
      SOC}(\mathbf{r}) +\int d^{3}r'\,
    \frac{\rho(\mathbf{r'})}{\vert\mathbf{r} - \mathbf{r'}\vert}
    \nonumber\\ & & + \frac{\delta E_{\rm xc}[\rho,\chi]}{\delta
      \rho(\mathbf{r})},
\end{eqnarray}
\begin{eqnarray}
   \Delta_{\rm eff}(\mathbf{r},\mathbf{r'}) & = & \sum_{j=0}^{3}
   D^{j}(\mathbf{r},\mathbf{r'})\, {\bf \sigma}_{j}\, i{\bf
     \sigma}_{2},
   \end{eqnarray}
and
\begin{eqnarray}
   D^{j}(\mathbf{r},\mathbf{r'}) & =& D^{j}_{\rm
     ext}(\mathbf{r},\mathbf{r'}) -\frac{\delta E_{\rm xc}[\rho,
       \chi]}{\delta \chi^{j*}(\mathbf{r},\mathbf{r'})}.
\end{eqnarray}
Here, the index $j$ refers to the singlet ($j=0$) and triplet
($j=1,2,3$) components of the anomalous density function and the
electron-electron interaction. The exchange-correlation functional is
a combination of the normal-state functional, which depends on the
normal electron density, and an anomalous part, which depends on the
interaction kernel $\lambda^{j}(\mathbf{r})$ and the anomalous density
$\chi^{j}(\mathbf{r})$.
\begin{eqnarray}
 E_{\rm xc}[\rho,\chi] & = & E^{0}_{\rm xc}[\rho, \chi] \nonumber \\ &
 & -\ \int d^{3}r\,
 \sum_{i}\chi^{j*}(\mathbf{r})\lambda^{j}(\mathbf{r})\chi^{j}(\mathbf{r}). \nonumber
\end{eqnarray}
The general expression for the effective order parameter is very
complicated, but Ref.~\cite{Suvasini1993} reduces it to a local
spin-generalized form, $\chi^{j}({\bf r},{\bf r'}) = \chi^{j}({\bf
  r})\delta({\bf r},{\bf r'})$ and $\Delta_{\rm eff}({\bf r})=
\lambda^{j} \chi^{j}({\bf r})$, where $\lambda^{j}({\bf r})$ is a
local pairing interaction (kernel) for singlet and triplet pairings
($j=0,1,2,3$).

SIESTA-BdG has three options for solving DFT-BdG secular equations,
which differ in the extent to which self-consistency is achieved in
obtaining $\Delta$: (i) no self-consistency (non SCF-BdG), (ii) fixed
$\Delta$, and (iii) full SCF-BdG. In all cases, the superconducting
pairing $\lambda$ does not include the contribution of
electron-electron pairing mechanism, which require a separate
computation.

Reference~\cite{Reho2024} benchmarked their simulations of the
superconducting order parameter, the proximity effect, and Andreev
reflections against another approach based on
Ref.~\cite{Oliveira1988}. R\"ussmann {\it et al.}  \cite{Russmann2022,
  Russmann2023} implemented DFT-BdG using a Korringa-Kohn-Rostoker
Green's function-based method, which is also a potential method to
complement NEGF simulations of surfaces and low-dimensional systems,
although here the electronic structure must be projected to an LCAO
basis. In a follow-up study, Reho {\it et al}.~\cite{Reho2025} applied
SIESTA-BdG approach to model interface-driven superconductivity in a
FeSe monolayer placed on a SrTiO$_{3}$ substrate. Their simulations
demonstrate anisotropic superconductivity with multiple coherence
peaks. They are also able to distinguish the atomic orbitals
responsible for superconductivity in the heterostructure. The ability
of SIESTA-BdG~\cite{Reho2024} to model superconductivity of an
interface with such a complicated electronic structure in an
atomic-orbital resolved form suggests that it could provide a powerful
tool for building ingredients of NEGF simulations of exotic JJs.


\section{Stationary ac Regime (Finite Bias)}
\label{sec:ac-regime}

As shown in Sec.~\ref{sec:NEGF}, the derivations involving SNS
junctions are parallel to those for SS junctions. Therefore, in the
interest of brevity, we will illustrate the ac regime by focusing on
the derivation for the SS junctions without spin-orbit coupling by
adopting the approach of Refs.~\cite{Yeyati1996} and
\cite{Cuevas1996}, and including previously omitted details. We begin
by rewriting Eq.~(\ref{eq:current-GF-LR}) as
\begin{eqnarray}
    I(t) & = & \frac{e}{\hbar} \int dt_1 \left\{ \mathrm{Tr} \left[
      V_{RL}(t,t_1)\, G_{LR}^<(t_1,t) \right] \right. \nonumber \\ & &
    \left. -\ \mathrm{Tr} \left[ V_{LR}(t,t_1) \, G_{RL}^<(t_1,t)
      \right] \right\},
    \label{eq:I_RL-ac}
\end{eqnarray}
where $V_{a,a'}(t,t_1) = U_{a,a'}(t)\, \tau_3\, \delta(t-t_1)$. As
discussed in Sec.~\ref{sec:gaugeout}, we gauge out the voltage
bias. Using Dyson's equation and the Langreth rules
\cite{Haug-Jauho2008}, we can obtain an equation for the lesser
Green's function,
\begin{eqnarray}
    G_{a,b}^< & = & g_{a,b}^< + \sum_{c,d} \left[ g_{a,c}^{\rm r}\,
      V_{c,d}^{\rm r}\, G_{d,b}^< + g_{a,c}^{\rm r}\, V_{c,d}^<\,
      G_{d,b}^{\rm a} \right. \nonumber \\ & & \left. +\ g_{a,c}^<\,
      V_{c,d}^{\rm a}\, G_{d,b}^{\rm a} \right],
    \label{eq:Glesser-ac}
\end{eqnarray}
where we dropped the time arguments and integrals for simplicity. In
Eq.~(\ref{eq:Glesser-ac}) and hereafter, lower case $g$ denotes a bare
Green's function (i.e., in the absence of coupling $U$), and $a$, $b$,
$c$, and $d$ are indices for lead $R$ and $L$ channels. For
simplicity, we will drop these indices and only indicate their domain.

One can show that the lesser Green's function satisfies the general
relation \cite{Haug-Jauho2008}
\begin{equation}
    G^< = \left( 1 + G^{\rm r}\, V^{\rm r} \right)\, g^{\rm r} \left(
    1+ V^{\rm a}\, G^{\rm a} \right) + G^{\rm r}\, V^< G^{\rm a}.
    \label{eq:G_lesser-ac}
\end{equation}
Because $V$ involves tunneling between different leads, it has
non-vanishing retarded and advanced components only when the lead
indices are from different domains ($R$ or $L$). Then, the lesser
Green's function between the left and right leads in
Eq.~(\ref{eq:I_RL-ac}) can be extracted from
Eq.~(\ref{eq:G_lesser-ac}):
\begin{eqnarray}
    G_{L,R}^< & = & g^{<}_{L,R} + \left[ G^{\rm r}\, V^{\rm r}\,
      g^{\rm r} \right]_{L,R} + \left[ g^{\rm r}\, V^{\rm a}\, G^{\rm
        a} \right]_{L,R} \nonumber \\ & & +\ \left[ G^{\rm r}\, V^{\rm
        r}\, g^{\rm r}\, V^{\rm a}\, G^{\rm a} \right]_{L,R}.
\end{eqnarray}
Note that the bare Green's functions vanish for different
lead domains since $g_{a,b} = g_{a}\, \delta_{a,b}$. Using this
property, we arrive at
\begin{eqnarray}
    G_{L,R}^< & = & G_{L,L}^{\rm r}\; V_{L,R}^{\rm r}\; g_{R}^< +
    g_{L}^<\; V_{L,R}^{\rm a}\; G_{R,R}^{\rm a} \nonumber \\ & &
    +\ G_{L,L}^{\rm r}\; V_{L,R}^{\rm r}\, g_{R}^<\; V_{R,L}^{\rm a}\;
    G_{L,R}^{\rm a} \nonumber \\ & & +\ G_{L,R}^{\rm r}\; V_{R,L}^{\rm
      r}\; g_{L}^<\; V_{L,R}^{\rm a}\; G_{R,R}^{\rm a}.
    \label{eq:G_LR_lesser-ac}
\end{eqnarray}
The retarded and advanced Green's functions can be written in a
similar way. Here, we provide two equivalent expressions for these
functions for the left lead:
\begin{eqnarray}
G_{L,L}^{\rm r/a} & = & g_{L}^{\rm r/a} + g_{L}^{\rm r/a}\;
V_{L,R}^{\rm r/a}\; G_{R,L}^{\rm r/a} \label{eq:GLL-1}\\ G_{L,L}^{\rm
  r/a} & = & g_{L}^{\rm r/a} + G_{L,R}^{\rm r/a}\; V_{R,L}^{\rm r/a}\;
g_{L}^{\rm r/a}. \label{eq:GLL-2}
\end{eqnarray}
By multiplying Eq.~(\ref{eq:GLL-1}) by $V_{LR}^{\rm r/a}$ from the
right and Eq.~(\ref{eq:GLL-2}) by $V_{RL}^{\rm r/a}$ from the left, we
obtain
\begin{eqnarray}
    G_{L,L}^{\rm r/a}\; V_{L,R}^{\rm r/a} & = & g_{L}^{\rm r/a} \; T^{\rm r/a}_{L,R}, \\
    V_{R,L}^{\rm r/a}\; G_{L,L}^{\rm r/a} & = & T^{\rm r/a}_{R,L}\; g_{L}^{\rm r/a},
\end{eqnarray}
where $T^{\rm r/a}_{a,b}$ represents the dressed tunneling matrix,
\begin{equation}
    T^{\rm r/a}_{a,b} \equiv V_{a,b}^{\rm r/a} + V_{a,b}^{\rm r/a} \;
    G_{b,a}^{\rm r/a}\; V_{a,b}^{\rm r/a},
    \label{eq:T_ab}
\end{equation}
and $a\neq b$. We also provide another expression derived from Dyson's
equation:
\begin{equation}
G_{R,L}^{{\rm r}/{\rm a}} = g_R^{{\rm r}/{\rm a}}\; V_{R,L}^{{\rm
    r}/{\rm a}}\; G_{L,L}^{{\rm r}/{\rm a}}.
\end{equation}
Using this relation and Eq.~(\ref{eq:GLL-1}) in Eq.~(\ref{eq:T_ab}),
we arrive at
\begin{equation}
T^{{\rm r}/{\rm a}}_{L,R} = V^{{\rm r}/{\rm a}}_{L,R} + V^{{\rm
    r}/{\rm a}}_{L,R}\; g_R^{{\rm r}/{\rm a}}\; V_{R,L}^{{\rm r}/{\rm
    a}}\; g_L^{{\rm r}/{\rm a}}\; T_{L,R}^{{\rm r}/{\rm a}}.
\label{eq:T-LR}
\end{equation}
Similarly, we can derive a self-consistent relation for the other
dressed tunneling matrix,
\begin{equation}
T^{{\rm r}/{\rm a}}_{R,L} = V^{{\rm r}/{\rm a}}_{R,L} + V^{{\rm
    r}/{\rm a}}_{R,L}\; g_L^{{\rm r}/{\rm a}}\; V_{L,R}^{{\rm r}/{\rm
    a}}\; g_R^{{\rm r}/{\rm a}}\; T_{R,L}^{{\rm r}/{\rm a}}.
\label{eq:T-RL}
\end{equation}

Going back to Eq.~(\ref{eq:G_LR_lesser-ac}) and inserting the dressed
tunneling, we obtain
\begin{eqnarray}
    G_{L,R}^< & = & g_{L}^{\rm r}\; T_{L,R}^{\rm r}\; g_{R}^< +
    g_{L}^<\; \hat{T}_{L,R}^{\rm a}\; g_{R}^{\rm a} \nonumber \\ & &
    +\ g_{L}^{\rm r}\; T_{L,R}^{\rm r}\; g_{R}^<\; V_{R,L}^{\rm a}\;
    \\ & & +\ g_{L}^{\rm r}\; T_{L,R}^{\rm r}\; g_{R}^<\; V_{R,L}^{\rm
      r}\; g_{L}^<\; T_{L,R}^{\rm a}\; g_{R}^{\rm a}.
    \label{eq:G_LR_lesser}
\end{eqnarray}
Using Eq.~(\ref{eq:T_ab}) as well as the Dyson equation for
$G_{a,b}^{\rm r,a}$ for $a\neq b$, Eq.~(\ref{eq:G_LR_lesser}) can be
simplified to
\begin{eqnarray}
    G_{L,R}^< & = & g_{L}^{\rm r}\; T_{L,R}^{\rm r} \; g_{R}^<\,
    T_{R,L}^{\rm a} \left[ V_{R,L}^{\rm a} \right]^{-1} \nonumber \\ &
    & +\ \left[ V_{R,L}^{\rm a} \right]^{-1}\; T_{R,L}^{\rm r} \;
    g_{L}^<\; T_{L,R}^{\rm a}\; g_{R}^{\rm a}.
    \label{eq:GLR-lesser}
\end{eqnarray}
The lesser Green's function $G_{R,L}^<$ can be obtained from
$G_{L,R}^<$ by exchanging $R$ and $L$, namely,
\begin{eqnarray}
    G_{R,L}^< & = & g_{R}^{\rm r}\; T_{R,L}^{\rm r}\; g_{L}^<\;
    T_{L,R}^{\rm a} \left[ V_{L,R}^a \right]^{-1} \nonumber \\ & &
    +\ \left[ V_{L,R}^{\rm a} \right]^{-1}\; T_{L,R}^{\rm r}\;
    g_{R}^<\; T_{R,L}^{\rm a}\; g_{L}^{\rm a}.
    \label{eq:GRL-lesser}
\end{eqnarray}
Finally, going back to Eq.~(\ref{eq:I_RL-ac}) and using
Eqs.~(\ref{eq:GLR-lesser}) and (\ref{eq:GRL-lesser}), we arrive at an
expression for the Josephson current in terms of dressed tunneling
matrices,
\begin{eqnarray}
    I(t) & = & \frac{e}{\hbar} \left\{ \mathrm{Tr} \left[ g_{L}^{\rm
        r}\, T_{L,R}^{\rm r}\, g_{R}^<\, T_{RL}^{\rm a}\, \right](t)
    \right. \nonumber \\ & & \left. +\ \mathrm{Tr} \left[ T_{R,L}^{\rm
        r}\; g_{L}^<\; T_{L,R}^{\rm a}\; g_{R}^{\rm a} \right](t)
    \right. \nonumber \\ & & \left. -\ \mathrm{Tr} \left[ g_{R}^{\rm
        r}\; T_{R,L}^{\rm r}\; g_{L}^<\; T_{L,R}^{\rm a} \right](t)
    \right. \nonumber \\ & & \left. -\ \mathrm{Tr} \left[ T_{L,R}^{\rm
        r}\; g_{R}^<\; T_{R,L}^{\rm a}\; g_{L}^{\rm a} \right] (t)
    \right\},
    \label{eq:I_ac_time}
\end{eqnarray}
where
\begin{eqnarray}
& &  {\rm Tr} \left[ A\, B\, C\, D \right](t) \nonumber \\ = & & {\rm
    Tr} \int dt_1\, dt_2\, dt_3\, A(t,t_1)\, B(t_1,t_2)\, C(t_2,t_3)\,
  D(t_3,t) \nonumber \\ & & 
\end{eqnarray}
and the trace runs over the spinor and site indices.

It is convenient to express the time-dependent functions in
Eq.~(\ref{eq:I_ac_time}) in terms of Fourier components. We start by
introducing a mixed time-energy representation,
\begin{equation}
G(t, \varepsilon) \equiv \int_{-\infty}^{\infty} dt' e^{i \varepsilon
  (t - t')}\, G( t, t'),
\end{equation}
In the stationary ac regime, we expect $G(t,\varepsilon)$ to be
periodic in time; let $T=2\pi/\omega$ be its period. Thus, a discrete
Fourier transform can be introduced:
\begin{equation}
G(t, \varepsilon) = \sum_{n=-\infty}^\infty G^{(n)}(\varepsilon)\,
e^{-in\omega t},
\end{equation}
resulting in
\begin{equation}
G(t, t') = \sum_{n=-\infty}^\infty \int_{-\infty}^{\infty}
\frac{d\varepsilon}{2\pi}\, e^{-i \varepsilon (t - t') - in\omega t}\,
G^{(n)}(\varepsilon).
\end{equation}
In this decomposition, the variable $\varepsilon$ can be confined
within the range set by $\omega$:
\begin{eqnarray}
G(t,t') & = & \sum_{n,m=-\infty}^\infty \int_{-\omega/2 +
  m\omega}^{\omega/2 + m\omega} \frac{d\varepsilon}{2\pi}\,
e^{-i(\varepsilon + m\omega)(t - t') - in\omega t} \nonumber \\ & &
\times G^{(n)}(\varepsilon + m\omega),
\end{eqnarray}
so that we can define a Floquet matrix representation of the $G$ function,
\begin{equation}
G_{n,m}^{\rm F}(\varepsilon ) \equiv G^{(n - m)}(\varepsilon  + m\omega ).
\end{equation}
The two-time function and the associated Floquet matrix are related
as follows:
\begin{eqnarray}
G\left(t,t^\prime\right) & = & \sum_{n,m=-\infty}^\infty \int_{-
  \omega /2}^{\omega /2} \frac{d\varepsilon}{2\pi}\, e^{ -
  i(\varepsilon + n\omega )t + i(\varepsilon + m\omega )t^\prime}
\nonumber \\ & & \times G_{n,m}^{\rm F}(\varepsilon )
\end{eqnarray}
and
\begin{eqnarray}
G_{n,m}^{\rm F}(\varepsilon ) & = & \int_{- \infty }^\infty dt^\prime
\int_{- \pi/\omega}^{\pi/\omega} \frac{dt}{(2\pi/\omega)}\,
e^{i(\varepsilon + n\omega )t - i(\varepsilon + m\omega )t^\prime}
\nonumber \\ & & \times G\left(t,t^\prime\right).
\label{eq:GF_double_floquet}
\end{eqnarray}
Since the Green's function oscillates with the same frequency as the
coupling amplitude $u_{a\sigma,a'\sigma'}(t)$ [see
Eq.~(\ref{eq:phase-transfer-S-S})], we identify $\omega=\omega_J/2$.


Returning to the Josephson current, we expect it to contain all
harmonics of the Josephson frequency $\omega_J$, allowing us to
perform the decomposition
\begin{equation}
    I(t) = \sum_{m=-\infty}^{\infty} I_m\, e^{i m \omega_J t}.
\end{equation}
Applying the double Fourier transformation developed above to
Eq.~(\ref{eq:I_ac_time}), the Josephson current harmonics can be
written in terms of a sum and an integration over the four Floquet
matrices $F_{n,m}^{(k)}(\varepsilon)$, $k=1,2,3,4$:
\begin{equation}
    I_m = \frac{e}{\hbar} \sum_{n=-\infty}^{\infty} \int \frac{d\varepsilon}{2\pi}
    \left[ \sum_{k=1}^4 F_{n,m}^{(k)}(\varepsilon)  \right],
    \label{eq:I_m_exact}
\end{equation}
where
\begin{eqnarray}
F_{n,m}^{(1)} & = & \left[ g_{L}^{\rm r} \right]^F_{0,0} \left[
  T_{L,R}^{\rm r} \right]^F_{0,n} \left[ g_{R}^< \right]^F_{n,n}
\left[ T_{R,L}^{\rm a} \right]^F_{n,m}, \\ F_{n,m}^{(2)} & = & \left[
  T_{R,L}^{\rm r} \right]^F_{0,n} \left[ g_{L}^< \right]^F_{n,n}
\left[ T_{L,R}^{\rm a} \right]^F_{n,m} \left[ g_{R}^{\rm a}
  \right]^F_{m,m}, \\ F_{n,m}^{(3)} & = & - \left[ g_{R}^{\rm r}
  \right]^F_{0,0} \left[ T_{R,L}^{\rm r} \right]^F_{0,n} \left[
  g_{L}^< \right]^F_{n,n} \left[ T_{L,R}^{\rm a} \right]^F_{n,m},
\\ F_{n,m}^{(4)} & = & - \left[ T_{L,R}^{\rm r} \right]^F_{0,n} \left[
  g_{R}^< \right]^F_{n,n} \left[ T_{R,L}^{\rm a} \right]^F_{n,m}
\left[ g_{L}^{\rm a} \right]^F_{m,m},
\end{eqnarray}
where the dependence on $\varepsilon$ present in each term is left
implicit.

Equation~(\ref{eq:I_m_exact}) represents a methodological departure
from the approach used to obtain the expression in
Eq.~(\ref{eq:supercurrent-2}) for the dc Josephson current. Instead of
dressing the Green's function of the non-superconducting region with
the coupling to the leads, we dressed the couplings themselves. This
approach is advantageous for the ac case, as we illustrate below.

Since the lead Green's functions $[g_a^{{\rm r},{\rm a},<}]^F_{n,n}$
can be computed analytically in the single-channel case or numerically
in the multi-channel case, the main challenge in obtaining the Floquet
matrices $F_{n,m}^{(k)}$ is to compute the dressed tunneling Floquet
matrices $[T_{a,b}^{{\rm a}/{\rm r}}]^F_{n,m}$. Below, following
Ref.~{\cite{Yeyati1996}}, we provide a method to obtain these
matrices.

\subsection{Computation of dressed tunneling matrices}

Let us consider the case of single-channel, identical superconducting
leads in the absence of spin-orbit coupling, when the lead Green's
function can be obtained from Eq.~(\ref{eq:GF-surface}) by an
appropriate analytical continuation:
\begin{equation}
\left[ g^{{\rm r},{\rm a}} \right]^F_{n,n}(\varepsilon)
=-\frac{\pi\rho(0)\, \left[ (\varepsilon + n\omega_J/2) \tau_0 +
    \Delta \tau_1 \right]}{\sqrt{\Delta^2 - (\varepsilon + n\omega_J/2
    \pm i0^+)^2}}.
\end{equation}
Assuming quasi-equilibrium, when $\hbar\omega_J \ll k_BT$, we can use
Eq.~(\ref{eq:fluct-dissip-theor}) to write the lead's lesser Green's
function in terms of the retarded and advanced ones, namely,
\begin{eqnarray}
\left[ g^< \right]^F_{n,n}(\varepsilon) & = & f(\varepsilon+n\omega_J/2)
\left\{ \left[ g^{\rm a} \right]^F_{n,n}(\varepsilon) - \left[ g^{\rm
    r} \right]^F_{n,n}(\varepsilon) \right\}, \nonumber \\ & & 
\end{eqnarray}
where $f(\varepsilon)$ is the Fermi-Dirac distribution. To obtain the
dressed tunneling matrices, we need to rewrite Eqs.~(\ref{eq:T-LR})
and (\ref{eq:T-RL}) in the Floquet representation. The calculation is
rather long; result is that those matrices satisfy the recurring
equation
\begin{eqnarray}
[T_{L,R}^{{\rm r}/{\rm a}}]_{n,m}^F(\varepsilon) & = & t_{n,m} +
\varepsilon^{{\rm r}/{\rm a}}_{n}(\varepsilon)\, [T_{L,R}^{{\rm
      r}/{\rm a}}]^F_{n,m}(\varepsilon) \nonumber \\ & & +\ v^{{\rm
    r}/{\rm a}}_{n,n+2}(\varepsilon)\, [T_{L,R}^{{\rm r}/{\rm
      a}}]^F_{n+2,m}(\omega) \nonumber \\ & & +\ v^{{\rm r}/{\rm
    a}}_{n,n-2}(\varepsilon)\, [T_{L,R}^{{\rm r}/{\rm
      a}}]^F_{n-2,m}(\varepsilon),
\label{eq:T-recurr}
\end{eqnarray}
where
\begin{equation}
t_{n,m} = u_+^\dagger \delta_{n,m+1} + u_-^\dagger \delta_{n,m-1},
\end{equation}
\begin{eqnarray}
\varepsilon^{{\rm r}/{\rm a}}_{n}(\varepsilon) & = & u_+\, [g^{{\rm
      r}/{\rm a}}]^F_{n-1,n-1}(\varepsilon)\, u_+\, [g^{{\rm r}/{\rm
      a}}]^F_{n,n}(\varepsilon) \nonumber \\ & & +\ u_-\, [g^{{\rm
      r}/{\rm a}}]^F_{n+1,n+1}(\varepsilon)\, u_-\, [g^{{\rm r}/{\rm
      a}}]^F_{n,n}(\varepsilon),
\end{eqnarray}
and
\begin{eqnarray}
v^{{\rm r}/{\rm a}}_{n,n\pm 2}(\varepsilon) & = & u_\mp\, [g^{{\rm r}/{\rm
      a}}]^F_{n\pm 1,n\pm 1}(\varepsilon)\, u_\pm\, [g^{{\rm r}/{\rm
      a}}]^F_{n\pm 2,n\pm 2}(\varepsilon), \nonumber \\ & & 
\end{eqnarray}
with the connecting matrices defined in terms of Pauli matrices,
\begin{equation}
u_\pm = (\tau_0 \pm \tau_3)\, u/2 .
\end{equation}
An equation similar to Eq.~(\ref{eq:T-recurr}) can be derived for
$[T_{R,L}^{{\rm r}/{\rm a}}]^F_{n,m}(\varepsilon)$, but it is not
necessary since
\begin{equation}
[T_{R,L}^{{\rm r}/{\rm a}}]^F_{n,m}(\varepsilon) = \left(
[T_{L,R}^{{\rm a}/{\rm r}}]^F_{m,n}(\varepsilon) \right)^\dagger.
\end{equation}

Notice that Eq.~(\ref{eq:T-recurr}) is equivalent to the standard
expression connecting wavefunction site amplitudes for a
one-dimensional tight-binding model, with $n$ being the site index,
$\varepsilon_n$ representing the on-site ``energy", and $v_{n,n\pm2}$
representing the ``hopping amplitudes". Equation~(\ref{eq:T-recurr})
can be solved using the ansatz \cite{Yeyati1996}
\begin{eqnarray}
\label{eq:T-ansatz1}
[T]_{n+2,m}^F(\varepsilon) & = & z^+_{n-1}(\varepsilon)\, [T]_{n,m}^F(\varepsilon), \ \ (n\geq 1), \\
\label{eq:T-ansatz2}
[T]_{n-2,m}^F(\varepsilon) & = & z^-_{n+1}(\varepsilon)\, [T]_{n,m}^F(\varepsilon), \ \ (n\leq-1),
\end{eqnarray}
where the transfer matrix satisfies the equation
\begin{eqnarray}
z_{n}^\pm(\varepsilon) & = & \left[ \tau_0 - \varepsilon_{n\pm
    3}(\varepsilon) - v_{n\pm 3,n\pm 5}(\varepsilon)\, z^\pm_{n\pm
    2}(\varepsilon) \right]^{-1} \nonumber \\ & & \times v_{n\pm
  3,n\pm 1}(\varepsilon).
\label{eq:transf-mat}
\end{eqnarray}
For brevity, we omitted several obvious superscripts and subscripts
above, but they can be easily reintroduced. To solve for $[T]_{n,m}^F$
for a given $m$, we first set $n=m\pm 1$ and use
Eq.~(\ref{eq:T-recurr}) and the ansatz to find two coupled linear
equations for $[T]_{m+1,m}^F$ and $[T]_{m-1,m}^F$, namely,
\begin{eqnarray}
[T]_{m+1,m}^F & = & u_+ + \varepsilon_{m+1}\, [T]^F_{m+1,m} \nonumber
\\ & & + v_{m+1,m+3}\, z^+_{m}\, [T]_{m+1,m}^F \nonumber \\ & & +
v_{m+1,m-1}\, [T]_{m-1,m}^F
\label{eq:T-sys1}
\end{eqnarray}
and
\begin{eqnarray}
[T]_{m-1,m}^F & = & u_- + \varepsilon_{m-1}\, [T]^F_{m-1,m} \nonumber
\\ & & + v_{m-1,m+1}\, [T]_{m-1,m}^F \nonumber \\ & & + v_{m-1,m-3}\,
z^-_{m}\, [T]_{m-1,m}^F.
\label{eq:T-sys2}
\end{eqnarray}
Once we obtain $[T]^F_{m\pm 1,m}$, the Floquet dressed matrices for
$n>m+1$ and $n<m-1$ can be obtained recursively using the ansatz. In
fact, we only need to solve this system of equations for the case
$m=0$ since $[T]^F_{n,m}(\varepsilon) =
[T]^F_{n-m,0}(\varepsilon+m\omega_J/2)$ [see
  Eq.~(\ref{eq:GF_double_floquet})]; the dressed matrices for $m>0$
can be obtained using this relation.

The main challenge is to solve Eq.~(\ref{eq:transf-mat}). In
Ref.~\cite{Cuevas1996}, the matrices $z^\pm_n(\varepsilon)$ were
assumed to be diagonal and expressed in terms of a set of scalar
functions $\{\lambda_k(\varepsilon)\}$ which satisfy a recurrence
relation. In the general case, this relation can only be solved
numerically after a truncation criterion has been established; in the
limits of very high and very low bias voltages, analytical solutions
have been obtained \cite{Cuevas1996}.

\subsection{Example: Single-channel ac Josephson junction}

Here we provide an example where Eq.~(\ref{eq:transf-mat}) is solved
numerically and the solution is used to obtain the dependence of the
ac Josephson current amplitudes on the bias voltage, temperature, and
tunneling amplitude in the single-channel case.

First, we notice that $|v_{n,n\pm 2}(\varepsilon)| \sim O(1/|n|)$ and
$|\varepsilon_n(\varepsilon)| \sim O(1)$ for $|n|\gg 1$, which leads
to $|z_n^\pm(\varepsilon)| \sim O(1/|n|)$. Therefore, for a fixed
energy $\varepsilon$ and a fixed harmonic index $m$, we can truncate
the recurrence for $z_n^+(\varepsilon)$ by setting
$v_{N+3,N+5}(\varepsilon)=0$ and $v_{N+4,N+2}(\varepsilon)=0$ in
Eq.~(\ref{eq:transf-mat}) for some $N\geq m$. Within this
approximation, $z^+_{n}(\varepsilon)=0$ for $n>N$,
\begin{equation}
z^+_{N}(\varepsilon) = \left[ \tau_0 - \varepsilon_{N+3}(\varepsilon)
  \right]^{-1} v_{N+3,N+1}(\varepsilon),
\end{equation}
and
\begin{equation}
z^+_{N-1}(\varepsilon) = \left[ \tau_0 -
  \varepsilon_{N+2}(\varepsilon) \right]^{-1} v_{N+2,N}(\varepsilon)
\end{equation}
From $z^+_{N}(\varepsilon)$ and $z^+_{N-1}(\varepsilon)$, we can
obtain all the other matrices down to $z^+_{0}(\varepsilon)$ by virtue
of Eq.~(\ref{eq:transf-mat}). Similarly, by setting
$v_{-N-3,-N-5}(\varepsilon)=0$ and $v_{-N-4,-N-2}(\varepsilon)=0$ in
Eq.~(\ref{eq:transf-mat}), results in $z^-_{-n}(\varepsilon)=0$ for
$n>N$,
\begin{equation}
z^-_{-N}(\varepsilon) = \left[ \tau_0 -
  \varepsilon_{-N-3}(\varepsilon) \right]^{-1}
v_{-N-3,-N-1}(\varepsilon),
\end{equation}
and
\begin{equation}
z^+_{-N+1}(\varepsilon) = \left[ \tau_0 -
  \varepsilon_{-N-2}(\varepsilon) \right]^{-1}
v_{-N-2,-N}(\varepsilon),
\end{equation}
from which we can obtain all $z^-_{-n}(\varepsilon)$ matrices down to
$z^-_0(\varepsilon)$. Inserting $z^\pm_{0}(\varepsilon)$ into
Eqs.~(\ref{eq:T-sys1}) and (\ref{eq:T-sys2}), we can solve these
equations to find $[T]^F_{\pm 1,0}(\varepsilon)$. All other dressed
tunneling matrices $[T]^F_{n,m}$ within the range $-N-2 \leq n \leq
N+2$ follow from the recurrence in Eqs.~(\ref{eq:T-ansatz1}) and
(\ref{eq:T-ansatz2}).

\begin{figure*}
  \centering
  \includegraphics[width=0.8\textwidth]{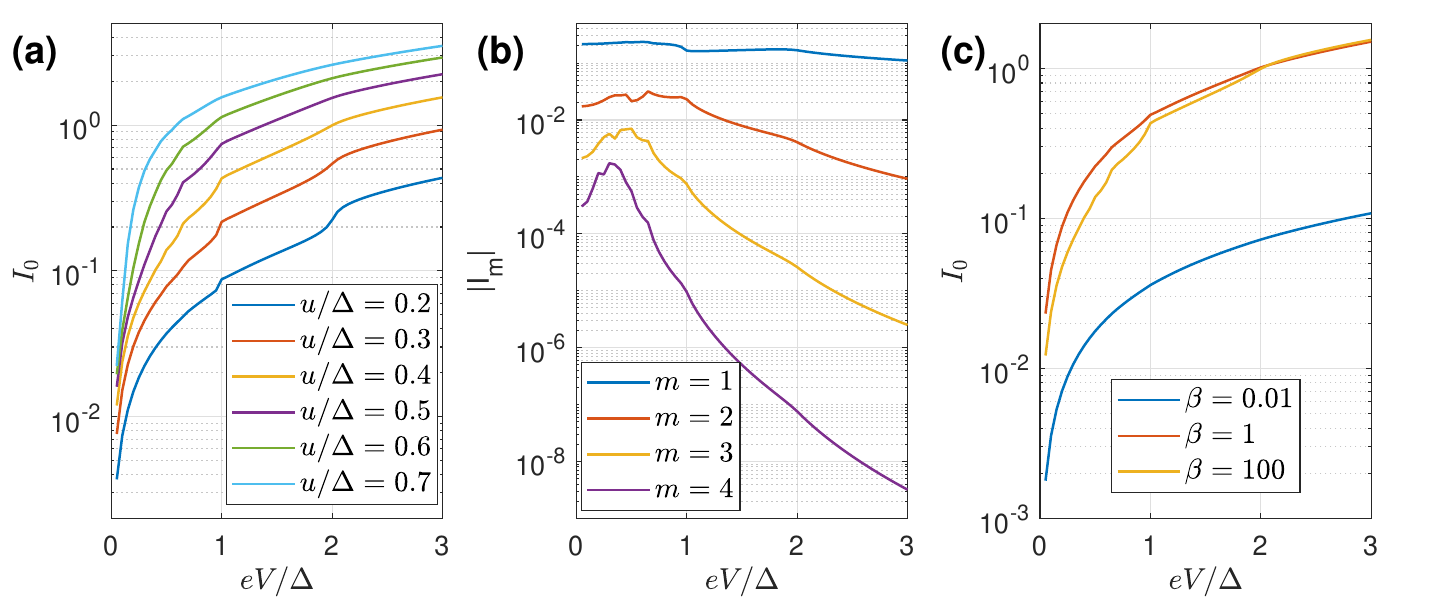}
  \caption{Josephson current in the ac (voltage biased) regime. (a) dc
    ($m=0$) component of the current as a function of the bias voltage
    for increasing tunneling coupling strengths $u/\Delta$ in
    increments of 0.1 at temperature $k_BT=0.01\Delta$. (b) First four
    harmonic components of the supercurrent as a function of the bias
    voltage for $u=0.4\Delta$ and $k_BT=0.01\Delta$. (c) The dc
    ($m=0$) component of the current as a function of the bias voltage
    for three different temperatures ($\beta\equiv\Delta/k_B T$) for
    $u=0.4\Delta$. All computations used an $N=6$ truncation.}
  \label{fig:AC}
\end{figure*}

Once the dressed matrices are obtained for a wide-enough range of
energies, the Floquet matrices $F_{n,m}^{(k)}$ are assembled and
Eq.~(\ref{eq:I_m_exact}) is used to compute the $m$-th component of
the ac Josephson current. In Fig.~\ref{fig:AC} we present the results
of this calculation. The dc ($m=0$) component is shown as a function
of the bias voltage for various values of the tunneling amplitude
(Fig.~\ref{fig:AC}a) and temperature
(Fig.~\ref{fig:AC}c). Figure~\ref{fig:AC}b shows the bias voltage
dependence of different components. Notice the fast decay of the
magnitude of these components with increasing $m$. In these numerical
calculations, the truncation number $N=6$ is used.

Before concluding this section, we briefly discuss the differences
between the NEGF and the scattering-matrix approaches
\cite{Averin_Bardas_1995,averin1998theoryacjosephsoneffect}.

In the scattering-matrix formulation, the ac Josephson current is
obtained by summing the contributions from all possible scattering
processes involving electron and hole quasiparticles. For a given
scattering configuration, the current can be evaluated from the
quasiparticle wavefunctions at a single interface between a
superconducting lead and the normal scattering region, owing to the
continuity of the current. The central task is therefore to determine
the electron and hole wavefunctions at the interface, from which the
current contribution associated with that scattering process can be
computed. These wave functions are fixed by two physical
constraints. (i) Andreev reflection at the superconductor-normal
interfaces: At each boundary, an incoming electron (hole) can be
converted into an outgoing hole (electron), with the corresponding
amplitudes related by the Andreev reflection processes
\cite{Beenakker1991}. This mechanism locally couples the electron and
hole modes at the interface. And (ii) voltage bias $V$ applied between
the left and right superconducting leads: The bias produces a
time-dependent superconducting phase difference, which results in
additional phase factors acquired during propagation between the two
interfaces. Specifically, an electron mode propagating from the left
to the right lead picks up a phase factor $e^{i\omega_J t/2 }$, while
a hole mode propagating in the same direction acquires the conjugate
phase factor $e^{-i\omega_J t/2}$. These opposite phase factors
reflect the opposite charges carried by electrons and holes.

Once the Andreev reflection conditions at the interfaces and the
voltage-induced phase accumulation during propagation are properly
accounted for, the wavefunctions of the electron and hole modes at the
interfaces are completely determined by the scattering matrix of the
normal region. As a result, all transport properties, including the
time-dependent Josephson current \cite{Averin_Bardas_1995} and shot
noise \cite{Naveh_Averin_1999}, can be expressed entirely in terms of
the normal-state scattering matrix and the Andreev reflection
amplitudes.

The scattering-matrix approach is typically formulated within the
Andreev approximation, which assumes that the Fermi energy is the
largest energy scale in the problem. In contrast, the non-equilibrium
Green's function approach does not rely on this approximation and
allows for a full microscopic treatment without assuming a separation
of energy scales. Because normal-region physics is encoded compactly
in the scattering matrix and the effects of the voltage bias enter
only through phase accumulation, the scattering-matrix approach is
particularly well suited for describing long Josephson junctions. In
comparison, an NEGF calculation for a long junction becomes
computationally expensive, as the Green's function of the normal
region grows into a large matrix.

Finally, while extending the scattering-matrix approach to systems
driven by general time-dependent (ac) voltages is technically
challenging, such extensions can be implemented naturally within the
NEGF formalism, which provides a flexible framework for treating
arbitrary time-dependent driving fields \cite{Xie_2018}.


\section{Summary and Outlook}
\label{sec:summary}

We provide an up-to-date review and an in-depth discussion of Green's
function methods for the modeling of Josephson junctions and the
computation of supercurrents in related systems. Formulations suitable
for tight-binding and other real-space representations, which are
particularly suitable for realistic, large-scale modeling of materials
and subsystems involved in junctions are discussed. Both the dc (zero
bias) and the ac (biased) regimes are covered. Details of how to build
comprehensive tight-binding models for the barrier region, the
superconducting leads, and the connecting interfaces are
delineated. The methods presented here would allow one to incorporate
effects of spin-orbit coupling and multiatomic on-site orbitals in
modeling and understanding the nature of supercurrents in practical
Josephson junction systems at a material-specific atomistic level.

Our dc and ac transport formulations rely on a single-particle
approximation and a mean-field description of the superconductor order
parameter. However, as discussed in Sec.~\ref{sec:beyondSK}, recent
progress in combining DFT and BdG provides more realistic
material-specific inputs to the mean-field description, potentially
leading to quantitatively more accurate modeling of the JJ
systems. While the mean-field description is usually adequate to
capture physical phenomena at a qualitative level, the single-particle
approximation can be limiting in cases where charging effects (e.g.,
Coulomb blockade) are important in the barrier region. This aspect is
particularly relevant to some recent experiments that explore
topological Andreev bound states in quantum dots coupled to multiple
superconducting leads \cite{Antonelli2025}. From a modeling
perspective, quantum-dot-based JJ systems often do not require
atomic-level detail, in which case an effective Hamiltonian approach
is sufficient, allowing one to go beyond the mean-field approximation
\cite{Levy-Yeyati1993,Meng_thesis}. However, when atomistic details
are important, the inclusion of electron-electron interactions effects
beyond the DFT-based parameterization of the tight-binding Hamiltonian
or beyond a mean-field approximation of charging effects remains an
open problem, especially for the JJ systems. In general, adding
electron-electron interactions to the scattering formulation of
electronic transport is a daunting task. An earlier attempt had to
rely on a Green's function formulation in intermediate steps of the
derivations \cite{Oehri2012}. Therefore, it is likely that any
progress in incorporating electron-electron correlations beyond the
mean-field approximation in JJs will require a Green's function-based
formulation. We expect this to remain a vigorous area of research with
the potential to uncover new correlation-driven transport phenomena at
the intersection of strong correlations and superconductivity in
nanoscale junctions.


\section{Acknowledgments}

We are grateful to D. Averin, T. Heikkil\"a, A. Kamenev, S. Kettemann,
A. Levy Yeyati, and P. Lyu for useful discussions. This work was
primarily supported by the National Science Foundation through the
Expand-QISE award NSF-OMA-2329067 and benefited from the resources of
Northeastern University's Advanced Scientific Computation Center, the
Discovery Cluster, the Massachusetts Technology Collaborative award
MTC-22032, and the Quantum Materials and Sensing Institute
(QMSI). J.~N. benefited from the resources of the Tampere Center for
Scientific Computing (TCSC).

\vspace{1cm}

\bibliographystyle{iopart-num-threeauthors}

\bibliography{references}


\end{document}